\documentclass[aip,reprint,amsmath,amssymb,graphicx]{revtex4-1} 

\usepackage{graphicx}
\usepackage{dcolumn}
\usepackage{bm}

\usepackage[utf8]{inputenc}
\usepackage[T1]{fontenc}
\usepackage{mathptmx}
\usepackage{etoolbox}
\usepackage{bm}
\usepackage{url}
\usepackage{hyperref}
\newcommand{\Cov}{\mathrm{Cov}}

\usepackage{mathtools}
\usepackage{xcolor}

\begin{document}


\title{Separating internal and externally-forced contributions to global temperature variability using a Bayesian stochastic energy balance framework} 

\author{Maybritt Schillinger}
\affiliation{Seminar for Statistics, Department of Mathematics, ETH Zurich, Rämistrasse 101, 8092 Zurich, Switzerland}

\author{Beatrice Ellerhoff}
\affiliation{Department of Physics and Department of Geosciences, Tübingen University, Schnarrenbergstr. 94-96, 72076 Tübingen, Germany}
\email{beatrice-marie.ellerhoff@uni-tuebingen.de}

\author{Robert Scheichl}
\affiliation{Institute of Applied Mathematics and Interdisciplinary Center for Scientific Computing (IWR), Heidelberg University, Im Neuenheimer Feld 205, 69120 Heidelberg, Germany}

\author{Kira Rehfeld}
\affiliation{Department of Physics and Department of Geosciences, Tübingen University, Schnarrenbergstr. 94-96, 72076 Tübingen, Germany}

\date{\today}

\begin{abstract} 
Earth's temperature variability can be partitioned into internal and externally-forced components. Yet, underlying mechanisms and their relative contributions remain insufficiently understood, especially on decadal to centennial timescales. Important reasons for this are difficulties in isolating internal and externally-forced variability. Here, we provide a physically-motivated emulation of global mean surface temperature (GMST) variability, which allows for the separation of internal and external variations. To this end, we introduce the ``ClimBayes'' software package, which infers climate parameters from a stochastic energy balance model (EBM) with a Bayesian approach. We apply our method to GMST data from temperature observations and 20 last millennium simulations from climate models of intermediate to high complexity. This yields the best estimates of the EBM's \textit{forced} and \textit{forced + internal} response, which we refer to as emulated variability. The timescale-dependent variance is obtained from spectral analysis. In particular, we contrast the emulated \textit{forced} and \textit{forced + internal} variance on interannual to centennial timescales with that of the GMST \textit{target}. Our findings show that a stochastic EBM closely approximates the power spectrum and timescale-dependent variance of GMST as simulated by modern climate models. Small deviations at interannual timescales can be attributed to the simplified representation of internal variability and, in particular, the absence of (pseudo-)oscillatory modes in the stochastic EBM. Altogether, we demonstrate the potential of combining Bayesian inference with conceptual climate models to emulate statistics of climate variables across timescales.
\end{abstract}


\maketitle 

\begin{quotation}
Understanding the statistical properties and sources of Earth's surface temperature variations is of great importance in climate science. To this end, we analyze the variability of global mean surface temperature (GMST) with a simple stochastic energy balance model (EBM). With Bayesian methods and spectral analysis, we separate internally-generated and externally-forced contributions to GMST variations on different timescales in state-of-the-art climate model simulations. Our results show that a stochastic EBM can emulate the variability of more complex climate models. The combined use of Bayesian inference and conceptual climate models therefore provides a versatile tool to advance the understanding of internal and forced variability in Earth's dynamical system.
\end{quotation}

\section{Introduction}

Climate variability describes the spatial and temporal variations in the mean and higher order statistics of climate parameters, and is of vital importance for living conditions on Earth \cite{katz1992}. While many sources of natural variability exist, anthropogenic influences clearly dominate the recent trend in global mean surface temperature (GMST). To characterize variability, it is typically partitioned into internal and external components. Internal variability arises from intrinsic climate system processes such as oceanic and atmospheric circulation. External sources include changes in radiative forcing, for example, from solar irradiance, volcanic eruptions, and greenhouse gases. Despite a general agreement of the total simulated and observed GMST variability over the Common Era (0-2000 CE) \cite{pages2kconsortium2019, ellerhoff2021}, uncertainties remain about the mechanisms and magnitude of internal and external variations \cite{hawkins2011, frankcombe2015, hebert2018}, especially on decadal to centennial timescales \cite{laepple2014, ellerhoff2021}.

Simple mathematical models help understand climate variability \cite{budyko1969, sellers1969, hasselmann1976} and can be used to emulate climate variables from more complex model simulations. Most general, the time evolution of a forced climate parameter $X(t)$ is described by $\dot{X}(t) = \mathcal{A}(t,X(t),F(t))$ for an arbitrary operator $\mathcal{A}$ and external driver $F(t)$. We consider $X$ as the GMST, for which many studies have formulated physically-motivated approximations of $\mathcal{A}$. One pivotal approach is centered around the idea of balancing incoming and outgoing radiation \cite{budyko1969,sellers1969}, later extended to a stochastic energy balance model (EBM) by \citet{hasselmann1976}. This approach assumes the climate system close to equilibrium, showing a linear and stationary response to perturbations. Then, $\mathcal{A}$ can be approximated by a linear stochastic operator \cite{budyko1969, sellers1969}:
\begin{equation}\label{eq:stochastic_ebm}
C \frac{\mathrm{d}}{{dt}} T(t) =  - \tilde{\lambda} T(t) + F(t) + \epsilon(t) 
\end{equation}

Formula~\eqref{eq:stochastic_ebm} describes the GMST anomaly $T(t)$ with respect to the equilibrium state, given Earth’s effective heat capacity $C$, a radiative forcing anomaly $F(t)$ and a term $\epsilon (t)$, representing stochastic dynamics such as weather fluctuations. The response parameter $\lambda \coloneqq \tilde{\lambda} / C$ is the reciprocal of the characteristic timescale $1/\lambda$. The response to radiative forcing $F(t)$ determines forced temperature variations. The response to the stochastic term $\epsilon(t)$ approximates internal variability. 

The stochastic EBM \eqref{eq:stochastic_ebm} is too simplistic to accurately represent long-term responses and, therefore, has been extended to so-called multibox EBMs \cite{fraedrich2004,held2010, geoffroy2013a, rypdal2014, fredriksen2017}. The latter are based on multiple ocean layers, referred to as boxes. The layers serve to approximate the vertical heat transfer and the integrated response to forcing over long periods. The EBM \eqref{eq:stochastic_ebm} laid the basis for attributing anthropogenic warming \cite{hasselmann1993, hegerl1996}. It was applied and modified to study climate sensitivity \cite{ghil1984, cox2018,  rypdal2014, rypdal2015}, climate and ice cap stability \cite{ghil1976, north1983, bodai2015, ostvand2014}, regional temperatures \cite{north1975, north1981, north2011}, Glacial/Interglacial cycles \cite{north1983, dortmans2019}, and future projections \cite{myrvoll-nilsen2020, hebert2021}. Key advantages of EBMs are their computational efficiency and comparatively easy interpretation.

To estimate uncertain parameters of conceptual climate models from data, Bayesian frameworks have become increasingly popular \cite{bodman2016, proistosescu2017, jonko2018, sherwood2020, skeie2018}. In comparison to other methods for inferring climate parameters, such as maximum likelihood estimation, Bayesian approaches have the advantage of providing full posterior distributions. The methods compute the posterior means and credible intervals (CIs) of uncertain parameters $\theta$  conditioned on target data $y$ while including prior knowledge on $\theta$. Central to this framework is applying Bayes theorem
\begin{equation}\label{eq:BayesTheorem}
    p (\theta | y) = \frac{p(y | \theta) p(\theta)}{p(y)}\,,
\end{equation}
with likelihood $p(y | \theta)$, prior $p(\theta)$, marginal $p(y)$ and posterior $p(\theta | y)$. We denote all probability densities by $p$ and distinguish them by their arguments. Combining Bayesian inference with conceptual climate models typically also yields the posterior of the model's fit to the data.

With the ability to quantify fluctuations across timescales, power spectral analysis has improved the understanding of climate variability \cite{mitchell1976, huybers2006, vonderheydt2021, franzke2020, ellerhoff2021}. Simple climate models have been combined with spectral analysis to explain timescale-dependent variability. For example, \citet{fredriksen2017} use a multibox EBM to study temporal scaling of temperature time series. Related works examine future projections \cite{myrvoll-nilsen2020} and climate sensitivity \cite{rypdal2018}. \citet{soldatenko2022} study the sensitivity of the power spectrum on uncertainties in the parameters of a two-box EBM, considering stochastic noise but neglecting deterministic forcing. Yet, the potential of combining stochastic multibox EBMs, Bayesian inference and spectral analysis to study the magnitude of unforced and forced variability across timescales remains untapped.

Here, we examine and separate timescale-dependent internal and externally-forced contributions to the GMST variations. In particular, we analyze GMST variability during the last millennium (850-1850 CE) as simulated by 20 climate models of intermediate to high complexity.
To this end, we combine a stochastic two-box EBM (section~\ref{sec:multiboxebm}) with Bayesian inference (section~\ref{sec:bayes}) and spectral analysis (section~\ref{sec:spectralmethods}). We present the ``ClimBayes'' software package\cite{schillinger2022} for Bayesian inference of climate parameters, which fits the stochastic EBM to GMST data. This results in the best estimate of the \textit{forced} and samples of the \textit{forced + internal} EBM's temperature response. First, we demonstrate our analysis on the example of historical observations (section~\ref{sec:results_hist}) and then apply it to the considered set of last millennium simulations (section~\ref{sec:results_parameters}). 
Section~\ref{sec:results_spectra} contrasts power spectra of the fitted EBM with and without internal variations. Comparing the internal and forced variance on interannual to centennial timescales (section~\ref{sec:separation}), a stochastic two-box EBM captures most variations of more comprehensive model simulations. We summarize and discuss the potential for physics-informed emulation of GMST data and separation of variance contributions across timescales in sections~\ref{sec:discussion} and~\ref{sec:conclusion}.

\section{Data}\label{sec:data}

Our study relies on annual GMST and corresponding radiative forcing time series. We use full-forced last millennium runs from climate models of varying complexity (Tab.~\ref{tab:table0}). We analyze 10 simulations with atmosphere-ocean general circulation models (AOGCMs), considered in the Coupled Model Intercomparison Project 5 (CMIP5)\cite{jungclaus2017}. Moreover, we use 10 simulations with Earth system models of intermediate complexity (EMICs) that are part of the IPCC's Fifth Assessment Report (AR5)\cite{flato2014} and described by \citet{eby2013}. The AR5 EMICs represent single simulations, except for CLIMBER2 and LOVECLIM V.1.2, which are ensemble means and denoted by ``(mean)'' in the following. To compare variability in single ensemble members to that of the ensemble mean, we use the five available ensemble members LOVECLIM V.1.2 (E1-E5). 

The transient radiative forcing applied to these simulations follows the Paleoclimate Modelling Intercomparison Project Phase III (PMIP3) protocol \cite{schmidt2012}. For AR5 EMICs, we take the total estimated radiative forcing provided by \citet{eby2013}. For CMIP5 simulations, we use the radiative forcing from reconstructions of well-mixed greenhouse gases (CO$_2$, CH$_4$, and N$_2$O), volcanic aerosols, total solar irradiance, and land use changes as provided by \citet{schmidt2012}. We neglect orbital forcing, which is assumed to play a negligible role for GMST variability over the last millennium. To remove potential unforced drifts of the simulated background climate \cite{tachiiri2010}, the simulated GMST is linearly detrended prior to analysis. For consistency, detrending is also applied to the corresponding forcing time series. This does not affect our results, as the simulations' forcing input exhibits no trend for the last millennium (850-1850 CE). Both, temperature and forcing time series are considered as anomalies with respect to the starting year.

We use the GMST from HadCRUT5 \cite{morice2021} observations (1850-2000 CE) to demonstrate the developed workflow of our Bayesian stochastic energy balance framework. As estimates for radiative forcing during the historical period, we consider the ``PEA'' land use (\citet{pongratz2008}), ``CEA'' volcanic forcing (\citet{crowley2008}), ``SBF'' solar irradiance reconstruction (\citet{steinhilber2009}) patched into \citet{wang2005}, and greenhouse gas concentrations from \citet{schmidt2012}.

\begin{table}
\caption{\label{tab:table0} Key forcing specifications and references of considered climate model simulations. The ``Forcing'' column gives the abbreviations from the PMIP3 protocol \cite{schmidt2012}, corresponding to the implemented land use, solar, and volcanic forcing reconstructions. The land use reconstruction PEA is taken from \citet{pongratz2008}. Solar forcing reconstructions correspond to DB: \citet{delaygue2011}, VSK: \citet{krivova2007, vieira2010}, and SBF: \citet{steinhilber2009}. They are calibrated to WLS modern values (1366.14 W/m$^2$) and continued by \citet{wang2005}. Volcanic forcing refers to CEA: \citet{crowley2008} and GRA: \citet{gao2008}. Trace gases are prescribed in all simulations and follow the PMIP3 protocol \cite{schmidt2012}.}
\scriptsize
\begin{ruledtabular}
\begin{tabular}{lcr}
Climate model  & Forcing              & Reference                \\ \hline
\textbf{AR5 EMICs} & & \\
Bern 3D                & PEA, DB, CEA                      &       \citet{ritz2011}  \\
CLIMBER-3alpha            & PEA, DB, CEA                      &      \citet{montoya2006} \\
CLIMBER2              & PEA, DB, CEA                      &        \citet{petoukhov2005}\\
DCESS ESM v1                 & PEA, DB, CEA                      &         \citet{shaffer2008}\\
IGSM 2.2                 & PEA, DB, CEA                      &        \citet{sokolov2005integrated}\\
LOVECLIM V.1.2             & PEA, DB, CEA                      &          \citet{goosse2010}  \\
MESMO 1.0                 & PEA, DB, CEA                      &          \citet{matsumoto2008} \\
MIROC3-lite           & PEA, DB, CEA                      &         \citet{tachiiri2010} \\
UMD                   & PEA, DB, CEA                      &    \citet{zeng2004} \\
UVic v2.9           & PEA, DB, CEA                      &   \citet{weaver2001}     \\ 
\textbf{CMIP5 models} & & \\
BBC-CSM1-1            & -, VSK, GRA                   & \citet{xiao-ge2013}\\
CCSM4                 &  PEA, VSK, GRA                   & \citet{landrum2013}\\
CSIRO-Mk3L-1-2          &  -, SBF, CEA                 & \citet{phipps2012}\\
FGOALS-s2             & -, VSK, GRA                   & \citet{bao2013}\\
GISS-E2-R             & PEA, SBF, CEA                & \citet{schmidt2006}\\
HadCM3                & PEA, SBF, CEA                & \citet{schurer2014} \\
HadGEM2-ES             & PEA, SBF, CEA                &      \citet{jones2011}\\
IPSL-CM5A-LR          & -, VSK, GRA                   &  \citet{dufresne2013, hourdin2013}  \\
MIROC-ESM             & -, DB, CEA               &        \citet{sueyoshi2013} \\
MPI-ESM-P             & PEA, VSK, CEA         & \citet{giorgetta2013, jungclaus2013} \\
\end{tabular}
\end{ruledtabular}
\end{table}

\section{Methods}\label{sec:methods}

Our analysis combines stochastic multibox EBMs, Bayesian inference and spectral analysis. We introduce the approach implemented in ``ClimBayes''\cite{schillinger2022} for the most generic case of a stochastic EBM with $N$ boxes in section~\ref{sec:multiboxebm}. All results are obtained from the special case $N=2$. 

\subsection{Stochastic two-box energy balance model (EBM)} \label{sec:multiboxebm}

The stochastic multibox EBM \cite{geoffroy2013a, fredriksen2017, myrvoll-nilsen2020} extends the one-dimensional linear operator from equation \eqref{eq:stochastic_ebm} by multiple vertical layers, approximating the heat exchange between surface and deep ocean layers. In matrix notation, the model reads \cite{fredriksen2017}
    \begin{equation}
    \label{eq:multibox-matrixform}
    \bm{C} \frac{d \bm{T}}{dt}(t)  =  \bm{K} \bm{T}(t) +\bm{F}(t) + \boldsymbol{\epsilon} (t) \,.
   \end{equation}
For $N$ boxes, $\bm{T}(t)$ is an $N$-dimensional vector, describing the temperature of each box. By convention, $T_1$ corresponds to the temperature of the uppermost and $T_N$ to the temperature of the lowermost box. Accordingly, $\bm{C}$ is a diagonal matrix with the effective heat capacity $C_{ii}$ of each layer $(i=1,...,N)$. $\bm{K}$ is a $N$-dimensional tridiagonal matrix, parameterizing the surface temperature response and vertical heat transfer (Appendix~\ref{app:solution_multibox_exp}). The time-dependent radiative forcing $\bm{F}(t)$ is only applied to the uppermost box, such that $F_1=F(t)$ and $F_k=0$ for $k = 2, ..., N$. The stochastic forcing $\boldsymbol{\epsilon} (t)$ is likewise implemented with non-zero entry $\epsilon_1 (t) = \sigma_W \xi(t)$. We motivate the white noise process $\xi(t)$ with standard deviation (SD) $\sigma_W$  by the found impact of weather fluctuations \cite{hasselmann1976, rypdal2014, cummins2020a}.

Integrating equation \eqref{eq:multibox-matrixform} yields the solution of the surface temperature $T_1(t)$, given by a sum of the forced response $T_{1,F}(t)$ and internal variations $T_{1,I}(t)$
\begin{eqnarray}\label{eq:multibox-solution}
         T_1(t) &&= T_{1,F}(t) + T_{1,I}(t) \nonumber \\
         &&= \int_{- \infty}^t R(t-s) \frac{1}{C_1} F(s) ds + \int_{- \infty}^t R(t-s) \frac{\sigma_W}{C_1} dW(s) \,,
\end{eqnarray}
where $C_1 = C_{11}$ is the heat capacity of the uppermost box. This assumes no interaction between forced and internal variability on a global scale. The response function 
\begin{equation}\label{eq:response}
    R(t) = \sum_{k=1}^N w_k e^{- \lambda_k t}
\end{equation}
is uniquely defined\cite{fredriksen2017, geoffroy2013a} by the response parameters $\lambda_k$ and weights $w_k \, (k=1,...,N)$ with $\sum_{k=1}^N w_k = 1$, which depend on the entries of  $\bm{C}$ and $\bm{K}$ (Appendix~\ref{app:solution_multibox_exp}). The internal fluctuations $T_{1,I}(t)$ in formula \eqref{eq:multibox-solution} represent an Itô-integral over the Wiener process $W(s)$. Therefore, $T_{1,I}(t)$ can be written as a weighted sum of Ornstein-Uhlenbeck (OU) processes, where the $k$-th OU process solves the stochastic differential equation $d U(t)=-\lambda_k U(t) dt + \frac{\sigma_W}{C_1} dW(t)$ and receives weight $w_k$. Accordingly, $T_{1,I}(t)$ is normally distributed with mean zero. Its covariance matrix is determined by $\sigma_W$, $C_1$, $w_k$, and $\lambda_k$ (Appendix~\ref{app:Covariance_OU_process}).

\subsection{Bayesian inference algorithm}\label{sec:bayes} 

\subsubsection{Joint emulation of forced and internal variations}

\begin{figure}[!htbp]
\includegraphics[width=0.5\textwidth]{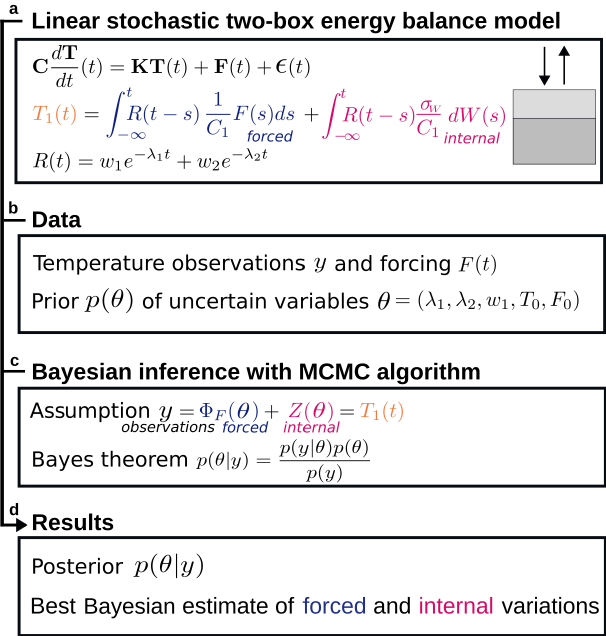} 
\caption{\label{fig:f1} Workflow of our Bayesian inference algorithm to emulate forced and internal GMST fluctuations. \textbf{a} The workflow builds on a linear stochastic two-box EBM, here in matrix notation. The solution for the surface temperature $T_1(t)$ is given by an integral with exponential response function $R(t)$. \textbf{b} Required input data include annually-resolved GMST and time-dependent forcing, as well as physics-informed prior information $p(\theta)$ on uncertain parameters $\theta$. \textbf{c} We infer the parameters of the two-box EBM using a Markov chain Monte Carlo (MCMC) algorithm and Bayes theorem, assuming that the target GMST can be described as a deterministic model $\Phi_F(\theta)$ and stochastic noise $Z(\theta)$. \textbf{d} The workflow yields posterior distributions $p(\theta|y)$ of uncertain parameters conditioned on temperature data $y$ and a physically-motivated emulation of the forced and internal variations.}
\end{figure}

To separate the internal and forced contributions to the GMST variance, we introduce the Bayesian inference algorithm implemented in ``ClimBayes''. We fit the linear stochastic two-box EBM to GMST data (Tab.~\ref{tab:table0}), as illustrated in Figure~\ref{fig:f1}. The Bayesian inference algorithm relies on annually-resolved temperature and forcing time series as input data. Moreover, it requires physics-informed prior information on $\theta =(\lambda_1, \lambda_2, w_1, C_1, T_0, F_0)$. The parameters $\lambda_1, \lambda_2$, and $w_1$ correspond to free parameters of the response function in equation \eqref{eq:response}. $C_1$ is the heat capacity of the upper ocean box. $T_0$ and $F_0$ are initial free parameters. $T_0$ allows for small deviations of the EBM solution to the input data in the starting year and is expected to be close to zero. Similarly, the additional parameter $F_0$ is needed to compensate for an initial forcing anomaly with respect to the equilibrium state.

We infer the posterior distributions of the uncertain parameters $\theta$ conditioned on target data $y$ via Bayes theorem (\ref{eq:BayesTheorem}), using a Markov chain Monte Carlo (MCMC) algorithm. To this end, we assume that the target data can be described by a deterministic model $\Phi_F$ and stochastic measurement or intrinsic noise $Z$ that is also allowed to depend on $\theta$:
\begin{equation}\label{eq:BayesModel}
 y = \Phi_F(\theta) + Z(\theta)
\end{equation}
Formula \eqref{eq:BayesModel} yields the likelihood $p(y | \theta)$. Combined with prior information $p(\theta)$, Bayes theorem \eqref{eq:BayesTheorem} defines the posterior distribution $p(\theta | y)$.
In our case, the deterministic model $\Phi_F(\theta)$ is given by a discretization of the temperature responses $T_{1,F}(t)$. The noise term $Z(\theta)$ corresponds to the internal fluctuations $T_{1,I}(t)$.

This approach provides a joint estimate of the internal and externally-forced response, based on the same physics-informed response function and  parameters. The best estimates of $\theta$ and the forced response $T_{1,F}(t)$ are defined as their posterior means $E[\theta|y]$ and  $E[\Phi_F(\theta)|y]$. The SD $\sigma_I$ of the internal variability $Z(\theta)$ determines $\sigma_W$ (Appendix~\ref{app:Covariance_OU_process}) and is approximated by the residuals' SD, that is, the difference between the target data and the forced response. Samples of the internal variations $T_{1,I}(t)$ can be drawn from its covariance matrix, using the best estimates of $\theta$ (Appendix~\ref{app:Covariance_OU_process}). The \textit{forced + internal} variations $T_{1,F}(t) + T_{1,I}(t)$ represent the full response of the stochastic two-box EBM and, thus, provide a model for the \textit{target} variability. To streamline our discussion, we will refer to the modeled \textit{forced} and \textit{forced + internal} response as emulation.

\subsubsection{Numerical implementation}

The ``ClimBayes'' package provides the numerical implementation of our approach (Fig.~\ref{fig:f1}) and allows for straightforward adjustments via a configuration file. Most importantly, this includes specifications of the number of boxes, prior distributions, MCMC sampling properties, and fixed parameters. We choose $N=2$ boxes (Appendix~\ref{sec:supp_comparison_boxes}) in line with \citet{held2010} and \citet{geoffroy2013a}. Our experiments use independent prior distributions. We choose beta distributions with shape parameters $\alpha=\beta=2$ for the marginal priors of $\lambda_1$, $\lambda_2$ as well as the initial parameters $T_0$ and $F_0$. This allows for a physics-informed, fixed parameter range, where the mean is preferred over boundary values. The algorithm's convergence is improved compared to uniform priors. The intervals for the response parameters $\lambda_1:(0.2, 2)\, \mathrm{yr^{-1}}$ and $\lambda_2:(0.005, 0.2) \, \mathrm{yr^{-1}}$ are similar to those from \citet{fredriksen2017}. Tailored to our goal to emulate interannual to centennial variability from last millennium simulations, our choice of priors assumes characteristic response times $1/\lambda_k$ smaller than 200 years. These response times implicitly set a characteristic depth scale for the two ocean boxes of our stochastic EBM.

The prior of $C_1: (4, 11) \mathrm{W yr m^{-2} K^{-1}}$ follows previous findings \cite{geoffroy2013, geoffroy2013a, fredriksen2017}. Initial values $T_0:(-0.5, 0.5) \, \mathrm{K}$ and $F_0:(-2, 2) \, \mathrm{Wm^{-2}}$ are centered around zero. The weight $w_1:(0,1)$ has an uniform prior. We consider no measurement noise, which is inherently fulfilled for simulated GMST. For observed GMST, we assume measurement errors to be small compared to internal fluctuations \cite{morice2021}. We verified that our findings are robust against reasonable variations of the MCMC and prior specifications.

To obtain the best parameter estimates, ``ClimBayes'' uses a Metropolis Hastings (MH) algorithm from the family of MCMC methods. To this end, we discretize the forward operator and compute annual temperature anomalies relative to the starting point $t=0$ by a midpoint rule\cite{myrvoll-nilsen2020}:
\begin{eqnarray}\label{eq:discretised-sol-ebm}
    T_{1,F}(t_m) &&= T_0 + \frac{1}{C_1} \sum_{k=1}^N w_k \sum_{j=1}^m e^{-(m-j + 1/2) \lambda_k} (F(t_j) + F_0)
\end{eqnarray}
Here, the time step $t_j$ corresponds to the $j$-th year, and $F(t_j)$ are forcing anomalies with respect to the starting point $t = 0$. $F_0$ represents an initial forcing anomaly. Discretizing $T_{1,I}$ leads to a normally-distributed weighted sum of AR(1)-processes with covariance matrix entries depending on $\lambda_k$ and $w_k$ (Appendix~\ref{app:Covariance_OU_process}). The likelihood $p(y | \theta)$ is given by the normal distribution of $T_{1,F}(t) + T_{1,I}(t)$. This requires the calculation of the covariance matrix for each sample in the Markov chain. For numerical robustness and computational efficiency, however, we approximate the likelihood function using an iterative scheme (Appendix~\ref{app:Covariance_OU_process}).

The MCMC algorithm uses four chains with 25,000 samples each, from which the first 5,000 are discarded as burn-in. The proposal distribution is initially set to a normal distribution with mean zero and variances $(0.2, 2, 1, 10, 1, 2)\cdot 10^{-5}$ for $\theta=(\lambda_1, \lambda_2, w_1, C_1, T_0,F_0$). After 2,500 samples, the proposals are distributed according to the weighted sum of the initial normal proposal distribution and the empirical covariance matrix of previous samples. 

We check the convergence of the Markov chains following two performance measures: First, the Gelman-Rubin diagnostics \cite{gelman1992, brooks1998} compares the inter-chain and between-chain variances. It is $\leq 1.1$ for most models, complying to recommendations\cite{gelman2014, plummer2020}. Second, we use the Monte Carlo standard error \cite{flegal2008}, which constructs an asymptotic confidence interval for the posterior mean \cite{flegal2008, haran2020}. In our experiments, the half-width of this interval is smaller than 5 \% of the prior mean. We found that this criterion guarantees robustness of the parameter estimates when the same run is repeated multiple times or additional samples are added. DCESS ESM v1 is the only outlier, showing a tendency for bimodal posterior densities which lead to less stable estimates with wide CIs. However, we confirmed that convergence can be achieved by fixing the heat capacity to the estimated parameter.

\subsection{Spectral analysis and variance ratios} \label{sec:spectralmethods}

Given a temperature time series $T(t)$, the power spectral density (PSD) at frequency $f$ corresponds to the Fourier transform of the autocovariance 
\begin{equation}\label{eq:spectrum_definition}
	S(f) = \int_{-\infty}^{\infty} \, \mathrm{e}^{-2 \pi i f k} \,E \left[ \left( T(t) -  \mu \right) \left( T (t+k) -  \mu  \right) \right] \, \,  dk \,,
\end{equation}
with lag $k=t_2 - t_1$ and mean $\mu \coloneqq E[T(t)]$. This assumes $X(t)$ to be weakly stationary, which is reasonably fulfilled after linear detrending the GMST data. Following \citet{ellerhoff2021}, we use the multitaper method with three windows to compute the PSD and $\chi^2-$distributed uncertainties. Mean spectra are obtained after interpolation to the lowest resolution and binning into equally spaced log-frequency intervals\cite{laepple2014a}. The spectra are visualized over periods $\tau=1/f$ and logarithmically smoothed using a Gaussian kernel of 0.04 decibels.

We compare the PSD for three types of time series: (1) the \textit{target} temperature data from historical observations or climate model simulations, AR5 EMIC and CMIP5 models, (2) the emulated \textit{forced} variations $T_{1,F}(t)$, and (3) the emulated \textit{forced + internal} temperature variations $T_{1,F}(t) + T_{1,I}(t)$ from the stochastic two-box EBM. To compute the \textit{forced + internal} PSD, we first sample 1000 realizations of the internal response $T_{1,I}(t)$. We add these to $T_{1,F}(t)$ and compute the PSD for all samples. Mean spectra and 95\% confidence bands are obtained from this ensemble.

Variance ratios are calculated by dividing the emulated by the \textit{target} variance. Following Parseval's theorem, we determine the timescale-dependent variance by integrating the PSD over frequencies. We consider frequency bands corresponding to interannual (2-5 yrs), decadal (5-20 yrs), multidecadal (20-50 yrs) and centennial (50-200 yrs) scales. 

\section{Results}\label{sec:results}

\subsection{Example application to historical observations}
\label{sec:results_hist}

\begin{figure*}[!htbp]
\includegraphics[width=\textwidth]{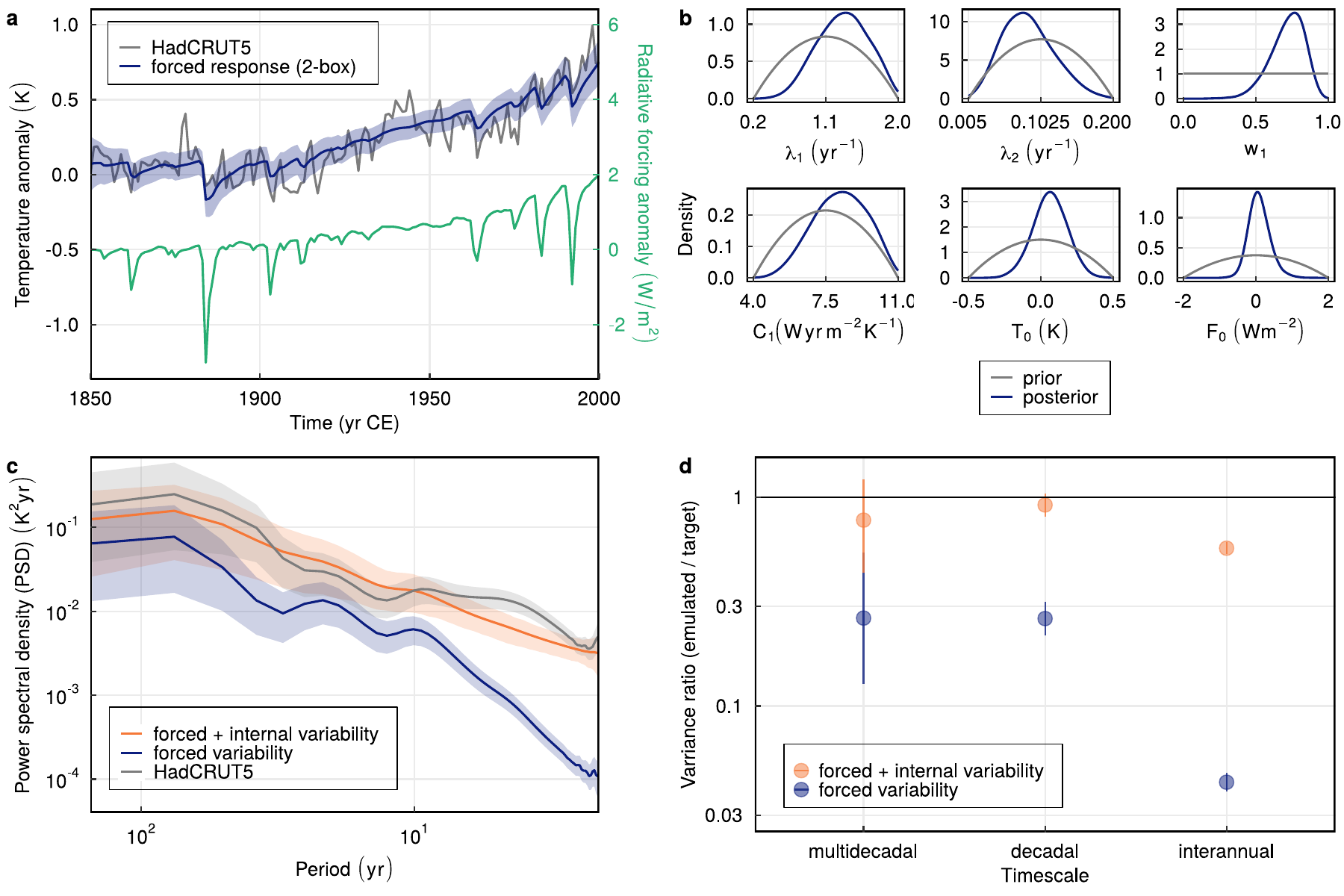}
\caption{\label{fig:demonstration} Example application of our developed approach to historical GMST observations from HadCRUT5. \textbf{a} \textit{Target} data $y$, forcing $F(t)$ (taken from \citet{schmidt2012, pongratz2008, crowley2008,steinhilber2009, wang2005}), and the best estimate of the \textit{forced} response $T_{1,F}(t)$, that are the posterior mean $E[\Phi_F(\theta) | y]$. Shaded areas correspond to 95 \% CIs of $T_{1,F}(t)$. \textbf{b} Marginal prior and posterior densities for uncertain parameters $\theta$. \textbf{c} PSD of the GMST observations, the \textit{forced} response (both with $\chi^2$-distributed confidence bands) and the sampled \textit{forced + internal} variations. The sampled \textit{forced + internal} PSD represents the mean and 95 \% confidence bands obtained from an ensemble of $T_{1,F}(t)+T_{1,I}(t)$ using 1000 realizations of $T_{1,I}(t)$. \textbf{d} Ratios of the emulated to observed GMST variance, computed by integration of the PSD on the multidecadal (20-100 yrs), decadal (5-20 yrs) and interannual (2-5 yrs) scales. Uncertainties (95\% CI) are calculated from a F-distribution based on the degrees of freedom of the variance estimate.}
\end{figure*}

We demonstrate the application of the Bayesian inference algorithm and our spectral analysis on the example of GMST observations from HadCRUT5\cite{morice2021}. Figure~\ref{fig:demonstration}a shows the forcing and temperature time series together with the best estimate of the forced response $T_{1,F}(t)$. The forced response follows the global warming trend and shows cooling periods after volcanic eruptions. Credible intervals (CIs) capture the uncertainties of the forced response, but not those due to internal variability. As a result, observations partly lie outside CIs. Uncertainties of the two-box forced response are largest at the time series' start.

Figure~\ref{fig:demonstration}b shows marginal prior and posterior distributions for the free parameters of the stochastic two-box EBM. The response parameters $\lambda_1= 1.31 \, (0.71, 1.86)\, $yr$^{-1}$ and $\lambda_2= 0.09 \, (0.03,0.16)\, $yr$^{-1}$ (Tab.~\ref{tab:hadcrut_params}) correspond to timescales of approximately 10 months and 10 years. The corresponding heat capacity $C_1$ of the upper ocean layer is $8.20\,(5.68, 10.43) \mathrm{W~yr~m^{-2}~K^{-1}}$. The initial values $T_0$ and $F_0$ are well constrained and close to zero. The weight $w_1= 0.72 \,(0.46, 0.90)$ tends to emphasize the fast response.

The power spectral density (Fig.~\ref{fig:demonstration}c) of the \textit{forced} response alone is smaller than that of the \textit{target} temperature. Conversely, the magnitude of the emulated \textit{forced + internal} PSD agrees with the \textit{target} PSD within uncertainties, except for the interannual scale. While the emulated PSD constantly increases from interannual to multidecadal scales, HadCRUT5 shows a modulation with increased power on periods of two to ten years, which is not captured by the emulated response. 

Figure~\ref{fig:demonstration}d compares the variance on interannual to multidecadal timescales. The variance ratios are formed by dividing the emulated \textit{forced} or \textit{forced + internal} variance by the variance obtained from the HadCRUT5 \textit{target}. The \textit{forced} variance is smaller than the \textit{target} variance on all timescales. Incorporating internal variability reduces this mismatch strongly, yet, is not enough to capture all fluctuations on interannual and multidecadal scales. On decadal scales, the emulated \textit{forced + internal} variability agrees well with the observations.

\subsection{Parameters estimated from last millennium simulations} 
\label{sec:results_parameters}

\begin{table*}[!htbp]
\caption{\label{tab:table1} Posterior means and 95\% CIs of uncertain parameters $\theta$ for the stochastic two-box EBM fitted to GMST from climate model simulations. The SD $\sigma_I$ of the internal variability equals the residual of the forced response (Appendix~\ref{app:Covariance_OU_process}) and is therefore given without CIs. Ensemble means are denoted by ``(mean)''. We exemplary show the LOVECLIM ensemble member ``E1'' as there are no major differences across the ensemble. Models with GRA\cite{gao2008} volcanic forcing are marked with an asterisk (*), all others use the CEA reconstruction \cite{crowley2008} (Tab.~\ref{tab:table0}). 
The weight $w_2=1-w_1$ (not shown) is uniquely defined by $w_1$.}
\scriptsize
\begin{ruledtabular}
\begin{tabular}{lccccccr}
Climate model    & $\lambda_1 ($yr$^{-1})$ & $\lambda_2 ($yr$^{-1})$ & $w_1$ (unitless) &  $C_1 \mathrm{(W yr m^{-2} K^{-1})}$ & $T_0$ (K) & $F_0 ($Wm$^{-2})$ & $\sigma_I$ (K)     \\ \hline
\textbf{AR5 EMICs} & & & & & & & \\
        Bern 3D & 0.74 (0.62,0.81) & 0.08 (0.06,0.15) & 0.91 (0.88,0.95) & 4.76 (4.54, 5.43) & -0.09 (-0.50,0.04) & 0.18 (-0.15, 1.48) & 0.02 \\
        CLIMBER-3alpha & 0.76 (0.72,0.80) & 0.06 (0.05,0.07) & 0.91 (0.90,0.93) & 6.18 (5.99, 6.35) & -0.02 (-0.05,0.02) & -0.04 (-0.13, 0.04) & 0.01 \\
        CLIMBER2 (mean) & 1.43 (0.25,1.97) & 0.19 (0.17,0.20) & 0.05 (0.00,0.19) & 10.96 (10.89,11.00) & 0.01 (-0.11,0.13) & -0.19 (-0.45, 0.07) & 0.05 \\
        DCESS ESM v1 & 0.31 (0.21,1.18) & 0.08 (0.03,0.18) & 0.64 (0.00,0.92) & 10.25 (9.95,10.52) & 0.02 (-0.05,0.08) & -0.18 (-0.27,-0.05) & 0.03 \\
        IGSM 2.2 & 0.64 (0.57,0.73) & 0.10 (0.07,0.15) & 0.89 (0.82,0.93) & 6.68 (6.38, 6.98) & 0.01 (-0.08,0.11) & -0.10 (-0.36, 0.15) & 0.03 \\ 
        LOVECLIM (E1) & 1.32 (0.99,1.69) & 0.13 (0.07,0.18) & 0.90 (0.85,0.94) & 6.03 (4.95,7.29) & -0.01 (-0.22,0.19) & 0.28 (-0.58,1.21) & 0.09 \\
        LOVECLIM (mean) & 1.12 (0.93,1.34) & 0.14 (0.09,0.19) & 0.88 (0.84,0.92) & 6.74 (6.06,7.42) & 0.04 (-0.08,0.16) & -0.50 (-1.03,0.01) & 0.04 \\
        MESMO 1.0 & 0.52 (0.49,0.56) & 0.03 (0.02,0.04) & 0.90 (0.88,0.92) & 7.96 (7.72, 8.19) & 0.02 (-0.03,0.06) & -0.32 (-0.39,-0.24) & 0.02 \\ 
        MIROC-lite & 0.44 (0.41,0.47) & 0.02 (0.01,0.04) & 0.96 (0.94,0.97) & 6.75 (6.56, 6.96) & 0.01 (-0.05,0.08) & -0.34 (-0.49,-0.25) & 0.03 \\ 
        UMD & 0.72 (0.68,0.77) & 0.01 (0.01,0.01) & 0.99 (0.99,0.99) & 8.81 (8.46, 9.15) & 0.00 (-0.04,0.02) & -0.09 (-0.21, 0.08) & 0.02 \\ 
        UVic v2.9 & 0.45 (0.41,0.51) & 0.05 (0.03,0.08) & 0.88 (0.81,0.93) & 10.88 (10.68,10.98) & -0.01 (-0.06,0.05) & -0.05 (-0.19, 0.09) & 0.02 \\
\textbf{CMIP5 models} & & & & & & & \\
        BCC-CSM1-1 * & 1.91 (1.76,1.99) & 0.06 (0.04,0.10) & 0.94 (0.91,0.96) & 10.49 (9.78,10.92) & -0.05 (-0.13,0.06) & 1.31 ( 0.53, 1.87) & 0.10 \\ 
        CCSM4 * & 1.62 (1.27,1.90) & 0.12 (0.08,0.16) & 0.88 (0.83,0.92) & 4.70 (4.12, 5.60) & -0.06 (-0.34,0.18) & 0.17 (-0.60, 1.05) & 0.14 \\ 
        CSIRO-Mk2L-1-2 & 0.48 (0.29,1.12) & 0.15 (0.07,0.19) & 0.65 (0.31,0.96) & 10.49 (9.78,10.92) & 0.02 (-0.17,0.21) & -0.23 (-0.76, 0.30) & 0.08 \\
        FGOALS-s2 * & 1.49 (1.13,1.88) & 0.03 (0.01,0.10) & 0.94 (0.89,0.97) & 8.41 (6.71,10.38) & -0.05 (-0.20,0.11) & -0.58 (-1.42,-0.06) & 0.14 \\ 
        GISS-E2-R & 0.56 (0.50,0.65) & 0.03 (0.01,0.15) & 0.99 (0.96,1.00) & 6.53 (5.94, 7.07) & 0.12 (-0.43,0.32) & -1.19 (-1.77, 0.48) & 0.09 \\ 
        HadCM3 & 0.49 (0.34,0.73) & 0.13 (0.05,0.19) & 0.76 (0.52,0.96) & 7.99 (7.15, 8.86) & -0.06 (-0.31,0.20) & 0.89 ( 0.27, 1.48) & 0.12 \\
        HadGEM2-ES & 0.64 (0.39,1.28) & 0.14 (0.05,0.19) & 0.74 (0.53,0.95) & 7.79 (6.52, 8.81) & -0.06 (-0.30,0.19) & 0.97 ( 0.36, 1.58) & 0.12 \\ 
        IPSL-CM5A-LR * & 1.69 (1.36,1.95) & 0.09 (0.06,0.14) & 0.85 (0.78,0.91) & 9.49 (7.89,10.73) & -0.16 (-0.34,0.00) & -0.99 (-1.77,-0.14) & 0.15 \\ 
        MIROC-ESM & 1.49 (1.04,1.90) & 0.14 (0.09,0.19) & 0.71 (0.60,0.81) & 9.62 (8.01,10.78) & -0.02 (-0.25,0.22) & 0.10 (-0.83, 1.03) & 0.11 \\ 
        MPI-ESM-P & 0.66 (0.45,0.97) & 0.14 (0.07,0.19) & 0.77 (0.61,0.93) & 6.11 (5.42, 6.80) & 0.04 (-0.22,0.29) & -0.57 (-1.15,-0.01) & 0.11 \\ 
\end{tabular}
\end{ruledtabular}
\end{table*}

We use Bayesian inference to fit the stochastic two-box EBM to GMST simulations from CMIP5 models and AR5 EMICs (section~\ref{sec:data}). Table~\ref{tab:table1} shows the best estimates, that is the posterior means, and 95\% CIs of  $\theta=(\lambda_1, \lambda_2, w_1, C_1, T_0, F_0)$ as well as the SD $\sigma_I$ of the internal variability $T_{1,I}(t)$. Across all simulations, the short-term response parameter $\lambda_1$ varies between 1.91 and 0.31  yr$^{-1}$, spanning the full prior range between six months and five years. The long-term response $\lambda_2$ varies between $0.19$ and $0.01$ yr$^{-1}$, corresponding to characteristic timescales of approximately five to one hundred years.

Differences between AR5 EMICs and CMIP5 models are most pronounced for $\lambda_1$. CMIP5 simulations exhibit larger CIs and inter-model differences, while AR5 EMICs show similar $\lambda_1$, except for CLIMBER 2 (mean) and LOVECLIM (E1 and mean). The weight $w_1>0.5$ is typically larger than $w_2=1-w_1$, emphasizing the relative importance of the fast compared to the slow response. $w_1$ often is closer to unity for AR5 EMICs than for CMIP5 models. CLIMBER2 (mean) shows exceptionally large $\lambda_1$ and small $w_1$. The heat capacity $C_1$ varies from 4.7 to 10.96 $\mathrm{Wyrm^{-1}K^{-1}}$. It spans the full prior range for both AR5 EMICs as well as CMIP5 models and shows no strong dependence on other parameters. The initial temperature $T_0$ is well constrained and close to zero. Inter-model differences are also large for $F_0$ and linked to varying temperature amplitudes at the beginning of the time series with respect to the mean (Fig.~\ref{fig:fit_supp}). The SD $\sigma_I$ of the internal variability from CMIP5 models lies between 0.09 and 0.15 K and is larger than for AR5 EMICs (0.01 to 0.09 K). 

\subsection{Emulation of power spectral density}
\label{sec:results_spectra}

\begin{figure*}[!htbp]
\includegraphics[width=\textwidth]{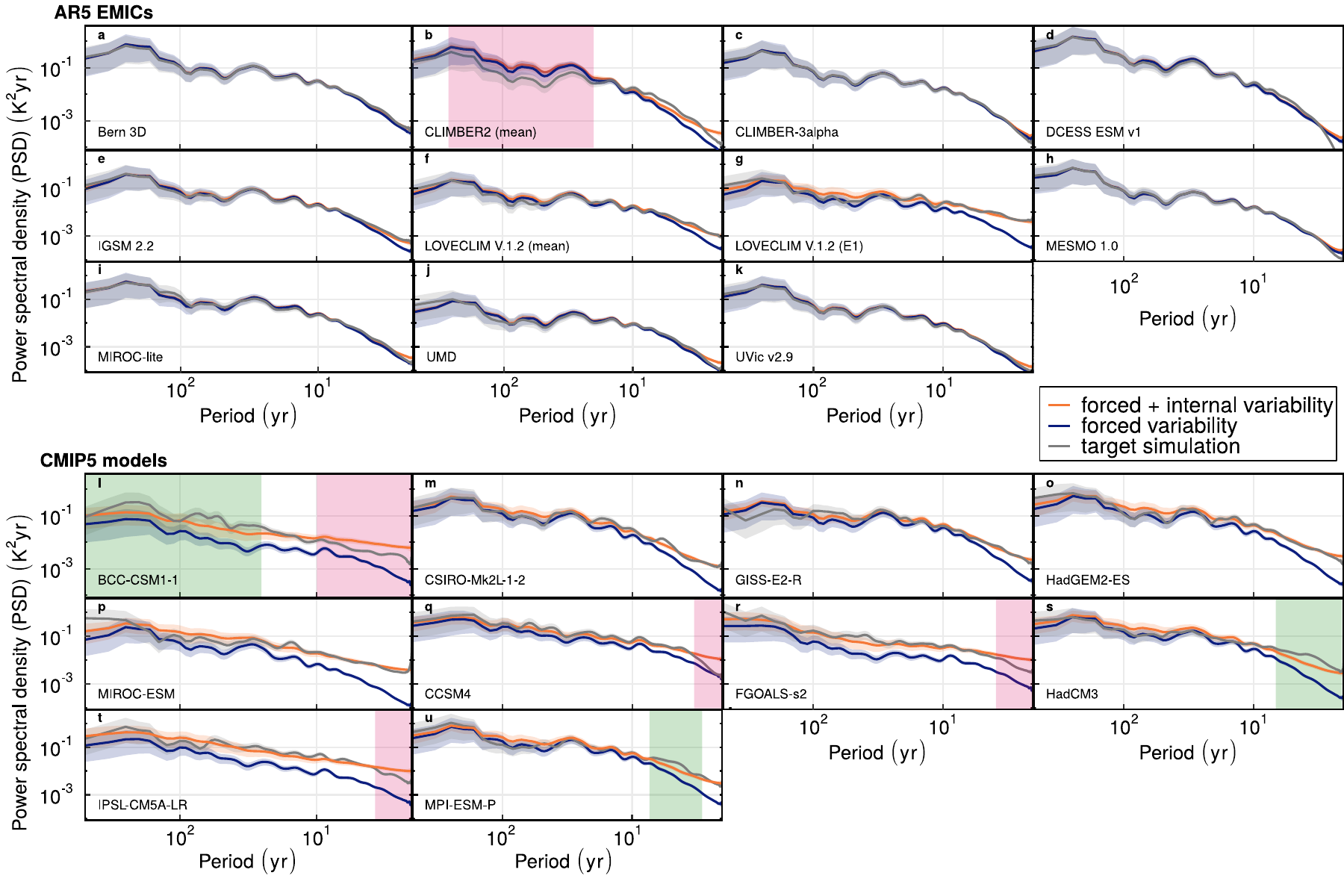}
\caption{\label{fig:spectra} \textit{Target} and emulated \textit{forced} and \textit{forced + internal} power spectral density (PSD) and 95\% confidence bands for AR5 EMICs (top:~\textbf{a-k}) and CMIP5 models (bottom:~\textbf{l-u}), as Fig.~\ref{fig:f1}c. Shaded intervals highlight examples of overestimation (pink) and underestimation (green) of the emulated PSD compared to the \textit{target}, which are discussed in the main text. We show the PSD for LOVECLIM V.1.2 (mean) and the first ensemble member LOVECLIM V.1.2 (E1), as there are no major differences across ensemble members.}
\end{figure*}

Figure~\ref{fig:spectra} compares the \textit{target} PSD to the emulated \textit{forced} and \textit{forced + internal} PSD. For most simulations, the emulated \textit{forced + internal} PSD agrees with the \textit{target} within uncertainties. AR5 EMICs show no major differences between \textit{forced} and \textit{forced + internal} PSD above decadal scales, except for the LOVECLIM ensemble members. Hence, the forced response is sufficient to emulate the long-term variability of most AR5 EMIC simulations. On interannual scales, considering internal variability in the emulation compensates for the mismatch between the \textit{forced} and \textit{target} PSD. Differences between emulated and \textit{target} PSD are most pronounced for CLIMBER2 (mean), showing an overestimation on multidecadal to centennial scales by the emulation.

For CMIP5 simulations, the emulated \textit{forced} PSD  underestimates the \textit{target} PSD on all timescales. Conversely, the \textit{forced + internal} variability matches the \textit{target} well for almost all models. Minor differences are found on interannual scales (Fig.~\ref{fig:spectra}(l) and 3(q)–3(u)). Here, the \textit{target} PSD of MPI-ESM-P and HadCM3 exhibit increased power on periods of two to eight years. CCSM4, FGOALS-s2 and IPSL-CM5A-LR overestimate the PSD on the shortest timescales of approximately two years. The emulated \textit{forced + internal} PSD for BCC-CSM1-1 deviates from that of the \textit{target} by showing increased power on interannual and decreased power on multidecadal to centennial scales.

\subsection{Separating internal and externally-forced variance}\label{sec:separation}

\begin{figure*}[!htbp]
\includegraphics[width=\textwidth]{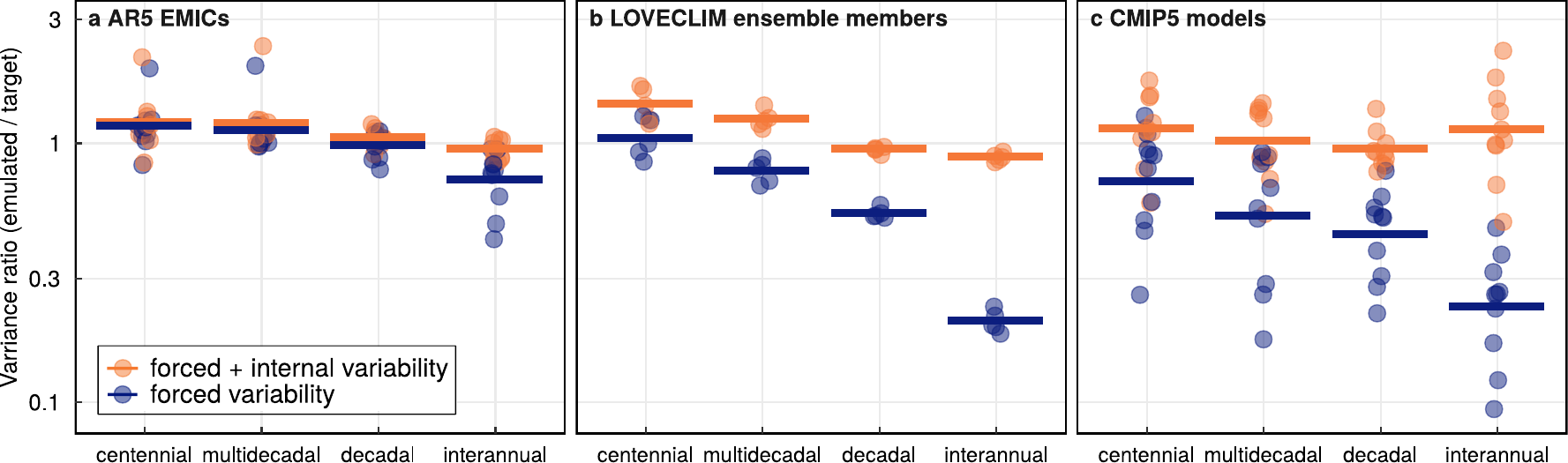}
\caption{\label{fig:ratios} Ratios of the emulated \textit{forced} and \textit{forced + internal} to the \textit{target} variance for AR5 EMICs (panel~\textbf{a}), LOVECLIM ensemble members (panel~\textbf{b}) and CMIP5 simulations (panel~\textbf{c}) on centennial (50-200 yrs), multidecadal (20-50 yrs), decadal (5-20 yrs) and interannual (2-5 yrs) scales (as Fig.~\ref{fig:f1}d). Bars indicate the mean variance ratio over the considered simulations for one model type. Circles correspond to individual simulations. Confidence bands for individual ratios (as in Fig.~\ref{fig:demonstration}) are not shown for better visibility.)}
\end{figure*}

Figure~\ref{fig:ratios} shows the mean and spread of variance ratios on interannual to centennial scales for the considered model types. Variance ratios smaller than one indicate less emulated \textit{forced} or \textit{forced+internal} than \textit{target} variance. We add a comparison of  the variance from the five LOVECLIM V.1.2 ensemble members (E1-E5) (Fig.~\ref{fig:ratios}b) to that of the remaining AR5 EMICs (Fig.~\ref{fig:ratios}a) and CMIP5 models (Fig.~\ref{fig:ratios}c). Here, AR5 EMICs explicitly include the LOVECLIM ensemble mean, but not its members. For all model types, the relative contribution of internal variability decreases with increasing timescale, as the ratios for \textit{forced} and \textit{forced + internal} variance become more similar. The contribution of internal variability is larger in LOVECLIM ensemble members and CMIP5 simulations compared to AR5 EMICs.

The emulated variance of AR5 EMICs (Figure~\ref{fig:ratios}a) is dominated by \textit{forced} variations and matches the \textit{target} variance on interannual and decadal scales. On longer timescales, the emulated variance tends to overestimate the \textit{target} variance. This is mostly due to outliers, which correspond to CLIMBER2 (mean), in line with the overestimated PSD in Figure~\ref{fig:spectra}b. Members of the LOVECLIM V.1.2. ensemble (Fig.~\ref{fig:ratios}b) exhibit more internal contributions to the variance on all timescales compared to AR5 EMICs. The ensemble's emulated \textit{forced + internal} variance approximates the \textit{target} well on the interannual and decadal scale. On larger timescales, we find an overestimation of the \textit{target} variance. 

The mean emulated \textit{forced + internal} variance of CMIP5 simulations (Fig.~\ref{fig:ratios}c) is close to one on all timescales. The relative contribution of internal compared to forced variations is similar to that of LOVECLIM ensemble members on interannual and decadal timescales. However, there is a larger spread of variance ratios. Moreover, we find a small tendency of the emulation to overestimate the interannual and centennial variance of CMIP5 models. BCC-CSM1-1 represents an outlier, with the uppermost variance ratio on the shorter and lowermost on the longer timescales, in line with the spectral analysis (Fig.~\ref{fig:spectra}l). 

\section{Discussion}\label{sec:discussion} 

We demonstrate the emulation of GMST variability as simulated by state-of-the-art climate models using a linear stochastic two-box EBM and Bayesian inference. Our analysis builds on the same, physically-motivated response function for internal and external processes, and allows for consistent separation of internal and externally-forced variability. Estimates of the timescale-dependent variance show that the relative contribution of internal variability increases with model complexity and decreases with timescale.  

Building on previous studies \cite{bodman2016, myrvoll-nilsen2020}, the strength of our Bayesian framework is that it yields the posterior means and CIs for the uncertain parameters of the stochastic two-box EBM fitted to GMST simulations. Due to our choice of priors and the convergence of the fit, the estimated heat capacity $C_1$ agrees by construction with previous findings \cite{geoffroy2013, geoffroy2013a, fredriksen2017}.
Our response parameters $\lambda_1$ and $\lambda_2$ are consistent with results from \citet{fredriksen2017}, estimated from observational data. However, the estimates differ from those obtained in 4x$\mathrm{CO_2}$ experiments \cite{geoffroy2013, geoffroy2013a}. This is because our framework accounts for high-frequency pulses and, thus, estimates response parameters associated with faster dynamics.
Furthermore, our findings reveal a dependence of the estimated response parameters on the imprint of intermittent volcanic eruptions on simulated temperatures. This is reflected in consistently high values for $w_1$, emphasizing the fast feedback. The inter-model spread of $\lambda_1$ in CMIP5 simulations suggests a link to the implemented volcanic forcing: CMIP5 simulations driven by a comparatively weak reconstruction (``GRA'') tend to show higher values for $\lambda_1$ compared to those driven by ``CEA'' (Tab.~\ref{tab:table0}), which has greater forcing amplitudes. $\lambda_1$ and $w_1$ are particularly high for BCC-CSM1-1, indicating a fast and weak forced response of the fitted EBM (Fig.~\ref{fig:fit_supp}). This is consistent with a weak forced response in BCC-CSM1-1\cite{bothe2013}. The parameter estimates for CLIMBER2 (mean) differ from the remaining AR5 EMICs. In particular, $\lambda_1$ is poorly constrained, as CIs span the full prior range. We find that the temperature response to volcanic eruptions in CLIMBER2 (mean) is delayed. The estimated $\lambda_1$ and $w_1$ can be reconciled with those of the other AR5 EMICs if the temperature data were shifted by 1 yr. Altogether, the sensitivity of the fit to intermittent volcanic forcing suggests that high-frequency forcing plays a crucial role for simulating temperature variability across scales correctly. 

Different methods have been developed to isolate forced and internal variations based on detrending \cite{zhang2013}, single-model ensembles \cite{frankcombe2015, olonscheck2017, lehner2020, mann2022}, and deterministic EBMs \cite{mann2014, crowley2000}, among others. The application of a Bayesian energy balance framework to the timescale-dependent quantification of forced and internal variance is novel. Our method provides a robust and joint separation of the variations at every step in time in a statistically sound way. Using data from the CESM Large Ensemble Community project \citep{kay2015}, we verify the robustness of our \textit{forced} response (Fig.~\ref{fig:LME}). The latter agrees well with the ensemble mean, which shows higher variability due to remaining internal variations that have not been averaged out over the 13 members. Hence, our method provides a robust tool to estimate the \textit{forced} response when large ensembles are not available. Moreover, we find a wide agreement of the emulated \textit{forced + internal} variance with that from CMIP5 simulations. The fact that the stochastic two-box EBM mimics the temperature variations well is in line with previous findings on a linear relation between external forcing and GMST \cite{geoffroy2013a, macmynowski2011, fredriksen2017, ellerhoff2021}. 

Small differences between the emulated \textit{forced+internal} and \textit{target} PSD for CMIP5 models on interannual timescales can be attributed to the simplified representation of internal variability as a weighted sum of AR(1) processes by the stochastic EBM. The latter cannot represent (pseudo-)oscillatory climate modes or modulations of internal variability by external forcing \cite{maher2018, bonnet2021}, which could result in the observed underestimation of the PSD by the emulation on these timescales. For HadCM3 and MPI-ESM (Fig.~\ref{fig:spectra}s and u), deviations on the interannual scale are similar to those in HadCRUT5 and likely due to the spectral imprint of the El Niño-Southern Oscillation \cite{ellerhoff2022a, ellerhoff2021}. Moreover, our approach assumes that the covariance structure of the internal variations is determined by the estimated feedback parameters. In CMIP5 simulations with ``GRA'' volcanic forcing, particularly high values for the fast response parameter $\lambda_1$ and the weight $w_1$ lead to internal variations with autocorrelations on short timescales. This can cause an overestimation of the emulated PSD (Fig.~\ref{fig:spectra}~l,~q,~r, and~t). Similarly, the overestimation of the \textit{target} PSD on longer timescales (Fig.~\ref{fig:spectra}b) in CLIMBER 2 can be explained by its estimated slow response. The latter is due to the biased delay between forcing and temperature time series (Tab.~\ref{tab:table1}). Additionally, mismatches on short timescales can propagate to longer timescales, as in the case of BCC-CSM1-1 (Fig.~\ref{fig:spectra}l). Hence, explaining spectral properties of temperature time series not only requires consideration of stochastic noise \cite{soldatenko2022}, but also precise knowledge of its correlation structure and the forced response. This highlights a need for simple, stochastic dynamical models \cite{lovejoy2021} to simulate temperature fluctuations on long timescales.

The components of AR5 EMICs show a reduced number of scales compared to the AOGCMs, which simplifies the complexity of the processes contributing to variability. Therefore, the relatively strong forced variability in EMICs (Fig.~\ref{fig:ratios}a and b) is not a main deficiency, but serves to explore the long-term coupling between different Earth system components in response to radiative forcing.
Compared to other AR5 EMICs, LOVECLIM V.1.2 features a more complex, three-layered atmosphere, which likely explains increased internal variations in the ensemble members. However, this variability is predominately short-term correlated. Due to the EBM's covariance structure, this can lead to an overestimation of the emulated \textit{forced+internal} variability (Fig.~\ref{fig:ratios}b) on longer timescales.
On interannual and decadal scales, the variance ratios based on the emulated \textit{forced} and \textit{forced + internal} variability from the LOVECLIM ensemble members (Fig.~\ref{fig:ratios}b) are similar to CMIP5 simulations (Fig.~\ref{fig:ratios}c). The similarity indicates that EMICs with a more realistic representation of atmospheric variability might better capture the relative contribution of forced and internal temperature variations on these timescales.

The contributions of internal variations on multidecadal scales and longer remain the largest in CMIP5 models, likely due to long-term variability mechanisms from the comprehensive ocean dynamics of AOGCMs. Compared to observations (Fig.~\ref{fig:demonstration}~d), however, internal variability on decadal and multidecadal scales is smaller in CMIP5 simulations (Fig.~\ref{fig:ratios}~c). This is particularly interesting given the agreement of observed and simulated total GMST variability on these scales \cite{pages2kconsortium2019, ellerhoff2021}. Smaller low-frequency internal variability in climate models than in observations \cite{yan2018} could be offset by enhanced forced variability in response to volcanic eruptions \cite{schurer2013, chylek2020}, such that the overall variance is largely conserved. We suspect that an incorrect ratio between internal and external variability could impact the long-term variability of simulated local temperatures. However, uncertainties in the interpretation of our HadCRUT5 findings arise due to the comparatively short time-span of the instrumental record and a possible change of the forced response under global warming \cite{hopcroft2018}. Further investigating the spatial variability structure \cite{kunz2021} and the link between local and global variability across climate states could help resolve mismatches between observed and simulated local variability on decadal and multidecadal scales.

One limitation of our study arises from the fact that the developed framework targets interannual to centennial timescales, and is therefore designed for annually-resolved GMST and forcing data. As a result, it cannot be readily applied to much shorter or longer timescales. Investigating the immediate effects of radiative forcing, for example, necessitates an extension to sub-annual resolution, and treatment of the seasonal cycle. An extension to coarser resolutions could be beneficial to study long-term changes such as millennial-scale variability. However, such applications require careful examination of the underlying assumptions of the current framework, and an extension and validation of the estimation algorithm.
The forced response is likely sensitive to model-specific rapid adjustments \cite{smith2008} and uncertainties in the forcing. Applying the workflow to paleoclimate reconstructions could therefore be challenging. On the one hand, ``ClimBayes'' does not yet run at the best possible speed, as there are faster Bayesian algorithms \cite{myrvoll-nilsen2020}. On the other hand, ``ClimBayes'' represents an accessible, transparent, and well-documented numerical framework that can be easily adapted and extended, for example by integrating ``Rstan'' \cite{guo2016rstan} or multilevel delayed acceptance MCMC \cite{lykkegaard2022}. Similar to \citet{fredriksen2017}, we use a response function of exponential form, solving the ordinary differential equation \eqref{eq:stochastic_ebm}. Future research could investigate the potential of Bayesian methods to find response functions describing the effects of climate forcing on different observables \cite{lucarini2018, torresmendonca2021}. Furthermore, future studies could test our findings with more advanced climate models including better representation of, for example, land surface processes, atmospheric dynamics and chemistry, and sea ice. The presented framework can be also applied to single forcing experiments\cite{schurer2014} for quantifying the contribution of single forcings to the spectrum. This will help better understand the climate system's response and interplay of intrinsic and external components in driving climate variability. 

\section{Conclusion}\label{sec:conclusion} 

We presented a physically-motivated emulation of GMST data using Bayesian inference and a stochastic energy balance model. Analyzing AR5 EMICs and CMIP5 simulations for the last millennium, we found that the power spectral density of the combined \textit{forced + internal} response approximates the \textit{target} spectrum well. We show that our emulation can be used to separate internal and forced contributions to GMST variability across timescales. The relative contribution of internal dynamics increases with model complexity and decreases with timescale. While AR5 EMICs predominately exhibit forced variations, simulations from CMIP5 models and the LOVECLIM ensemble members exhibit major contributions from the forced and internal response. This suggests that EMICs with more realistic atmospheric variability can simulate statistical properties of interannual to decadal climate fluctuations more reliably. Our results show that precise knowledge of the forced response and correlation structure of internal variability is necessary to explain variability across scales, needed to assess future variability and potentially associated risks with long-term projections. Our developed framework is robust, readily available and can thus be widely applied to describe, emulate and diagnose observed and simulated temperature variability.

\section*{Data Availability Statement}
The ClimBayes \cite{schillinger2022} software package in R is released at \url{https://github.com/paleovar/ClimBayes} and published under the Zenodo identifier \url{https://doi.org/10.5281/zenodo.7317984}. Code and data to reproduce all figures is available at \url{https://github.com/paleovar/EmulatingVariability}. The data that support the findings of this study are openly available from the data holdings of the Climate Research Programme’s Working Group on Coupled Modelling (e.g., \url{https://esgf-node.llnl.gov/search/cmip5/}), responsible for CMIP and PMIP, and from \citet{schmidt2012}, \citet{eby2013} (\url{https://climate.uvic.ca/EMICAR5/participants.html}), and \citet{morice2021}. 

\section*{Author Contributions}
MS and BE contributed equally to this work. They carried out the analysis, created the figures and wrote the paper, supervised by KR and RS. MS led the development of the ClimBayes package. BE led the variability analysis based on spectral methods. All authors designed the study, contributed to revisions and approved the final version of the manuscript.

\begin{acknowledgements}
We acknowledge the World Climate Research Programme’s Working Group on Coupled Modelling, responsible for PMIP and CMIP. We thank the research groups listed in Tab.~\ref{tab:table0} and the Met Office for producing and making available their model output, measurements, and forcing reconstructions. We thank M. Eby for EMIC discussions, E. Myrvoll-Nilsen for discussion of Bayesian inference and the INLA package, and T. Gasenzer for discussion of conceptual climate models. We are grateful to N. Meinshausen, N. Weitzel, and E. Ziegler for helpful comments on the manuscript. We thank two anonymous referees for their constructive and valuable review. This study has been supported by funds of the Deutsche Forschungsgemeinschaft (DFG, German Research Foundation) Project No.~395588486, by the PalMod project (subproject No.~01LP1926C), the Heinrich-Böll-Stiftung (Heinrich Böll Foundation), and the Studienstiftung des deutschen Volkes (German Academic Scholarship Foundation). R. Scheichl is supported by the Deutsche Forschungsgemeinschaft under Germany’s Excellence Strategy EXC 2181/1 - 390900948 (the Heidelberg STRUCTURES Excellence Cluster). The study benefited from discussions within the CVAS working group, a working group of the Past Global Changes (PAGES) project.
\end{acknowledgements}

\appendix

\section{Solution to the multibox EBM}\label{app:solution_multibox_exp}

The tri-diagonal matrix $\bm{K}$ of the multibox EBM in matrix notation  \eqref{eq:multibox-matrixform} is given by 
\begin{equation*}
    \bm{K} = \begin{pmatrix}
        -(\tilde{\lambda} + \kappa_2) & \kappa_2 & 0 & \cdots & & 0 \\
         \kappa_2 & -(\kappa_2 + \kappa_3) & \kappa_3 & \ddots & &  \\
        0 & \kappa_3 & \ddots & \ddots &  & \vdots \\
        \vdots & \ddots & \ddots &  & &  \\
         &  &  &  & & \kappa_N \\
        0 & \cdots &  & \kappa_N & & -\kappa_N, \\
        \end{pmatrix} \,.
\end{equation*}
The parameter $\tilde{\lambda}$ controls the feedback of the surface layer. The coefficients  $\kappa_2,...,\kappa_N > 0$ describe the vertical heat transfer between ocean layers. The full solution to the multibox EBM\cite{fredriksen2017} reads
\begin{equation*}
    \bm{T} = \int_{- \infty}^t e^{(t-s) \bm{C}^{-1} \bm{K}} \bm{C}^{-1} \bm{F}(s) ds.
\end{equation*}
$\bm{K}$ is symmetric and negative definite, and, thus, diagonalizable. Multiplication with the positive diagonal matrix $\bm{C}^{-1}$ does not change this property. Accordingly, the matrix exponential
\begin{eqnarray*}
    e^{\bm{C}^{-1} \bm{K}} &= \bm{V^T} \begin{pmatrix}
                                        e^{-\lambda_1} & 0 & 0 & \cdots & 0 \\
                                        0 & e^{-\lambda_2} & 0 & \cdots & 0 \\
                                        0 & \ddots & \ddots & \ddots & 0 \\
                                        0 & \cdots & \cdots & 0 & e^{-\lambda_N} \\
                                        \end{pmatrix} \bm{V} 
\end{eqnarray*}
exists for an orthonormal matrix $\bm{V}$ and eigenvalues $-\lambda_k$ of $\bm{C}^{-1} \bm{K}$. Since only the first component of the forcing vector $\bm{F}(t)$ is non-zero, the matrix entry $(e^{\bm{C}^{-1} \bm{K}})_{11}$ defines the surface temperature response. The response function reads
\begin{equation*}
    R(t) = (e^{t \bm{C}^{-1} \bm{K}})_{11} = \sum_{k=1}^N w_k e^{- \lambda_k t} \,.
\end{equation*}
The normalization of the weights results from $\sum_{k=1}^N w_k = \sum_{k=1}^N (V_{k1})^2 = 1$ for any orthonormal matrix $\bm{V}$.

\section{Covariance of the Ornstein-Uhlenbeck and AR(1) process}\label{app:Covariance_OU_process}
\subsection{Covariance matrix}
The noise term in equation \eqref{eq:multibox-solution} is a weighted sum of Ornstein-Uhlenbeck (OU) processes. Consequently, its covariance structure results in:
    \begin{equation}
    \label{eq:cov-multibox}
        \Cov(T_{1,I}(t), T_{1,I}(t + s)) = \frac{\sigma_W^2}{C_1^2} \sum_{k=1}^N \sum_{l=1}^N w_k w_l \frac{e^{- \lambda_l \vert s \vert}}{\lambda_k + \lambda_l}
    \end{equation}
In the special case of $N=1$, the covariance reduces to the covariance of a single OU process \cite{rypdal2018} $\Cov(T_{1,I}(t), T_{1,I}(t + s))_{N=1} = \frac{\sigma_W^2}{2 C_1^2 \lambda_1}  e^{- \lambda_1 \vert s \vert}$. Formula~\eqref{eq:cov-multibox} follows from a generalization of this special case to arbitrary $N$. Discretizing the noise term in equation \eqref{eq:multibox-solution} 
results in a weighted sum of AR(1) processes. This sum $Z$ is normally-distributed with mean zero and covariance matrix
    \begin{equation}
    \label{eq:cov-ar1}
        \Cov(Z)_{ij} \coloneqq \Cov(Z_i, Z_j) = \frac{\sigma_W^2}{C_1^2} \sum_{k=1}^N \sum_{l=1}^N w_k w_l \frac{e^{- \lambda_l \vert i - j \vert}}{\lambda_k + \lambda_l} \,.
    \end{equation}
For given $\lambda_k$, $w_k$ and $C_1$, the SD of the stochastic forcing, $\sigma_W$, uniquely defines the SD of the internal variations $\sigma_I \coloneqq \sqrt{\Cov(Z_i, Z_i)}$, and vice versa. 
It is possible to estimate $\sigma_I$ within the Bayesian framework. However, additional uncertainties arise from the fact that the covariance matrix in equation \ref{eq:cov-ar1} is only an approximation to the true correlation structure of the residuals. As a result, the Bayesian estimation of $\sigma_I$ might not preserve the total variance. Therefore, we determine $\sigma_I$ from the residuals, that is the data minus the estimated forced response.

\subsection{Iterative computation of the likelihood}
\label{app:iterative-scheme}

Theoretically, the likelihood is given by a normal distribution with mean $T_{1,F}$ and covariance matrix $\Cov(Z)$, depending on $\theta$. However, computing the covariance matrix dynamically for each sample in the Markov chain can lead to difficulties: In particular, $\Cov(Z)$ needs to be inverted for every sample, which is computationally expensive. Moreover, the determinant $\mathrm{det}(\Cov(Z))$ can be close to zero, which can make numerical calculations unstable. Potential biases include decreasing goodness of fit and accuracy of estimated posteriors. 

To solve this problem, we propose an iterative approach. This keeps $\lambda_k$ and $w_k$ in the covariance matrix fixed for each iteration of the algorithm. The first iteration uses the prior means for $\lambda_k$ and $w_k$ as well as a starting value for the ratio $\sigma_W / C_1$. It is not necessary to consider $\sigma_W$ and $C_1$ separately, since equations (\ref{eq:cov-multibox}) and (\ref{eq:cov-ar1}) depend only on their ratio. This ratio is chosen such that $\sigma_I=0.1 \mathrm{K}$ for CMIP5 simulations and LOVECLIM ensemble members, and $\sigma_I = 0.05~\mathrm{K}$ for AR5 EMICs. For the second iteration, the estimated posterior means of $\lambda_k$ and $w_k$ define the covariance matrix entries. Additionally, $\sigma_I$ is set to the SD of the residuals, which defines $\sigma_W / C_1$. The results of this second iteration are the posterior distributions for $\lambda_1,..,\lambda_N$, weights $w_2,...,w_N$, heat capacity $C_1$, initial forcing $F_0$, and initial temperature $T_0$. These iterations can be repeated and adjusted with ``ClimBayes''. We find that two iterations are enough to fit the \textit{forced + internal} response to the considered data well, and that further iterations do not improve the goodness of fit.

\subsection{Sampling from internal variability}
\begin{figure}[!htbp]
\includegraphics[width=0.5\textwidth]{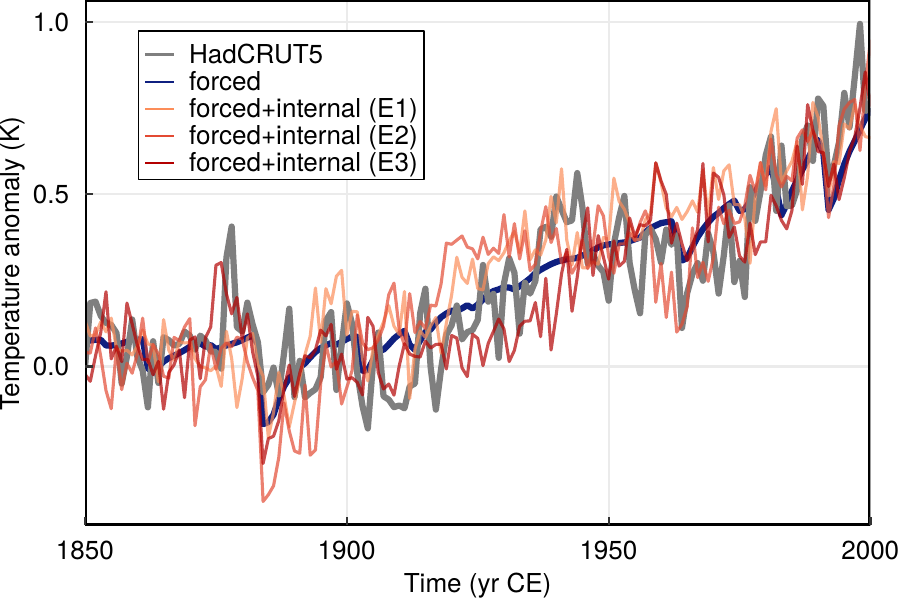}
\caption{\label{fig:fit_supp_fitnoise} HadCRUT5 \textit{target} observation, the posterior mean of the \textit{forced} response of the stochastic two-box EBM and three example realizations of the emulated \textit{forced + internal} variability.}
\end{figure}

Sampling internal variations $T_{1,I}(t)$ requires the values of $\lambda_k, w_k$, and $\sigma_W$ in $\Cov(Z)$ (equation \ref{eq:cov-multibox}). $\lambda_k$ and $w_k$ are set to the posterior means. We calculate $\sigma_W$ from $\sigma_I$, which we assume to equal the SD of the residuals. As an example, Figure \ref{fig:fit_supp_fitnoise} shows realizations of $T_{1,I}(t)$ drawn from the estimated covariance of HadCRUT5 observations.

\section{Comparison of one-, two- and three-box EBM}
\label{sec:supp_comparison_boxes}

\begin{table*}[!htbp]
\caption{\label{tab:hadcrut_params} Posterior means and 95\% CI of estimated feedback parameters and weights for the one-, two- and three-box EBM fitted to the HadCRUT5 GMST.}
\scriptsize
\begin{ruledtabular}
\begin{tabular}{lcccccr}
Number of boxes      & $\lambda_1 ($yr$^{-1})$ & $\lambda_2 ($yr$^{-1})$ & $\lambda_3 ($yr$^{-1})$ & $C_1$ (W yr m$^{-2}$ K$^{-1}$)  & $w_1$ (unitless) & $w_2$ (unitless)  \\ \hline
        1-box & 0.35 (0.21,0.63) & -  & -  &  9.08 (6.83,10.68) & - & -  \\ 
        2-box & 1.31 (0.71,1.86) & 0.09 (0.03,0.16) & - & 8.20 (5.68,10.43) & 0.72 (0.46,0.9) & -  \\ 
        3-box & 1.29 (0.70,1.86) & 0.11 (0.04,0.18) & 0.01 (0.01,0.02) & 8.46 (5.86,10.57) & 0.74 (0.46,0.90) & 0.22 (0.02,0.53)  \\
\end{tabular}
\end{ruledtabular}
\end{table*}

\begin{figure}[!htbp]
\includegraphics[width=0.5\textwidth]{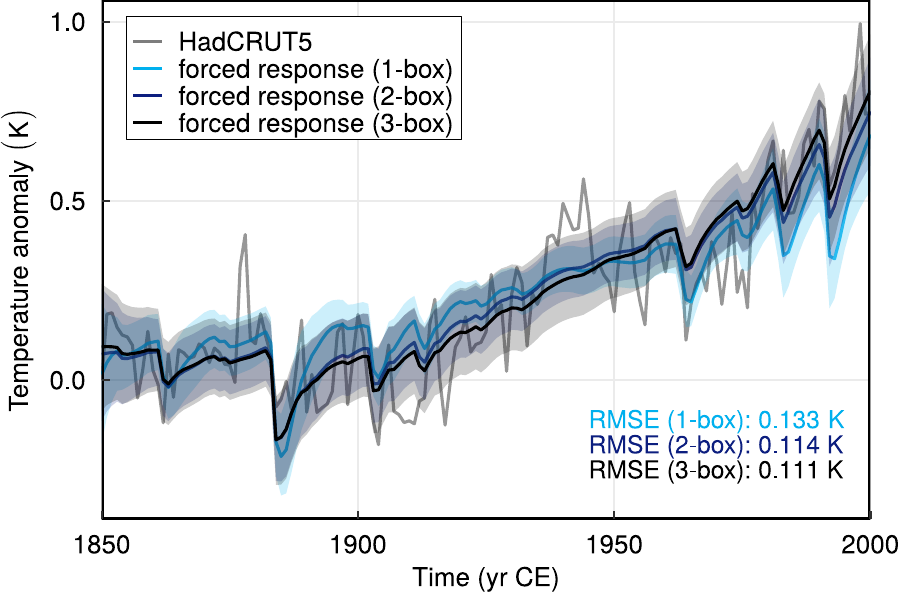} 
\caption{\label{fig:fit_supp_had} HadCRUT5 \textit{target} observation (grey), the \textit{forced} response from the one-, two- and three-box EBM fit to the data. We show their posterior means and the CIs (shaded) as well as their root mean square errors (RMSE).}
\end{figure}

We have verified that our results are robust against reasonable variations of the number of boxes. Here, we examine the difference between $N\in (1,2,3)$ boxes on the example of the HadCRUT5 GMST (Fig.~\ref{fig:fit_supp_had}). We choose the priors of the response parameters to cover the same overall range as for the two-box model ($N=1$: $\lambda_1 \in (1/200, 2) \, \mathrm{yr^{-1}}$, and $N=3$: $\lambda_1 \in (1/5, 2) \mathrm{yr^{-1}}, \lambda_2 \in (1/50, 1/5) \mathrm{yr^{-1}}, \lambda_3 \in (1/200, 1/50)  \mathrm{yr^{-1}}$).

The stochastic two-box EBM fits the data more accurately (root mean square error: RMSE = 0.114~K) than the one-box EBM (RMSE = 0.133~K) (Appendix Fig.~\ref{fig:fit_supp_had}). The three-box EBM yields only minor improvements (RMSE = 0.0111~K). This pattern is consistent for AR5 EMICs and CMIP5 simulations, and reflected in similar forced responses and power spectral densities for $N \in (1,2,3)$. Adding boxes, however, increases the risk of overfitting due to increasing degrees of freedom. This is reflected in increasing CIs for the forced response and parameters (Tab.~\ref{tab:hadcrut_params}) with more boxes. That is why $N=2$ represents the best compromise between goodness of fit, identifiability of parameters, and number of free parameters in our experiments.

\section{Emulated forced temperature response for considered simulations}

\begin{figure*}[!htbp]
\includegraphics[width=\textwidth]{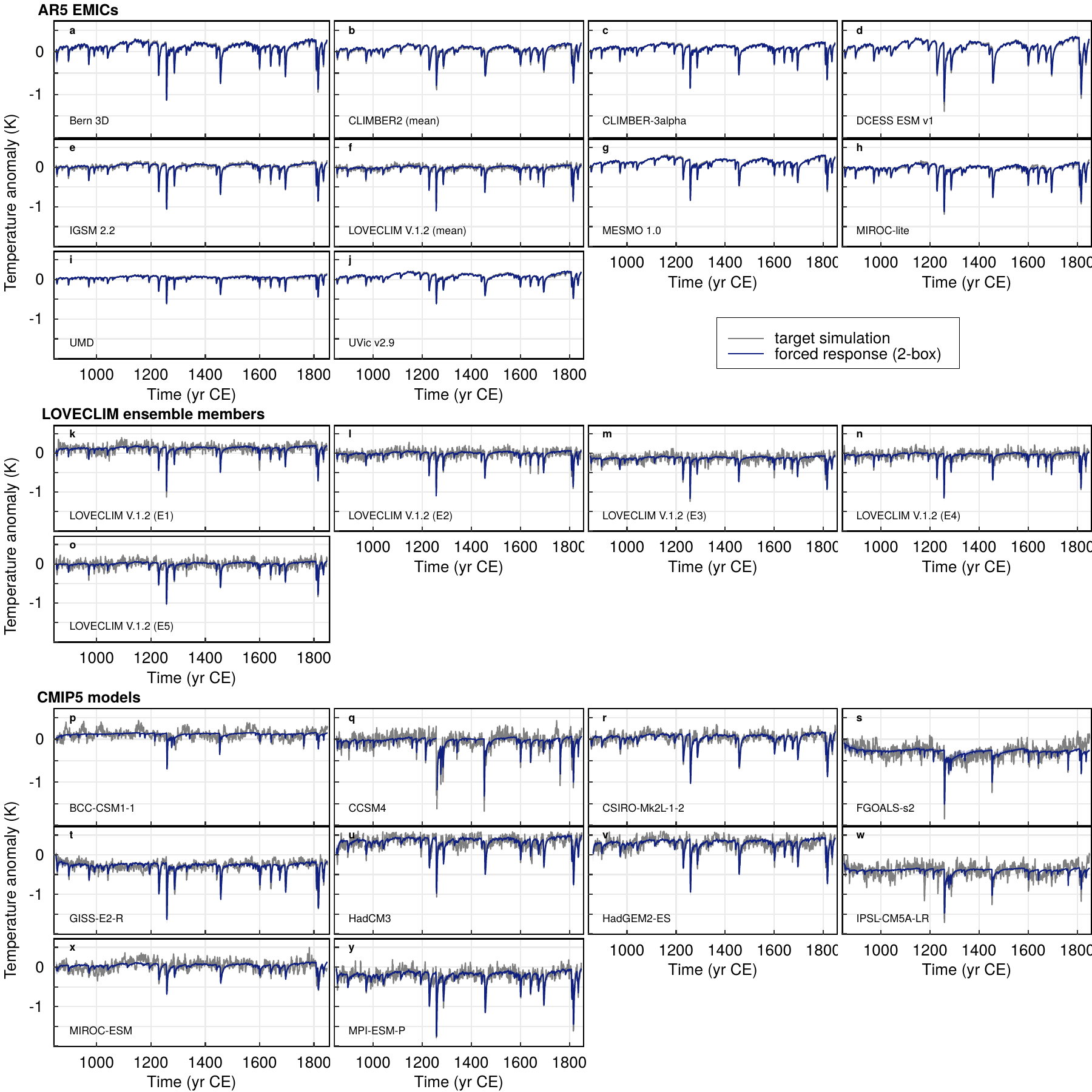} 
\caption{\label{fig:fit_supp} Emulated \textit{forced} response $T_{1,F}(t)$, that is the posterior mean, of the stochastic two-box EBM fitted to the GMST target data from AR5 EMICs (panel \textbf{a-j}), LOVECLIM  V.1.2 ensemble members (panel \textbf{k-o}) and CMIP5 simulations (panel \textbf{p-y}).}
\end{figure*}

Figure~\ref{fig:fit_supp} shows the best estimate of the EBM's forced response, fitted to the target simulations from all considered models. CIs are much narrower and almost vanishing compared to HadCRUT (Fig.~\ref{fig:demonstration}a). This is due to the fact that with increasing length of the time series the posterior uncertainties of the parameters and the forced response decrease. 

\begin{figure}
    \centering
    \includegraphics[width=0.5\textwidth]{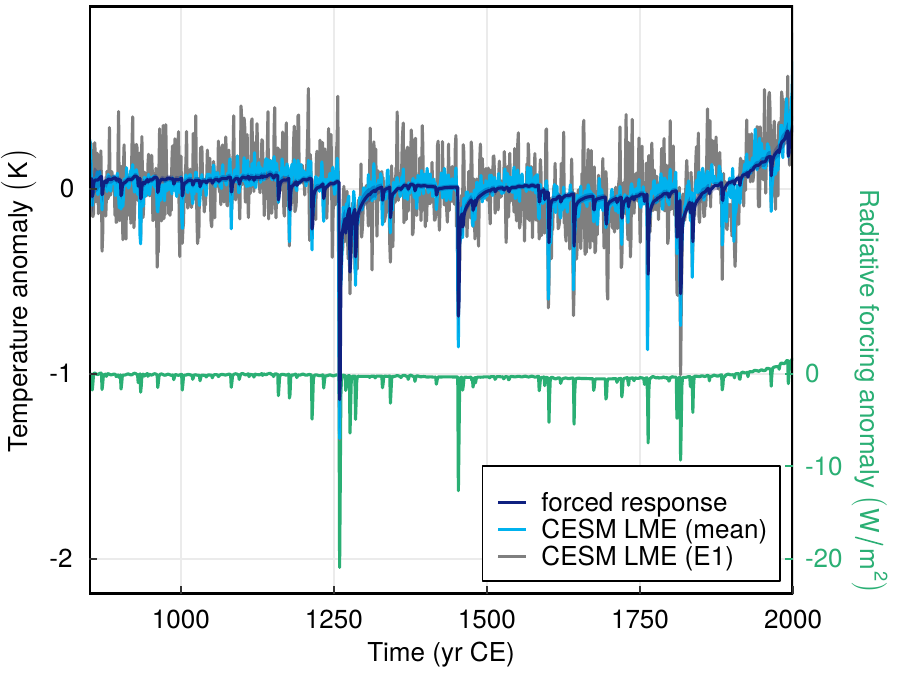}
    \caption{Ensemble mean of the CESM Large Ensemble \citep{otto-bliesner2016} (13 members) and the emulated \textit{forced} response $T_{1,F}(t)$ of the stochastic two-box EBM fitted to GMST target data from one of the ensemble members (E1). The corresponding radiative forcing time series is shown in green.}
    \label{fig:LME}
\end{figure}

Figure ~\ref{fig:LME} compares the emulated \textit{forced} response against simulation data from the Last Millennium Ensemble of the Community Earth System Model (CESM) \cite{otto-bliesner2016}. Forming the ensemble mean over the available 13 members serves to average out uncorrelated internal variability. Our emulated \textit{forced} response, fitted to the first ensemble member (E1), shows a large overlap with the ensemble mean despite remaining internal variability that has not been averaged out.

\clearpage


\begin{thebibliography}{110}%
\makeatletter
\providecommand \@ifxundefined [1]{%
 \@ifx{#1\undefined}
}%
\providecommand \@ifnum [1]{%
 \ifnum #1\expandafter \@firstoftwo
 \else \expandafter \@secondoftwo
 \fi
}%
\providecommand \@ifx [1]{%
 \ifx #1\expandafter \@firstoftwo
 \else \expandafter \@secondoftwo
 \fi
}%
\providecommand \natexlab [1]{#1}%
\providecommand \enquote  [1]{``#1''}%
\providecommand \bibnamefont  [1]{#1}%
\providecommand \bibfnamefont [1]{#1}%
\providecommand \citenamefont [1]{#1}%
\providecommand \href@noop [0]{\@secondoftwo}%
\providecommand \href [0]{\begingroup \@sanitize@url \@href}%
\providecommand \@href[1]{\@@startlink{#1}\@@href}%
\providecommand \@@href[1]{\endgroup#1\@@endlink}%
\providecommand \@sanitize@url [0]{\catcode `\\12\catcode `\$12\catcode
  `\&12\catcode `\#12\catcode `\^12\catcode `\_12\catcode `\%12\relax}%
\providecommand \@@startlink[1]{}%
\providecommand \@@endlink[0]{}%
\providecommand \url  [0]{\begingroup\@sanitize@url \@url }%
\providecommand \@url [1]{\endgroup\@href {#1}{\urlprefix }}%
\providecommand \urlprefix  [0]{URL }%
\providecommand \Eprint [0]{\href }%
\providecommand \doibase [0]{http://dx.doi.org/}%
\providecommand \selectlanguage [0]{\@gobble}%
\providecommand \bibinfo  [0]{\@secondoftwo}%
\providecommand \bibfield  [0]{\@secondoftwo}%
\providecommand \translation [1]{[#1]}%
\providecommand \BibitemOpen [0]{}%
\providecommand \bibitemStop [0]{}%
\providecommand \bibitemNoStop [0]{.\EOS\space}%
\providecommand \EOS [0]{\spacefactor3000\relax}%
\providecommand \BibitemShut  [1]{\csname bibitem#1\endcsname}%
\let\auto@bib@innerbib\@empty
\bibitem [{\citenamefont {Katz}\ and\ \citenamefont {Brown}(1992)}]{katz1992}%
  \BibitemOpen
  \bibfield  {author} {\bibinfo {author} {\bibfnamefont {R.~W.}\ \bibnamefont
  {Katz}}\ and\ \bibinfo {author} {\bibfnamefont {B.~G.}\ \bibnamefont
  {Brown}},\ }\bibfield  {title} {\enquote {\bibinfo {title} {Extreme events in
  a changing climate: {{Variability}} is more important than averages},}\
  }\href {\doibase 10.1007/BF00139728} {\bibfield  {journal} {\bibinfo
  {journal} {Climatic Change}\ }\textbf {\bibinfo {volume} {21}},\ \bibinfo
  {pages} {289--302} (\bibinfo {year} {1992})}\BibitemShut {NoStop}%
\bibitem [{\citenamefont {{PAGES 2k
  Consortium}}(2019)}]{pages2kconsortium2019}%
  \BibitemOpen
  \bibfield  {author} {\bibinfo {author} {\bibnamefont {{PAGES 2k
  Consortium}}},\ }\bibfield  {title} {\enquote {\bibinfo {title} {Consistent
  multidecadal variability in global temperature reconstructions and
  simulations over the {{Common Era}}},}\ }\href {\doibase
  10.1038/s41561-019-0400-0} {\bibfield  {journal} {\bibinfo  {journal} {Nature
  Geoscience}\ }\textbf {\bibinfo {volume} {12}},\ \bibinfo {pages} {643--649}
  (\bibinfo {year} {2019})}\BibitemShut {NoStop}%
\bibitem [{\citenamefont {Ellerhoff}\ and\ \citenamefont
  {Rehfeld}(2021)}]{ellerhoff2021}%
  \BibitemOpen
  \bibfield  {author} {\bibinfo {author} {\bibfnamefont {B.}~\bibnamefont
  {Ellerhoff}}\ and\ \bibinfo {author} {\bibfnamefont {K.}~\bibnamefont
  {Rehfeld}},\ }\bibfield  {title} {\enquote {\bibinfo {title} {Probing the
  timescale dependency of local and global variations in surface air
  temperature from climate simulations and reconstructions of the last
  millennia},}\ }\href {\doibase 10.1103/PhysRevE.104.064136} {\bibfield
  {journal} {\bibinfo  {journal} {Physical Review E}\ }\textbf {\bibinfo
  {volume} {104}},\ \bibinfo {pages} {064136} (\bibinfo {year}
  {2021})}\BibitemShut {NoStop}%
\bibitem [{\citenamefont {Hawkins}\ and\ \citenamefont
  {Sutton}(2011)}]{hawkins2011}%
  \BibitemOpen
  \bibfield  {author} {\bibinfo {author} {\bibfnamefont {E.}~\bibnamefont
  {Hawkins}}\ and\ \bibinfo {author} {\bibfnamefont {R.}~\bibnamefont
  {Sutton}},\ }\bibfield  {title} {\enquote {\bibinfo {title} {The potential to
  narrow uncertainty in projections of regional precipitation change},}\ }\href
  {\doibase 10.1007/s00382-010-0810-6} {\bibfield  {journal} {\bibinfo
  {journal} {Climate Dynamics}\ }\textbf {\bibinfo {volume} {37}},\ \bibinfo
  {pages} {407--418} (\bibinfo {year} {2011})}\BibitemShut {NoStop}%
\bibitem [{\citenamefont {Frankcombe}\ \emph {et~al.}(2015)\citenamefont
  {Frankcombe}, \citenamefont {England}, \citenamefont {Mann},\ and\
  \citenamefont {Steinman}}]{frankcombe2015}%
  \BibitemOpen
  \bibfield  {author} {\bibinfo {author} {\bibfnamefont {L.~M.}\ \bibnamefont
  {Frankcombe}}, \bibinfo {author} {\bibfnamefont {M.~H.}\ \bibnamefont
  {England}}, \bibinfo {author} {\bibfnamefont {M.~E.}\ \bibnamefont {Mann}}, \
  and\ \bibinfo {author} {\bibfnamefont {B.~A.}\ \bibnamefont {Steinman}},\
  }\bibfield  {title} {\enquote {\bibinfo {title} {Separating {{Internal
  Variability}} from the {{Externally Forced Climate Response}}},}\ }\href
  {\doibase 10.1175/JCLI-D-15-0069.1} {\bibfield  {journal} {\bibinfo
  {journal} {Journal of Climate}\ }\textbf {\bibinfo {volume} {28}},\ \bibinfo
  {pages} {8184--8202} (\bibinfo {year} {2015})}\BibitemShut {NoStop}%
\bibitem [{\citenamefont {H{\'e}bert}\ and\ \citenamefont
  {Lovejoy}(2018)}]{hebert2018}%
  \BibitemOpen
  \bibfield  {author} {\bibinfo {author} {\bibfnamefont {R.}~\bibnamefont
  {H{\'e}bert}}\ and\ \bibinfo {author} {\bibfnamefont {S.}~\bibnamefont
  {Lovejoy}},\ }\bibfield  {title} {\enquote {\bibinfo {title} {Regional
  {{Climate Sensitivity-}} and {{Historical-Based Projections}} to 2100},}\
  }\href {\doibase 10.1002/2017GL076649} {\bibfield  {journal} {\bibinfo
  {journal} {Geophysical Research Letters}\ }\textbf {\bibinfo {volume} {45}},\
  \bibinfo {pages} {4248--4254} (\bibinfo {year} {2018})}\BibitemShut {NoStop}%
\bibitem [{\citenamefont {Laepple}\ and\ \citenamefont
  {Huybers}(2014{\natexlab{a}})}]{laepple2014}%
  \BibitemOpen
  \bibfield  {author} {\bibinfo {author} {\bibfnamefont {T.}~\bibnamefont
  {Laepple}}\ and\ \bibinfo {author} {\bibfnamefont {P.}~\bibnamefont
  {Huybers}},\ }\bibfield  {title} {\enquote {\bibinfo {title} {Global and
  regional variability in marine surface temperatures},}\ }\href {\doibase
  10.1002/2014GL059345} {\bibfield  {journal} {\bibinfo  {journal} {Geophysical
  Research Letters}\ }\textbf {\bibinfo {volume} {41}},\ \bibinfo {pages}
  {2528--2534} (\bibinfo {year} {2014}{\natexlab{a}})}\BibitemShut {NoStop}%
\bibitem [{\citenamefont {Budyko}(1969)}]{budyko1969}%
  \BibitemOpen
  \bibfield  {author} {\bibinfo {author} {\bibfnamefont {M.~I.}\ \bibnamefont
  {Budyko}},\ }\bibfield  {title} {\enquote {\bibinfo {title} {The effect of
  solar radiation variations on the climate of the {{Earth}}},}\ }\href
  {\doibase 10.3402/tellusa.v21i5.10109} {\bibfield  {journal} {\bibinfo
  {journal} {Tellus}\ }\textbf {\bibinfo {volume} {21}},\ \bibinfo {pages}
  {611--619} (\bibinfo {year} {1969})}\BibitemShut {NoStop}%
\bibitem [{\citenamefont {Sellers}(1969)}]{sellers1969}%
  \BibitemOpen
  \bibfield  {author} {\bibinfo {author} {\bibfnamefont {W.~D.}\ \bibnamefont
  {Sellers}},\ }\bibfield  {title} {\enquote {\bibinfo {title} {A {{Global
  Climatic Model Based}} on the {{Energy Balance}} of the {{Earth-Atmosphere
  System}}},}\ }\href {\doibase
  10.1175/1520-0450(1969)008<0392:AGCMBO>2.0.CO;2} {\bibfield  {journal}
  {\bibinfo  {journal} {Journal of Applied Meteorology}\ }\textbf {\bibinfo
  {volume} {8}},\ \bibinfo {pages} {392--400} (\bibinfo {year}
  {1969})}\BibitemShut {NoStop}%
\bibitem [{\citenamefont {Hasselmann}(1976)}]{hasselmann1976}%
  \BibitemOpen
  \bibfield  {author} {\bibinfo {author} {\bibfnamefont {K.}~\bibnamefont
  {Hasselmann}},\ }\bibfield  {title} {\enquote {\bibinfo {title} {Stochastic
  climate models {{Part I}}. {{Theory}}},}\ }\href {\doibase
  10.3402/tellusa.v28i6.11316} {\bibfield  {journal} {\bibinfo  {journal}
  {Tellus}\ }\textbf {\bibinfo {volume} {28}},\ \bibinfo {pages} {473--485}
  (\bibinfo {year} {1976})}\BibitemShut {NoStop}%
\bibitem [{\citenamefont {Fraedrich}, \citenamefont {Luksch},\ and\
  \citenamefont {Blender}(2004)}]{fraedrich2004}%
  \BibitemOpen
  \bibfield  {author} {\bibinfo {author} {\bibfnamefont {K.}~\bibnamefont
  {Fraedrich}}, \bibinfo {author} {\bibfnamefont {U.}~\bibnamefont {Luksch}}, \
  and\ \bibinfo {author} {\bibfnamefont {R.}~\bibnamefont {Blender}},\
  }\bibfield  {title} {\enquote {\bibinfo {title} {1 / f model for long-time
  memory of the ocean surface temperature},}\ }\href {\doibase
  10.1103/PhysRevE.70.037301} {\bibfield  {journal} {\bibinfo  {journal}
  {Physical Review E}\ }\textbf {\bibinfo {volume} {70}},\ \bibinfo {pages}
  {037301} (\bibinfo {year} {2004})}\BibitemShut {NoStop}%
\bibitem [{\citenamefont {Held}\ \emph {et~al.}(2010)\citenamefont {Held},
  \citenamefont {Winton}, \citenamefont {Takahashi}, \citenamefont {Delworth},
  \citenamefont {Zeng},\ and\ \citenamefont {Vallis}}]{held2010}%
  \BibitemOpen
  \bibfield  {author} {\bibinfo {author} {\bibfnamefont {I.~M.}\ \bibnamefont
  {Held}}, \bibinfo {author} {\bibfnamefont {M.}~\bibnamefont {Winton}},
  \bibinfo {author} {\bibfnamefont {K.}~\bibnamefont {Takahashi}}, \bibinfo
  {author} {\bibfnamefont {T.}~\bibnamefont {Delworth}}, \bibinfo {author}
  {\bibfnamefont {F.}~\bibnamefont {Zeng}}, \ and\ \bibinfo {author}
  {\bibfnamefont {G.~K.}\ \bibnamefont {Vallis}},\ }\bibfield  {title}
  {\enquote {\bibinfo {title} {Probing the {{Fast}} and {{Slow Components}} of
  {{Global Warming}} by {{Returning Abruptly}} to {{Preindustrial Forcing}}},}\
  }\href {\doibase 10.1175/2009JCLI3466.1} {\bibfield  {journal} {\bibinfo
  {journal} {Journal of Climate}\ }\textbf {\bibinfo {volume} {23}},\ \bibinfo
  {pages} {2418--2427} (\bibinfo {year} {2010})}\BibitemShut {NoStop}%
\bibitem [{\citenamefont {Geoffroy}\ \emph
  {et~al.}(2013{\natexlab{a}})\citenamefont {Geoffroy}, \citenamefont
  {{Saint-Martin}}, \citenamefont {Olivi{\'e}}, \citenamefont {Voldoire},
  \citenamefont {Bellon},\ and\ \citenamefont {Tyt{\'e}ca}}]{geoffroy2013a}%
  \BibitemOpen
  \bibfield  {author} {\bibinfo {author} {\bibfnamefont {O.}~\bibnamefont
  {Geoffroy}}, \bibinfo {author} {\bibfnamefont {D.}~\bibnamefont
  {{Saint-Martin}}}, \bibinfo {author} {\bibfnamefont {D.~J.~L.}\ \bibnamefont
  {Olivi{\'e}}}, \bibinfo {author} {\bibfnamefont {A.}~\bibnamefont
  {Voldoire}}, \bibinfo {author} {\bibfnamefont {G.}~\bibnamefont {Bellon}}, \
  and\ \bibinfo {author} {\bibfnamefont {S.}~\bibnamefont {Tyt{\'e}ca}},\
  }\bibfield  {title} {\enquote {\bibinfo {title} {Transient {{Climate
  Response}} in a {{Two-Layer Energy-Balance Model}}. {{Part I}}: {{Analytical
  Solution}} and {{Parameter Calibration Using CMIP5 AOGCM Experiments}}},}\
  }\href {\doibase 10.1175/JCLI-D-12-00195.1} {\bibfield  {journal} {\bibinfo
  {journal} {Journal of Climate}\ }\textbf {\bibinfo {volume} {26}},\ \bibinfo
  {pages} {1841--1857} (\bibinfo {year} {2013}{\natexlab{a}})}\BibitemShut
  {NoStop}%
\bibitem [{\citenamefont {Rypdal}\ and\ \citenamefont
  {Rypdal}(2014)}]{rypdal2014}%
  \BibitemOpen
  \bibfield  {author} {\bibinfo {author} {\bibfnamefont {M.}~\bibnamefont
  {Rypdal}}\ and\ \bibinfo {author} {\bibfnamefont {K.}~\bibnamefont
  {Rypdal}},\ }\bibfield  {title} {\enquote {\bibinfo {title} {Long-{{Memory
  Effects}} in {{Linear Response Models}} of {{Earth}}'s {{Temperature}} and
  {{Implications}} for {{Future Global Warming}}},}\ }\href {\doibase
  10.1175/JCLI-D-13-00296.1} {\bibfield  {journal} {\bibinfo  {journal}
  {Journal of Climate}\ }\textbf {\bibinfo {volume} {27}},\ \bibinfo {pages}
  {5240--5258} (\bibinfo {year} {2014})}\BibitemShut {NoStop}%
\bibitem [{\citenamefont {Fredriksen}\ and\ \citenamefont
  {Rypdal}(2017)}]{fredriksen2017}%
  \BibitemOpen
  \bibfield  {author} {\bibinfo {author} {\bibfnamefont {H.-B.}\ \bibnamefont
  {Fredriksen}}\ and\ \bibinfo {author} {\bibfnamefont {M.}~\bibnamefont
  {Rypdal}},\ }\bibfield  {title} {\enquote {\bibinfo {title} {Long-{{Range
  Persistence}} in {{Global Surface Temperatures Explained}} by {{Linear
  Multibox Energy Balance Models}}},}\ }\href {\doibase
  10.1175/JCLI-D-16-0877.1} {\bibfield  {journal} {\bibinfo  {journal} {Journal
  of Climate}\ }\textbf {\bibinfo {volume} {30}},\ \bibinfo {pages}
  {7157--7168} (\bibinfo {year} {2017})}\BibitemShut {NoStop}%
\bibitem [{\citenamefont {Hasselmann}(1993)}]{hasselmann1993}%
  \BibitemOpen
  \bibfield  {author} {\bibinfo {author} {\bibfnamefont {K.}~\bibnamefont
  {Hasselmann}},\ }\bibfield  {title} {\enquote {\bibinfo {title} {Optimal
  {{Fingerprints}} for the {{Detection}} of {{Time-dependent Climate
  Change}}},}\ }\href {\doibase
  10.1175/1520-0442(1993)006<1957:OFFTDO>2.0.CO;2} {\bibfield  {journal}
  {\bibinfo  {journal} {Journal of Climate}\ }\textbf {\bibinfo {volume} {6}},\
  \bibinfo {pages} {1957--1971} (\bibinfo {year} {1993})}\BibitemShut {NoStop}%
\bibitem [{\citenamefont {Hegerl}\ \emph {et~al.}(1996)\citenamefont {Hegerl},
  \citenamefont {{von Storch}}, \citenamefont {Hasselmann}, \citenamefont
  {Santer}, \citenamefont {Cubasch},\ and\ \citenamefont {Jones}}]{hegerl1996}%
  \BibitemOpen
  \bibfield  {author} {\bibinfo {author} {\bibfnamefont {G.~C.}\ \bibnamefont
  {Hegerl}}, \bibinfo {author} {\bibfnamefont {H.}~\bibnamefont {{von
  Storch}}}, \bibinfo {author} {\bibfnamefont {K.}~\bibnamefont {Hasselmann}},
  \bibinfo {author} {\bibfnamefont {B.~D.}\ \bibnamefont {Santer}}, \bibinfo
  {author} {\bibfnamefont {U.}~\bibnamefont {Cubasch}}, \ and\ \bibinfo
  {author} {\bibfnamefont {P.~D.}\ \bibnamefont {Jones}},\ }\bibfield  {title}
  {\enquote {\bibinfo {title} {Detecting {{Greenhouse-Gas-Induced Climate
  Change}} with an {{Optimal Fingerprint Method}}},}\ }\href {\doibase
  10.1175/1520-0442(1996)009<2281:DGGICC>2.0.CO;2} {\bibfield  {journal}
  {\bibinfo  {journal} {Journal of Climate}\ }\textbf {\bibinfo {volume} {9}},\
  \bibinfo {pages} {2281--2306} (\bibinfo {year} {1996})}\BibitemShut {NoStop}%
\bibitem [{\citenamefont {Ghil}(1984)}]{ghil1984}%
  \BibitemOpen
  \bibfield  {author} {\bibinfo {author} {\bibfnamefont {M.}~\bibnamefont
  {Ghil}},\ }\bibfield  {title} {\enquote {\bibinfo {title} {Climate
  sensitivity, energy balance models, and oscillatory climate models},}\ }\href
  {\doibase 10.1029/JD089iD01p01280} {\bibfield  {journal} {\bibinfo  {journal}
  {Journal of Geophysical Research}\ }\textbf {\bibinfo {volume} {89}},\
  \bibinfo {pages} {1280} (\bibinfo {year} {1984})}\BibitemShut {NoStop}%
\bibitem [{\citenamefont {Cox}, \citenamefont {Huntingford},\ and\
  \citenamefont {Williamson}(2018)}]{cox2018}%
  \BibitemOpen
  \bibfield  {author} {\bibinfo {author} {\bibfnamefont {P.~M.}\ \bibnamefont
  {Cox}}, \bibinfo {author} {\bibfnamefont {C.}~\bibnamefont {Huntingford}}, \
  and\ \bibinfo {author} {\bibfnamefont {M.~S.}\ \bibnamefont {Williamson}},\
  }\bibfield  {title} {\enquote {\bibinfo {title} {Emergent constraint on
  equilibrium climate sensitivity from global temperature variability},}\
  }\href {\doibase 10.1038/nature25450} {\bibfield  {journal} {\bibinfo
  {journal} {Nature}\ }\textbf {\bibinfo {volume} {553}},\ \bibinfo {pages}
  {319--322} (\bibinfo {year} {2018})}\BibitemShut {NoStop}%
\bibitem [{\citenamefont {Rypdal}, \citenamefont {Rypdal},\ and\ \citenamefont
  {Fredriksen}(2015)}]{rypdal2015}%
  \BibitemOpen
  \bibfield  {author} {\bibinfo {author} {\bibfnamefont {K.}~\bibnamefont
  {Rypdal}}, \bibinfo {author} {\bibfnamefont {M.}~\bibnamefont {Rypdal}}, \
  and\ \bibinfo {author} {\bibfnamefont {H.-B.}\ \bibnamefont {Fredriksen}},\
  }\bibfield  {title} {\enquote {\bibinfo {title} {Spatiotemporal {{Long-Range
  Persistence}} in {{Earth}}'s {{Temperature Field}}: {{Analysis}} of
  {{Stochastic}}\textendash{{Diffusive Energy Balance Models}}},}\ }\href
  {\doibase 10.1175/JCLI-D-15-0183.1} {\bibfield  {journal} {\bibinfo
  {journal} {Journal of Climate}\ }\textbf {\bibinfo {volume} {28}},\ \bibinfo
  {pages} {8379--8395} (\bibinfo {year} {2015})}\BibitemShut {NoStop}%
\bibitem [{\citenamefont {Ghil}(1976)}]{ghil1976}%
  \BibitemOpen
  \bibfield  {author} {\bibinfo {author} {\bibfnamefont {M.}~\bibnamefont
  {Ghil}},\ }\bibfield  {title} {\enquote {\bibinfo {title} {Climate
  {{Stability}} for a {{Sellers-Type Model}}},}\ }\href {\doibase
  10.1175/1520-0469(1976)033<0003:CSFAST>2.0.CO;2} {\bibfield  {journal}
  {\bibinfo  {journal} {Journal of the Atmospheric Sciences}\ }\textbf
  {\bibinfo {volume} {33}},\ \bibinfo {pages} {3--20} (\bibinfo {year}
  {1976})}\BibitemShut {NoStop}%
\bibitem [{\citenamefont {North}, \citenamefont {Mengel},\ and\ \citenamefont
  {Short}(1983)}]{north1983}%
  \BibitemOpen
  \bibfield  {author} {\bibinfo {author} {\bibfnamefont {G.~R.}\ \bibnamefont
  {North}}, \bibinfo {author} {\bibfnamefont {J.~G.}\ \bibnamefont {Mengel}}, \
  and\ \bibinfo {author} {\bibfnamefont {D.~A.}\ \bibnamefont {Short}},\
  }\bibfield  {title} {\enquote {\bibinfo {title} {Simple energy balance model
  resolving the seasons and the continents: {{Application}} to the astronomical
  theory of the ice ages},}\ }\href {\doibase 10.1029/JC088iC11p06576}
  {\bibfield  {journal} {\bibinfo  {journal} {Journal of Geophysical Research}\
  }\textbf {\bibinfo {volume} {88}},\ \bibinfo {pages} {6576} (\bibinfo {year}
  {1983})}\BibitemShut {NoStop}%
\bibitem [{\citenamefont {B{\'o}dai}\ \emph {et~al.}(2015)\citenamefont
  {B{\'o}dai}, \citenamefont {Lucarini}, \citenamefont {Lunkeit},\ and\
  \citenamefont {Boschi}}]{bodai2015}%
  \BibitemOpen
  \bibfield  {author} {\bibinfo {author} {\bibfnamefont {T.}~\bibnamefont
  {B{\'o}dai}}, \bibinfo {author} {\bibfnamefont {V.}~\bibnamefont {Lucarini}},
  \bibinfo {author} {\bibfnamefont {F.}~\bibnamefont {Lunkeit}}, \ and\
  \bibinfo {author} {\bibfnamefont {R.}~\bibnamefont {Boschi}},\ }\bibfield
  {title} {\enquote {\bibinfo {title} {Global instability in the
  {{Ghil}}\textendash{{Sellers}} model},}\ }\href {\doibase
  10.1007/s00382-014-2206-5} {\bibfield  {journal} {\bibinfo  {journal}
  {Climate Dynamics}\ }\textbf {\bibinfo {volume} {44}},\ \bibinfo {pages}
  {3361--3381} (\bibinfo {year} {2015})}\BibitemShut {NoStop}%
\bibitem [{\citenamefont {{\O}stvand}\ \emph {et~al.}(2014)\citenamefont
  {{\O}stvand}, \citenamefont {Nilsen}, \citenamefont {Rypdal}, \citenamefont
  {Divine},\ and\ \citenamefont {Rypdal}}]{ostvand2014}%
  \BibitemOpen
  \bibfield  {author} {\bibinfo {author} {\bibfnamefont {L.}~\bibnamefont
  {{\O}stvand}}, \bibinfo {author} {\bibfnamefont {T.}~\bibnamefont {Nilsen}},
  \bibinfo {author} {\bibfnamefont {K.}~\bibnamefont {Rypdal}}, \bibinfo
  {author} {\bibfnamefont {D.}~\bibnamefont {Divine}}, \ and\ \bibinfo {author}
  {\bibfnamefont {M.}~\bibnamefont {Rypdal}},\ }\bibfield  {title} {\enquote
  {\bibinfo {title} {Long-range memory in internal and forced dynamics of
  millennium-long climate model simulations},}\ }\href {\doibase
  10.5194/esd-5-295-2014} {\bibfield  {journal} {\bibinfo  {journal} {Earth
  System Dynamics}\ }\textbf {\bibinfo {volume} {5}},\ \bibinfo {pages}
  {295--308} (\bibinfo {year} {2014})}\BibitemShut {NoStop}%
\bibitem [{\citenamefont {North}(1975)}]{north1975}%
  \BibitemOpen
  \bibfield  {author} {\bibinfo {author} {\bibfnamefont {G.~R.}\ \bibnamefont
  {North}},\ }\bibfield  {title} {\enquote {\bibinfo {title} {Analytical
  {{Solution}} to a {{Simple Climate Model}} with {{Diffusive Heat
  Transport}}},}\ }\href {\doibase
  10.1175/1520-0469(1975)032<1301:ASTASC>2.0.CO;2} {\bibfield  {journal}
  {\bibinfo  {journal} {Journal of the Atmospheric Sciences}\ }\textbf
  {\bibinfo {volume} {32}},\ \bibinfo {pages} {1301--1307} (\bibinfo {year}
  {1975})}\BibitemShut {NoStop}%
\bibitem [{\citenamefont {North}\ and\ \citenamefont
  {Cahalan}(1981)}]{north1981}%
  \BibitemOpen
  \bibfield  {author} {\bibinfo {author} {\bibfnamefont {G.~R.}\ \bibnamefont
  {North}}\ and\ \bibinfo {author} {\bibfnamefont {R.~F.}\ \bibnamefont
  {Cahalan}},\ }\bibfield  {title} {\enquote {\bibinfo {title} {Predictability
  in a {{Solvable Stochastic Climate Model}}},}\ }\href {\doibase
  10.1175/1520-0469(1981)038<0504:PIASSC>2.0.CO;2} {\bibfield  {journal}
  {\bibinfo  {journal} {Journal of the Atmospheric Sciences}\ }\textbf
  {\bibinfo {volume} {38}},\ \bibinfo {pages} {504--513} (\bibinfo {year}
  {1981})}\BibitemShut {NoStop}%
\bibitem [{\citenamefont {North}, \citenamefont {Wang},\ and\ \citenamefont
  {Genton}(2011)}]{north2011}%
  \BibitemOpen
  \bibfield  {author} {\bibinfo {author} {\bibfnamefont {G.~R.}\ \bibnamefont
  {North}}, \bibinfo {author} {\bibfnamefont {J.}~\bibnamefont {Wang}}, \ and\
  \bibinfo {author} {\bibfnamefont {M.~G.}\ \bibnamefont {Genton}},\ }\bibfield
   {title} {\enquote {\bibinfo {title} {Correlation {{Models}} for
  {{Temperature Fields}}},}\ }\href {\doibase 10.1175/2011JCLI4199.1}
  {\bibfield  {journal} {\bibinfo  {journal} {Journal of Climate}\ }\textbf
  {\bibinfo {volume} {24}},\ \bibinfo {pages} {5850--5862} (\bibinfo {year}
  {2011})}\BibitemShut {NoStop}%
\bibitem [{\citenamefont {Dortmans}, \citenamefont {Langford},\ and\
  \citenamefont {Willms}(2019)}]{dortmans2019}%
  \BibitemOpen
  \bibfield  {author} {\bibinfo {author} {\bibfnamefont {B.}~\bibnamefont
  {Dortmans}}, \bibinfo {author} {\bibfnamefont {W.~F.}\ \bibnamefont
  {Langford}}, \ and\ \bibinfo {author} {\bibfnamefont {A.~R.}\ \bibnamefont
  {Willms}},\ }\bibfield  {title} {\enquote {\bibinfo {title} {An energy
  balance model for paleoclimate transitions},}\ }\href {\doibase
  10.5194/cp-15-493-2019} {\bibfield  {journal} {\bibinfo  {journal} {Climate
  of the Past}\ }\textbf {\bibinfo {volume} {15}},\ \bibinfo {pages} {493--520}
  (\bibinfo {year} {2019})}\BibitemShut {NoStop}%
\bibitem [{\citenamefont {{Myrvoll-Nilsen}}\ \emph {et~al.}(2020)\citenamefont
  {{Myrvoll-Nilsen}}, \citenamefont {S{\o}rbye}, \citenamefont {Fredriksen},
  \citenamefont {Rue},\ and\ \citenamefont {Rypdal}}]{myrvoll-nilsen2020}%
  \BibitemOpen
  \bibfield  {author} {\bibinfo {author} {\bibfnamefont {E.}~\bibnamefont
  {{Myrvoll-Nilsen}}}, \bibinfo {author} {\bibfnamefont {S.~H.}\ \bibnamefont
  {S{\o}rbye}}, \bibinfo {author} {\bibfnamefont {H.-B.}\ \bibnamefont
  {Fredriksen}}, \bibinfo {author} {\bibfnamefont {H.}~\bibnamefont {Rue}}, \
  and\ \bibinfo {author} {\bibfnamefont {M.}~\bibnamefont {Rypdal}},\
  }\bibfield  {title} {\enquote {\bibinfo {title} {Statistical estimation of
  global surface temperature response to forcing under the assumption of
  temporal scaling},}\ }\href {\doibase 10.5194/esd-11-329-2020} {\bibfield
  {journal} {\bibinfo  {journal} {Earth System Dynamics}\ }\textbf {\bibinfo
  {volume} {11}},\ \bibinfo {pages} {329--345} (\bibinfo {year}
  {2020})}\BibitemShut {NoStop}%
\bibitem [{\citenamefont {H{\'e}bert}, \citenamefont {Lovejoy},\ and\
  \citenamefont {Tremblay}(2021)}]{hebert2021}%
  \BibitemOpen
  \bibfield  {author} {\bibinfo {author} {\bibfnamefont {R.}~\bibnamefont
  {H{\'e}bert}}, \bibinfo {author} {\bibfnamefont {S.}~\bibnamefont {Lovejoy}},
  \ and\ \bibinfo {author} {\bibfnamefont {B.}~\bibnamefont {Tremblay}},\
  }\bibfield  {title} {\enquote {\bibinfo {title} {An observation-based scaling
  model for climate sensitivity estimates and global projections to 2100},}\
  }\href {\doibase 10.1007/s00382-020-05521-x} {\bibfield  {journal} {\bibinfo
  {journal} {Climate Dynamics}\ }\textbf {\bibinfo {volume} {56}},\ \bibinfo
  {pages} {1105--1129} (\bibinfo {year} {2021})}\BibitemShut {NoStop}%
\bibitem [{\citenamefont {Bodman}\ and\ \citenamefont
  {Jones}(2016)}]{bodman2016}%
  \BibitemOpen
  \bibfield  {author} {\bibinfo {author} {\bibfnamefont {R.~W.}\ \bibnamefont
  {Bodman}}\ and\ \bibinfo {author} {\bibfnamefont {R.~N.}\ \bibnamefont
  {Jones}},\ }\bibfield  {title} {\enquote {\bibinfo {title} {Bayesian
  estimation of climate sensitivity using observationally constrained simple
  climate models},}\ }\href {\doibase 10.1002/wcc.397} {\bibfield  {journal}
  {\bibinfo  {journal} {WIREs Climate Change}\ }\textbf {\bibinfo {volume}
  {7}},\ \bibinfo {pages} {461--473} (\bibinfo {year} {2016})}\BibitemShut
  {NoStop}%
\bibitem [{\citenamefont {Proistosescu}\ and\ \citenamefont
  {Huybers}(2017)}]{proistosescu2017}%
  \BibitemOpen
  \bibfield  {author} {\bibinfo {author} {\bibfnamefont {C.}~\bibnamefont
  {Proistosescu}}\ and\ \bibinfo {author} {\bibfnamefont {P.~J.}\ \bibnamefont
  {Huybers}},\ }\bibfield  {title} {\enquote {\bibinfo {title} {Slow climate
  mode reconciles historical and model-based estimates of climate
  sensitivity},}\ }\href {\doibase 10.1126/sciadv.1602821} {\bibfield
  {journal} {\bibinfo  {journal} {Science Advances}\ }\textbf {\bibinfo
  {volume} {3}},\ \bibinfo {pages} {e1602821} (\bibinfo {year}
  {2017})}\BibitemShut {NoStop}%
\bibitem [{\citenamefont {Jonko}, \citenamefont {Urban},\ and\ \citenamefont
  {Nadiga}(2018)}]{jonko2018}%
  \BibitemOpen
  \bibfield  {author} {\bibinfo {author} {\bibfnamefont {A.}~\bibnamefont
  {Jonko}}, \bibinfo {author} {\bibfnamefont {N.~M.}\ \bibnamefont {Urban}}, \
  and\ \bibinfo {author} {\bibfnamefont {B.}~\bibnamefont {Nadiga}},\
  }\bibfield  {title} {\enquote {\bibinfo {title} {Towards {{Bayesian}}
  hierarchical inference of equilibrium climate sensitivity from a combination
  of {{CMIP5}} climate models and observational data},}\ }\href {\doibase
  10.1007/s10584-018-2232-0} {\bibfield  {journal} {\bibinfo  {journal}
  {Climatic Change}\ }\textbf {\bibinfo {volume} {149}},\ \bibinfo {pages}
  {247--260} (\bibinfo {year} {2018})}\BibitemShut {NoStop}%
\bibitem [{\citenamefont {Sherwood}\ \emph {et~al.}(2020)\citenamefont
  {Sherwood}, \citenamefont {Webb}, \citenamefont {Annan}, \citenamefont
  {Armour}, \citenamefont {Forster}, \citenamefont {Hargreaves}, \citenamefont
  {Hegerl}, \citenamefont {Klein}, \citenamefont {Marvel}, \citenamefont
  {Rohling}, \citenamefont {Watanabe}, \citenamefont {Andrews}, \citenamefont
  {Braconnot}, \citenamefont {Bretherton}, \citenamefont {Foster},
  \citenamefont {Hausfather}, \citenamefont {Heydt}, \citenamefont {Knutti},
  \citenamefont {Mauritsen}, \citenamefont {Norris}, \citenamefont
  {Proistosescu}, \citenamefont {Rugenstein}, \citenamefont {Schmidt},
  \citenamefont {Tokarska},\ and\ \citenamefont {Zelinka}}]{sherwood2020}%
  \BibitemOpen
  \bibfield  {author} {\bibinfo {author} {\bibfnamefont {S.~C.}\ \bibnamefont
  {Sherwood}}, \bibinfo {author} {\bibfnamefont {M.~J.}\ \bibnamefont {Webb}},
  \bibinfo {author} {\bibfnamefont {J.~D.}\ \bibnamefont {Annan}}, \bibinfo
  {author} {\bibfnamefont {K.~C.}\ \bibnamefont {Armour}}, \bibinfo {author}
  {\bibfnamefont {P.~M.}\ \bibnamefont {Forster}}, \bibinfo {author}
  {\bibfnamefont {J.~C.}\ \bibnamefont {Hargreaves}}, \bibinfo {author}
  {\bibfnamefont {G.}~\bibnamefont {Hegerl}}, \bibinfo {author} {\bibfnamefont
  {S.~A.}\ \bibnamefont {Klein}}, \bibinfo {author} {\bibfnamefont {K.~D.}\
  \bibnamefont {Marvel}}, \bibinfo {author} {\bibfnamefont {E.~J.}\
  \bibnamefont {Rohling}}, \bibinfo {author} {\bibfnamefont {M.}~\bibnamefont
  {Watanabe}}, \bibinfo {author} {\bibfnamefont {T.}~\bibnamefont {Andrews}},
  \bibinfo {author} {\bibfnamefont {P.}~\bibnamefont {Braconnot}}, \bibinfo
  {author} {\bibfnamefont {C.~S.}\ \bibnamefont {Bretherton}}, \bibinfo
  {author} {\bibfnamefont {G.~L.}\ \bibnamefont {Foster}}, \bibinfo {author}
  {\bibfnamefont {Z.}~\bibnamefont {Hausfather}}, \bibinfo {author}
  {\bibfnamefont {A.~S.}\ \bibnamefont {Heydt}}, \bibinfo {author}
  {\bibfnamefont {R.}~\bibnamefont {Knutti}}, \bibinfo {author} {\bibfnamefont
  {T.}~\bibnamefont {Mauritsen}}, \bibinfo {author} {\bibfnamefont {J.~R.}\
  \bibnamefont {Norris}}, \bibinfo {author} {\bibfnamefont {C.}~\bibnamefont
  {Proistosescu}}, \bibinfo {author} {\bibfnamefont {M.}~\bibnamefont
  {Rugenstein}}, \bibinfo {author} {\bibfnamefont {G.~A.}\ \bibnamefont
  {Schmidt}}, \bibinfo {author} {\bibfnamefont {K.~B.}\ \bibnamefont
  {Tokarska}}, \ and\ \bibinfo {author} {\bibfnamefont {M.~D.}\ \bibnamefont
  {Zelinka}},\ }\bibfield  {title} {\enquote {\bibinfo {title} {An
  {{Assessment}} of {{Earth}}'s {{Climate Sensitivity Using Multiple Lines}} of
  {{Evidence}}},}\ }\href {\doibase 10.1029/2019RG000678} {\bibfield  {journal}
  {\bibinfo  {journal} {Reviews of Geophysics}\ }\textbf {\bibinfo {volume}
  {58}} (\bibinfo {year} {2020}),\ 10.1029/2019RG000678}\BibitemShut {NoStop}%
\bibitem [{\citenamefont {Skeie}\ \emph {et~al.}(2018)\citenamefont {Skeie},
  \citenamefont {Berntsen}, \citenamefont {Aldrin}, \citenamefont {Holden},\
  and\ \citenamefont {Myhre}}]{skeie2018}%
  \BibitemOpen
  \bibfield  {author} {\bibinfo {author} {\bibfnamefont {R.~B.}\ \bibnamefont
  {Skeie}}, \bibinfo {author} {\bibfnamefont {T.}~\bibnamefont {Berntsen}},
  \bibinfo {author} {\bibfnamefont {M.}~\bibnamefont {Aldrin}}, \bibinfo
  {author} {\bibfnamefont {M.}~\bibnamefont {Holden}}, \ and\ \bibinfo {author}
  {\bibfnamefont {G.}~\bibnamefont {Myhre}},\ }\bibfield  {title} {\enquote
  {\bibinfo {title} {Climate sensitivity estimates \textendash{} sensitivity to
  radiative forcing time series and observational data},}\ }\href {\doibase
  10.5194/esd-9-879-2018} {\bibfield  {journal} {\bibinfo  {journal} {Earth
  System Dynamics}\ }\textbf {\bibinfo {volume} {9}},\ \bibinfo {pages}
  {879--894} (\bibinfo {year} {2018})}\BibitemShut {NoStop}%
\bibitem [{\citenamefont {Mitchell}(1976)}]{mitchell1976}%
  \BibitemOpen
  \bibfield  {author} {\bibinfo {author} {\bibfnamefont {J.~M.}\ \bibnamefont
  {Mitchell}},\ }\bibfield  {title} {\enquote {\bibinfo {title} {An
  {{Overview}} of {{Climatic Variability}} and its {{Causal Mechanisms}}},}\
  }\href {\doibase 10.1016/0033-5894(76)90021-1} {\bibfield  {journal}
  {\bibinfo  {journal} {Quaternary Research}\ }\textbf {\bibinfo {volume}
  {6}},\ \bibinfo {pages} {481--493} (\bibinfo {year} {1976})}\BibitemShut
  {NoStop}%
\bibitem [{\citenamefont {Huybers}\ and\ \citenamefont
  {Curry}(2006)}]{huybers2006}%
  \BibitemOpen
  \bibfield  {author} {\bibinfo {author} {\bibfnamefont {P.}~\bibnamefont
  {Huybers}}\ and\ \bibinfo {author} {\bibfnamefont {W.}~\bibnamefont
  {Curry}},\ }\bibfield  {title} {\enquote {\bibinfo {title} {Links between
  annual, {{Milankovitch}} and continuum temperature variability},}\ }\href
  {\doibase 10.1038/nature04745} {\bibfield  {journal} {\bibinfo  {journal}
  {Nature}\ }\textbf {\bibinfo {volume} {441}},\ \bibinfo {pages} {329--332}
  (\bibinfo {year} {2006})}\BibitemShut {NoStop}%
\bibitem [{\citenamefont {{von der Heydt}}\ \emph {et~al.}(2021)\citenamefont
  {{von der Heydt}}, \citenamefont {Ashwin}, \citenamefont {Camp},
  \citenamefont {Crucifix}, \citenamefont {Dijkstra}, \citenamefont
  {Ditlevsen},\ and\ \citenamefont {Lenton}}]{vonderheydt2021}%
  \BibitemOpen
  \bibfield  {author} {\bibinfo {author} {\bibfnamefont {A.~S.}\ \bibnamefont
  {{von der Heydt}}}, \bibinfo {author} {\bibfnamefont {P.}~\bibnamefont
  {Ashwin}}, \bibinfo {author} {\bibfnamefont {C.~D.}\ \bibnamefont {Camp}},
  \bibinfo {author} {\bibfnamefont {M.}~\bibnamefont {Crucifix}}, \bibinfo
  {author} {\bibfnamefont {H.~A.}\ \bibnamefont {Dijkstra}}, \bibinfo {author}
  {\bibfnamefont {P.}~\bibnamefont {Ditlevsen}}, \ and\ \bibinfo {author}
  {\bibfnamefont {T.~M.}\ \bibnamefont {Lenton}},\ }\bibfield  {title}
  {\enquote {\bibinfo {title} {Quantification and interpretation of the climate
  variability record},}\ }\href {\doibase 10.1016/j.gloplacha.2020.103399}
  {\bibfield  {journal} {\bibinfo  {journal} {Global and Planetary Change}\
  }\textbf {\bibinfo {volume} {197}},\ \bibinfo {pages} {103399} (\bibinfo
  {year} {2021})}\BibitemShut {NoStop}%
\bibitem [{\citenamefont {Franzke}\ \emph {et~al.}(2020)\citenamefont
  {Franzke}, \citenamefont {Barbosa}, \citenamefont {Blender}, \citenamefont
  {Fredriksen}, \citenamefont {Laepple}, \citenamefont {Lambert}, \citenamefont
  {Nilsen}, \citenamefont {Rypdal}, \citenamefont {Rypdal}, \citenamefont
  {Scotto}, \citenamefont {Vannitsem}, \citenamefont {Watkins}, \citenamefont
  {Yang},\ and\ \citenamefont {Yuan}}]{franzke2020}%
  \BibitemOpen
  \bibfield  {author} {\bibinfo {author} {\bibfnamefont {C.~L.~E.}\
  \bibnamefont {Franzke}}, \bibinfo {author} {\bibfnamefont {S.}~\bibnamefont
  {Barbosa}}, \bibinfo {author} {\bibfnamefont {R.}~\bibnamefont {Blender}},
  \bibinfo {author} {\bibfnamefont {H.-B.}\ \bibnamefont {Fredriksen}},
  \bibinfo {author} {\bibfnamefont {T.}~\bibnamefont {Laepple}}, \bibinfo
  {author} {\bibfnamefont {F.}~\bibnamefont {Lambert}}, \bibinfo {author}
  {\bibfnamefont {T.}~\bibnamefont {Nilsen}}, \bibinfo {author} {\bibfnamefont
  {K.}~\bibnamefont {Rypdal}}, \bibinfo {author} {\bibfnamefont
  {M.}~\bibnamefont {Rypdal}}, \bibinfo {author} {\bibfnamefont {M.~G.}\
  \bibnamefont {Scotto}}, \bibinfo {author} {\bibfnamefont {S.}~\bibnamefont
  {Vannitsem}}, \bibinfo {author} {\bibfnamefont {N.~W.}\ \bibnamefont
  {Watkins}}, \bibinfo {author} {\bibfnamefont {L.}~\bibnamefont {Yang}}, \
  and\ \bibinfo {author} {\bibfnamefont {N.}~\bibnamefont {Yuan}},\ }\bibfield
  {title} {\enquote {\bibinfo {title} {The {{Structure}} of {{Climate
  Variability Across Scales}}},}\ }\href {\doibase 10.1029/2019RG000657}
  {\bibfield  {journal} {\bibinfo  {journal} {Reviews of Geophysics}\ }\textbf
  {\bibinfo {volume} {58}} (\bibinfo {year} {2020}),\
  10.1029/2019RG000657}\BibitemShut {NoStop}%
\bibitem [{\citenamefont {Rypdal}\ \emph {et~al.}(2018)\citenamefont {Rypdal},
  \citenamefont {Fredriksen}, \citenamefont {{Myrvoll-Nilsen}}, \citenamefont
  {Rypdal},\ and\ \citenamefont {S{\o}rbye}}]{rypdal2018}%
  \BibitemOpen
  \bibfield  {author} {\bibinfo {author} {\bibfnamefont {M.}~\bibnamefont
  {Rypdal}}, \bibinfo {author} {\bibfnamefont {H.-B.}\ \bibnamefont
  {Fredriksen}}, \bibinfo {author} {\bibfnamefont {E.}~\bibnamefont
  {{Myrvoll-Nilsen}}}, \bibinfo {author} {\bibfnamefont {K.}~\bibnamefont
  {Rypdal}}, \ and\ \bibinfo {author} {\bibfnamefont {S.}~\bibnamefont
  {S{\o}rbye}},\ }\bibfield  {title} {\enquote {\bibinfo {title} {Emergent
  {{Scale Invariance}} and {{Climate Sensitivity}}},}\ }\href {\doibase
  10.3390/cli6040093} {\bibfield  {journal} {\bibinfo  {journal} {Climate}\
  }\textbf {\bibinfo {volume} {6}},\ \bibinfo {pages} {93} (\bibinfo {year}
  {2018})}\BibitemShut {NoStop}%
\bibitem [{\citenamefont {Soldatenko}\ and\ \citenamefont
  {Colman}(2022)}]{soldatenko2022}%
  \BibitemOpen
  \bibfield  {author} {\bibinfo {author} {\bibfnamefont {S.~A.}\ \bibnamefont
  {Soldatenko}}\ and\ \bibinfo {author} {\bibfnamefont {R.~A.}\ \bibnamefont
  {Colman}},\ }\bibfield  {title} {\enquote {\bibinfo {title} {Power {{Spectrum
  Sensitivity Analysis}} of the {{Global Mean Surface Temperature Fluctuations
  Simulated}} in a {{Two-Box Stochastic Energy Balance Model}}},}\ }\href
  {\doibase 10.16993/tellusa.40} {\bibfield  {journal} {\bibinfo  {journal}
  {Tellus A: Dynamic Meteorology and Oceanography}\ }\textbf {\bibinfo {volume}
  {74}},\ \bibinfo {pages} {68} (\bibinfo {year} {2022})}\BibitemShut {NoStop}%
\bibitem [{\citenamefont {Schillinger}\ and\ \citenamefont
  {Ellerhoff}(2022)}]{schillinger2022}%
  \BibitemOpen
  \bibfield  {author} {\bibinfo {author} {\bibfnamefont {M.}~\bibnamefont
  {Schillinger}}\ and\ \bibinfo {author} {\bibfnamefont {B.}~\bibnamefont
  {Ellerhoff}},\ }\href {https://github.com/paleovar/ClimBayes} {\enquote
  {\bibinfo {title} {R package "{{ClimBayes}}"},}\ } (\bibinfo {year}
  {2022})\BibitemShut {NoStop}%
\bibitem [{\citenamefont {Jungclaus}\ \emph {et~al.}(2017)\citenamefont
  {Jungclaus}, \citenamefont {Bard}, \citenamefont {Baroni}, \citenamefont
  {Braconnot}, \citenamefont {Cao}, \citenamefont {Chini}, \citenamefont
  {Egorova}, \citenamefont {Evans}, \citenamefont {{Gonz{\'a}lez-Rouco}},
  \citenamefont {Goosse}, \citenamefont {Hurtt}, \citenamefont {Joos},
  \citenamefont {Kaplan}, \citenamefont {Khodri}, \citenamefont
  {Klein~Goldewijk}, \citenamefont {Krivova}, \citenamefont {LeGrande},
  \citenamefont {Lorenz}, \citenamefont {Luterbacher}, \citenamefont {Man},
  \citenamefont {Maycock}, \citenamefont {Meinshausen}, \citenamefont {Moberg},
  \citenamefont {Muscheler}, \citenamefont {{Nehrbass-Ahles}}, \citenamefont
  {{Otto-Bliesner}}, \citenamefont {Phipps}, \citenamefont {Pongratz},
  \citenamefont {Rozanov}, \citenamefont {Schmidt}, \citenamefont {Schmidt},
  \citenamefont {Schmutz}, \citenamefont {Schurer}, \citenamefont {Shapiro},
  \citenamefont {Sigl}, \citenamefont {Smerdon}, \citenamefont {Solanki},
  \citenamefont {Timmreck}, \citenamefont {Toohey}, \citenamefont {Usoskin},
  \citenamefont {Wagner}, \citenamefont {Wu}, \citenamefont {Yeo},
  \citenamefont {Zanchettin}, \citenamefont {Zhang},\ and\ \citenamefont
  {Zorita}}]{jungclaus2017}%
  \BibitemOpen
  \bibfield  {author} {\bibinfo {author} {\bibfnamefont {J.~H.}\ \bibnamefont
  {Jungclaus}}, \bibinfo {author} {\bibfnamefont {E.}~\bibnamefont {Bard}},
  \bibinfo {author} {\bibfnamefont {M.}~\bibnamefont {Baroni}}, \bibinfo
  {author} {\bibfnamefont {P.}~\bibnamefont {Braconnot}}, \bibinfo {author}
  {\bibfnamefont {J.}~\bibnamefont {Cao}}, \bibinfo {author} {\bibfnamefont
  {L.~P.}\ \bibnamefont {Chini}}, \bibinfo {author} {\bibfnamefont
  {T.}~\bibnamefont {Egorova}}, \bibinfo {author} {\bibfnamefont
  {M.}~\bibnamefont {Evans}}, \bibinfo {author} {\bibfnamefont {J.~F.}\
  \bibnamefont {{Gonz{\'a}lez-Rouco}}}, \bibinfo {author} {\bibfnamefont
  {H.}~\bibnamefont {Goosse}}, \bibinfo {author} {\bibfnamefont {G.~C.}\
  \bibnamefont {Hurtt}}, \bibinfo {author} {\bibfnamefont {F.}~\bibnamefont
  {Joos}}, \bibinfo {author} {\bibfnamefont {J.~O.}\ \bibnamefont {Kaplan}},
  \bibinfo {author} {\bibfnamefont {M.}~\bibnamefont {Khodri}}, \bibinfo
  {author} {\bibfnamefont {K.}~\bibnamefont {Klein~Goldewijk}}, \bibinfo
  {author} {\bibfnamefont {N.}~\bibnamefont {Krivova}}, \bibinfo {author}
  {\bibfnamefont {A.~N.}\ \bibnamefont {LeGrande}}, \bibinfo {author}
  {\bibfnamefont {S.~J.}\ \bibnamefont {Lorenz}}, \bibinfo {author}
  {\bibfnamefont {J.}~\bibnamefont {Luterbacher}}, \bibinfo {author}
  {\bibfnamefont {W.}~\bibnamefont {Man}}, \bibinfo {author} {\bibfnamefont
  {A.~C.}\ \bibnamefont {Maycock}}, \bibinfo {author} {\bibfnamefont
  {M.}~\bibnamefont {Meinshausen}}, \bibinfo {author} {\bibfnamefont
  {A.}~\bibnamefont {Moberg}}, \bibinfo {author} {\bibfnamefont
  {R.}~\bibnamefont {Muscheler}}, \bibinfo {author} {\bibfnamefont
  {C.}~\bibnamefont {{Nehrbass-Ahles}}}, \bibinfo {author} {\bibfnamefont
  {B.~I.}\ \bibnamefont {{Otto-Bliesner}}}, \bibinfo {author} {\bibfnamefont
  {S.~J.}\ \bibnamefont {Phipps}}, \bibinfo {author} {\bibfnamefont
  {J.}~\bibnamefont {Pongratz}}, \bibinfo {author} {\bibfnamefont
  {E.}~\bibnamefont {Rozanov}}, \bibinfo {author} {\bibfnamefont {G.~A.}\
  \bibnamefont {Schmidt}}, \bibinfo {author} {\bibfnamefont {H.}~\bibnamefont
  {Schmidt}}, \bibinfo {author} {\bibfnamefont {W.}~\bibnamefont {Schmutz}},
  \bibinfo {author} {\bibfnamefont {A.}~\bibnamefont {Schurer}}, \bibinfo
  {author} {\bibfnamefont {A.~I.}\ \bibnamefont {Shapiro}}, \bibinfo {author}
  {\bibfnamefont {M.}~\bibnamefont {Sigl}}, \bibinfo {author} {\bibfnamefont
  {J.~E.}\ \bibnamefont {Smerdon}}, \bibinfo {author} {\bibfnamefont {S.~K.}\
  \bibnamefont {Solanki}}, \bibinfo {author} {\bibfnamefont {C.}~\bibnamefont
  {Timmreck}}, \bibinfo {author} {\bibfnamefont {M.}~\bibnamefont {Toohey}},
  \bibinfo {author} {\bibfnamefont {I.~G.}\ \bibnamefont {Usoskin}}, \bibinfo
  {author} {\bibfnamefont {S.}~\bibnamefont {Wagner}}, \bibinfo {author}
  {\bibfnamefont {C.-J.}\ \bibnamefont {Wu}}, \bibinfo {author} {\bibfnamefont
  {K.~L.}\ \bibnamefont {Yeo}}, \bibinfo {author} {\bibfnamefont
  {D.}~\bibnamefont {Zanchettin}}, \bibinfo {author} {\bibfnamefont
  {Q.}~\bibnamefont {Zhang}}, \ and\ \bibinfo {author} {\bibfnamefont
  {E.}~\bibnamefont {Zorita}},\ }\bibfield  {title} {\enquote {\bibinfo {title}
  {The {{PMIP4}} contribution to {{CMIP6}} \textendash{} {{Part}} 3: {{The}}
  last millennium, scientific objective, and experimental design for the
  {{PMIP4}} \&lt;i\&gt;past1000\&lt;/i\&gt; simulations},}\ }\href {\doibase
  10.5194/gmd-10-4005-2017} {\bibfield  {journal} {\bibinfo  {journal}
  {Geoscientific Model Development}\ }\textbf {\bibinfo {volume} {10}},\
  \bibinfo {pages} {4005--4033} (\bibinfo {year} {2017})}\BibitemShut {NoStop}%
\bibitem [{\citenamefont {Flato}\ \emph {et~al.}(2014)\citenamefont {Flato},
  \citenamefont {Marotzke}, \citenamefont {Abiodun}, \citenamefont {Braconnot},
  \citenamefont {Chou}, \citenamefont {Collins}, \citenamefont {Cox},
  \citenamefont {Driouech}, \citenamefont {Emori}, \citenamefont {Eyring} \emph
  {et~al.}}]{flato2014}%
  \BibitemOpen
  \bibfield  {author} {\bibinfo {author} {\bibfnamefont {G.}~\bibnamefont
  {Flato}}, \bibinfo {author} {\bibfnamefont {J.}~\bibnamefont {Marotzke}},
  \bibinfo {author} {\bibfnamefont {B.}~\bibnamefont {Abiodun}}, \bibinfo
  {author} {\bibfnamefont {P.}~\bibnamefont {Braconnot}}, \bibinfo {author}
  {\bibfnamefont {S.~C.}\ \bibnamefont {Chou}}, \bibinfo {author}
  {\bibfnamefont {W.}~\bibnamefont {Collins}}, \bibinfo {author} {\bibfnamefont
  {P.}~\bibnamefont {Cox}}, \bibinfo {author} {\bibfnamefont {F.}~\bibnamefont
  {Driouech}}, \bibinfo {author} {\bibfnamefont {S.}~\bibnamefont {Emori}},
  \bibinfo {author} {\bibfnamefont {V.}~\bibnamefont {Eyring}},  \emph
  {et~al.},\ }\bibfield  {title} {\enquote {\bibinfo {title} {Evaluation of
  climate models},}\ }in\ \href
  {https://www.ipcc.ch/site/assets/uploads/2018/02/WG1AR5_Chapter09_FINAL.pdf}
  {\emph {\bibinfo {booktitle} {Climate Change 2013: The Physical Science
  Basis. {{Contribution}} of Working Group {{I}} to the Fifth Assessment Report
  of the Intergovernmental Panel on Climate Change}}}\ (\bibinfo  {publisher}
  {{Cambridge University Press}},\ \bibinfo {year} {2014})\ pp.\ \bibinfo
  {pages} {741--866}\BibitemShut {NoStop}%
\bibitem [{\citenamefont {Eby}\ \emph {et~al.}(2013)\citenamefont {Eby},
  \citenamefont {Weaver}, \citenamefont {Alexander}, \citenamefont {Zickfeld},
  \citenamefont {{Abe-Ouchi}}, \citenamefont {Cimatoribus}, \citenamefont
  {Crespin}, \citenamefont {Drijfhout}, \citenamefont {Edwards}, \citenamefont
  {Eliseev}, \citenamefont {Feulner}, \citenamefont {Fichefet}, \citenamefont
  {Forest}, \citenamefont {Goosse}, \citenamefont {Holden}, \citenamefont
  {Joos}, \citenamefont {Kawamiya}, \citenamefont {Kicklighter}, \citenamefont
  {Kienert}, \citenamefont {Matsumoto}, \citenamefont {Mokhov}, \citenamefont
  {Monier}, \citenamefont {Olsen}, \citenamefont {Pedersen}, \citenamefont
  {Perrette}, \citenamefont {{Philippon-Berthier}}, \citenamefont {Ridgwell},
  \citenamefont {Schlosser}, \citenamefont {{Schneider von Deimling}},
  \citenamefont {Shaffer}, \citenamefont {Smith}, \citenamefont {Spahni},
  \citenamefont {Sokolov}, \citenamefont {Steinacher}, \citenamefont
  {Tachiiri}, \citenamefont {Tokos}, \citenamefont {Yoshimori}, \citenamefont
  {Zeng},\ and\ \citenamefont {Zhao}}]{eby2013}%
  \BibitemOpen
  \bibfield  {author} {\bibinfo {author} {\bibfnamefont {M.}~\bibnamefont
  {Eby}}, \bibinfo {author} {\bibfnamefont {A.~J.}\ \bibnamefont {Weaver}},
  \bibinfo {author} {\bibfnamefont {K.}~\bibnamefont {Alexander}}, \bibinfo
  {author} {\bibfnamefont {K.}~\bibnamefont {Zickfeld}}, \bibinfo {author}
  {\bibfnamefont {A.}~\bibnamefont {{Abe-Ouchi}}}, \bibinfo {author}
  {\bibfnamefont {A.~A.}\ \bibnamefont {Cimatoribus}}, \bibinfo {author}
  {\bibfnamefont {E.}~\bibnamefont {Crespin}}, \bibinfo {author} {\bibfnamefont
  {S.~S.}\ \bibnamefont {Drijfhout}}, \bibinfo {author} {\bibfnamefont {N.~R.}\
  \bibnamefont {Edwards}}, \bibinfo {author} {\bibfnamefont {A.~V.}\
  \bibnamefont {Eliseev}}, \bibinfo {author} {\bibfnamefont {G.}~\bibnamefont
  {Feulner}}, \bibinfo {author} {\bibfnamefont {T.}~\bibnamefont {Fichefet}},
  \bibinfo {author} {\bibfnamefont {C.~E.}\ \bibnamefont {Forest}}, \bibinfo
  {author} {\bibfnamefont {H.}~\bibnamefont {Goosse}}, \bibinfo {author}
  {\bibfnamefont {P.~B.}\ \bibnamefont {Holden}}, \bibinfo {author}
  {\bibfnamefont {F.}~\bibnamefont {Joos}}, \bibinfo {author} {\bibfnamefont
  {M.}~\bibnamefont {Kawamiya}}, \bibinfo {author} {\bibfnamefont
  {D.}~\bibnamefont {Kicklighter}}, \bibinfo {author} {\bibfnamefont
  {H.}~\bibnamefont {Kienert}}, \bibinfo {author} {\bibfnamefont
  {K.}~\bibnamefont {Matsumoto}}, \bibinfo {author} {\bibfnamefont {I.~I.}\
  \bibnamefont {Mokhov}}, \bibinfo {author} {\bibfnamefont {E.}~\bibnamefont
  {Monier}}, \bibinfo {author} {\bibfnamefont {S.~M.}\ \bibnamefont {Olsen}},
  \bibinfo {author} {\bibfnamefont {J.~O.~P.}\ \bibnamefont {Pedersen}},
  \bibinfo {author} {\bibfnamefont {M.}~\bibnamefont {Perrette}}, \bibinfo
  {author} {\bibfnamefont {G.}~\bibnamefont {{Philippon-Berthier}}}, \bibinfo
  {author} {\bibfnamefont {A.}~\bibnamefont {Ridgwell}}, \bibinfo {author}
  {\bibfnamefont {A.}~\bibnamefont {Schlosser}}, \bibinfo {author}
  {\bibfnamefont {T.}~\bibnamefont {{Schneider von Deimling}}}, \bibinfo
  {author} {\bibfnamefont {G.}~\bibnamefont {Shaffer}}, \bibinfo {author}
  {\bibfnamefont {R.~S.}\ \bibnamefont {Smith}}, \bibinfo {author}
  {\bibfnamefont {R.}~\bibnamefont {Spahni}}, \bibinfo {author} {\bibfnamefont
  {A.~P.}\ \bibnamefont {Sokolov}}, \bibinfo {author} {\bibfnamefont
  {M.}~\bibnamefont {Steinacher}}, \bibinfo {author} {\bibfnamefont
  {K.}~\bibnamefont {Tachiiri}}, \bibinfo {author} {\bibfnamefont
  {K.}~\bibnamefont {Tokos}}, \bibinfo {author} {\bibfnamefont
  {M.}~\bibnamefont {Yoshimori}}, \bibinfo {author} {\bibfnamefont
  {N.}~\bibnamefont {Zeng}}, \ and\ \bibinfo {author} {\bibfnamefont
  {F.}~\bibnamefont {Zhao}},\ }\bibfield  {title} {\enquote {\bibinfo {title}
  {Historical and idealized climate model experiments: An intercomparison of
  {{Earth}} system models of intermediate complexity},}\ }\href {\doibase
  10.5194/cp-9-1111-2013} {\bibfield  {journal} {\bibinfo  {journal} {Climate
  of the Past}\ }\textbf {\bibinfo {volume} {9}},\ \bibinfo {pages}
  {1111--1140} (\bibinfo {year} {2013})}\BibitemShut {NoStop}%
\bibitem [{\citenamefont {Schmidt}\ \emph {et~al.}(2012)\citenamefont
  {Schmidt}, \citenamefont {Jungclaus}, \citenamefont {Ammann}, \citenamefont
  {Bard}, \citenamefont {Braconnot}, \citenamefont {Crowley}, \citenamefont
  {Delaygue}, \citenamefont {Joos}, \citenamefont {Krivova}, \citenamefont
  {Muscheler}, \citenamefont {{Otto-Bliesner}}, \citenamefont {Pongratz},
  \citenamefont {Shindell}, \citenamefont {Solanki}, \citenamefont
  {Steinhilber},\ and\ \citenamefont {Vieira}}]{schmidt2012}%
  \BibitemOpen
  \bibfield  {author} {\bibinfo {author} {\bibfnamefont {G.~A.}\ \bibnamefont
  {Schmidt}}, \bibinfo {author} {\bibfnamefont {J.~H.}\ \bibnamefont
  {Jungclaus}}, \bibinfo {author} {\bibfnamefont {C.~M.}\ \bibnamefont
  {Ammann}}, \bibinfo {author} {\bibfnamefont {E.}~\bibnamefont {Bard}},
  \bibinfo {author} {\bibfnamefont {P.}~\bibnamefont {Braconnot}}, \bibinfo
  {author} {\bibfnamefont {T.~J.}\ \bibnamefont {Crowley}}, \bibinfo {author}
  {\bibfnamefont {G.}~\bibnamefont {Delaygue}}, \bibinfo {author}
  {\bibfnamefont {F.}~\bibnamefont {Joos}}, \bibinfo {author} {\bibfnamefont
  {N.~A.}\ \bibnamefont {Krivova}}, \bibinfo {author} {\bibfnamefont
  {R.}~\bibnamefont {Muscheler}}, \bibinfo {author} {\bibfnamefont {B.~L.}\
  \bibnamefont {{Otto-Bliesner}}}, \bibinfo {author} {\bibfnamefont
  {J.}~\bibnamefont {Pongratz}}, \bibinfo {author} {\bibfnamefont {D.~T.}\
  \bibnamefont {Shindell}}, \bibinfo {author} {\bibfnamefont {S.~K.}\
  \bibnamefont {Solanki}}, \bibinfo {author} {\bibfnamefont {F.}~\bibnamefont
  {Steinhilber}}, \ and\ \bibinfo {author} {\bibfnamefont {L.~E.~A.}\
  \bibnamefont {Vieira}},\ }\bibfield  {title} {\enquote {\bibinfo {title}
  {Climate forcing reconstructions for use in {{PMIP}} simulations of the
  {{Last Millennium}} (v1.1)},}\ }\href {\doibase 10.5194/gmd-5-185-2012}
  {\bibfield  {journal} {\bibinfo  {journal} {Geoscientific Model Development}\
  }\textbf {\bibinfo {volume} {5}},\ \bibinfo {pages} {185--191} (\bibinfo
  {year} {2012})}\BibitemShut {NoStop}%
\bibitem [{\citenamefont {Tachiiri}\ \emph {et~al.}(2010)\citenamefont
  {Tachiiri}, \citenamefont {Hargreaves}, \citenamefont {Annan}, \citenamefont
  {Oka}, \citenamefont {{Abe-Ouchi}},\ and\ \citenamefont
  {Kawamiya}}]{tachiiri2010}%
  \BibitemOpen
  \bibfield  {author} {\bibinfo {author} {\bibfnamefont {K.}~\bibnamefont
  {Tachiiri}}, \bibinfo {author} {\bibfnamefont {J.~C.}\ \bibnamefont
  {Hargreaves}}, \bibinfo {author} {\bibfnamefont {J.~D.}\ \bibnamefont
  {Annan}}, \bibinfo {author} {\bibfnamefont {A.}~\bibnamefont {Oka}}, \bibinfo
  {author} {\bibfnamefont {A.}~\bibnamefont {{Abe-Ouchi}}}, \ and\ \bibinfo
  {author} {\bibfnamefont {M.}~\bibnamefont {Kawamiya}},\ }\bibfield  {title}
  {\enquote {\bibinfo {title} {Development of a system emulating the global
  carbon cycle in {{Earth}} system models},}\ }\href {\doibase
  10.5194/gmd-3-365-2010} {\bibfield  {journal} {\bibinfo  {journal}
  {Geoscientific Model Development}\ }\textbf {\bibinfo {volume} {3}},\
  \bibinfo {pages} {365--376} (\bibinfo {year} {2010})}\BibitemShut {NoStop}%
\bibitem [{\citenamefont {Morice}\ \emph {et~al.}(2021)\citenamefont {Morice},
  \citenamefont {Kennedy}, \citenamefont {Rayner}, \citenamefont {Winn},
  \citenamefont {Hogan}, \citenamefont {Killick}, \citenamefont {Dunn},
  \citenamefont {Osborn}, \citenamefont {Jones},\ and\ \citenamefont
  {Simpson}}]{morice2021}%
  \BibitemOpen
  \bibfield  {author} {\bibinfo {author} {\bibfnamefont {C.~P.}\ \bibnamefont
  {Morice}}, \bibinfo {author} {\bibfnamefont {J.~J.}\ \bibnamefont {Kennedy}},
  \bibinfo {author} {\bibfnamefont {N.~A.}\ \bibnamefont {Rayner}}, \bibinfo
  {author} {\bibfnamefont {J.~P.}\ \bibnamefont {Winn}}, \bibinfo {author}
  {\bibfnamefont {E.}~\bibnamefont {Hogan}}, \bibinfo {author} {\bibfnamefont
  {R.~E.}\ \bibnamefont {Killick}}, \bibinfo {author} {\bibfnamefont
  {R.~J.~H.}\ \bibnamefont {Dunn}}, \bibinfo {author} {\bibfnamefont {T.~J.}\
  \bibnamefont {Osborn}}, \bibinfo {author} {\bibfnamefont {P.~D.}\
  \bibnamefont {Jones}}, \ and\ \bibinfo {author} {\bibfnamefont {I.~R.}\
  \bibnamefont {Simpson}},\ }\bibfield  {title} {\enquote {\bibinfo {title} {An
  {{Updated Assessment}} of {{Near}}-{{Surface Temperature Change From}} 1850:
  {{The HadCRUT5 Data Set}}},}\ }\href {\doibase 10.1029/2019JD032361}
  {\bibfield  {journal} {\bibinfo  {journal} {Journal of Geophysical Research:
  Atmospheres}\ }\textbf {\bibinfo {volume} {126}} (\bibinfo {year} {2021}),\
  10.1029/2019JD032361}\BibitemShut {NoStop}%
\bibitem [{\citenamefont {Pongratz}\ \emph {et~al.}(2008)\citenamefont
  {Pongratz}, \citenamefont {Reick}, \citenamefont {Raddatz},\ and\
  \citenamefont {Claussen}}]{pongratz2008}%
  \BibitemOpen
  \bibfield  {author} {\bibinfo {author} {\bibfnamefont {J.}~\bibnamefont
  {Pongratz}}, \bibinfo {author} {\bibfnamefont {C.}~\bibnamefont {Reick}},
  \bibinfo {author} {\bibfnamefont {T.}~\bibnamefont {Raddatz}}, \ and\
  \bibinfo {author} {\bibfnamefont {M.}~\bibnamefont {Claussen}},\ }\bibfield
  {title} {\enquote {\bibinfo {title} {A reconstruction of global agricultural
  areas and land cover for the last millennium},}\ }\href {\doibase
  10.1029/2007GB003153} {\bibfield  {journal} {\bibinfo  {journal} {Global
  Biogeochemical Cycles}\ }\textbf {\bibinfo {volume} {22}} (\bibinfo {year}
  {2008}),\ 10.1029/2007GB003153}\BibitemShut {NoStop}%
\bibitem [{\citenamefont {Crowley}\ \emph {et~al.}(2008)\citenamefont
  {Crowley}, \citenamefont {Zielinski}, \citenamefont {Vinther}, \citenamefont
  {Udisti}, \citenamefont {Kreutz}, \citenamefont {{Cole-Dai}},\ and\
  \citenamefont {Castellano}}]{crowley2008}%
  \BibitemOpen
  \bibfield  {author} {\bibinfo {author} {\bibfnamefont {T.~J.}\ \bibnamefont
  {Crowley}}, \bibinfo {author} {\bibfnamefont {G.}~\bibnamefont {Zielinski}},
  \bibinfo {author} {\bibfnamefont {B.}~\bibnamefont {Vinther}}, \bibinfo
  {author} {\bibfnamefont {R.}~\bibnamefont {Udisti}}, \bibinfo {author}
  {\bibfnamefont {K.}~\bibnamefont {Kreutz}}, \bibinfo {author} {\bibfnamefont
  {J.}~\bibnamefont {{Cole-Dai}}}, \ and\ \bibinfo {author} {\bibfnamefont
  {E.}~\bibnamefont {Castellano}},\ }\bibfield  {title} {\enquote {\bibinfo
  {title} {Volcanism and the {{Little Ice Age}}},}\ }\href {\doibase
  10.22498/pages.16.2.22} {\bibfield  {journal} {\bibinfo  {journal} {PAGES
  news}\ }\textbf {\bibinfo {volume} {16}},\ \bibinfo {pages} {22--23}
  (\bibinfo {year} {2008})}\BibitemShut {NoStop}%
\bibitem [{\citenamefont {Steinhilber}, \citenamefont {Beer},\ and\
  \citenamefont {Fr{\"o}hlich}(2009)}]{steinhilber2009}%
  \BibitemOpen
  \bibfield  {author} {\bibinfo {author} {\bibfnamefont {F.}~\bibnamefont
  {Steinhilber}}, \bibinfo {author} {\bibfnamefont {J.}~\bibnamefont {Beer}}, \
  and\ \bibinfo {author} {\bibfnamefont {C.}~\bibnamefont {Fr{\"o}hlich}},\
  }\bibfield  {title} {\enquote {\bibinfo {title} {Total solar irradiance
  during the {{Holocene}}},}\ }\href {\doibase 10.1029/2009GL040142} {\bibfield
   {journal} {\bibinfo  {journal} {Geophysical Research Letters}\ }\textbf
  {\bibinfo {volume} {36}},\ \bibinfo {pages} {L19704} (\bibinfo {year}
  {2009})}\BibitemShut {NoStop}%
\bibitem [{\citenamefont {Wang}, \citenamefont {Lean},\ and\ \citenamefont
  {N.~R.~Sheeley}(2005)}]{wang2005}%
  \BibitemOpen
  \bibfield  {author} {\bibinfo {author} {\bibfnamefont {Y.-M.}\ \bibnamefont
  {Wang}}, \bibinfo {author} {\bibfnamefont {J.~L.}\ \bibnamefont {Lean}}, \
  and\ \bibinfo {author} {\bibfnamefont {J.}~\bibnamefont {N.~R.~Sheeley}},\
  }\bibfield  {title} {\enquote {\bibinfo {title} {Modeling the sun's magnetic
  field and irradiance since 1713},}\ }\href {\doibase 10.1086/429689}
  {\bibfield  {journal} {\bibinfo  {journal} {The Astrophysical Journal}\
  }\textbf {\bibinfo {volume} {625}},\ \bibinfo {pages} {522--538} (\bibinfo
  {year} {2005})}\BibitemShut {NoStop}%
\bibitem [{\citenamefont {Delaygue}\ and\ \citenamefont
  {Bard}(2011)}]{delaygue2011}%
  \BibitemOpen
  \bibfield  {author} {\bibinfo {author} {\bibfnamefont {G.}~\bibnamefont
  {Delaygue}}\ and\ \bibinfo {author} {\bibfnamefont {E.}~\bibnamefont
  {Bard}},\ }\bibfield  {title} {\enquote {\bibinfo {title} {An {{Antarctic}}
  view of {{Beryllium-10}} and solar activity for the past millennium},}\
  }\href {\doibase 10.1007/s00382-010-0795-1} {\bibfield  {journal} {\bibinfo
  {journal} {Climate Dynamics}\ }\textbf {\bibinfo {volume} {36}},\ \bibinfo
  {pages} {2201--2218} (\bibinfo {year} {2011})}\BibitemShut {NoStop}%
\bibitem [{\citenamefont {Krivova}, \citenamefont {Balmaceda},\ and\
  \citenamefont {Solanki}(2007)}]{krivova2007}%
  \BibitemOpen
  \bibfield  {author} {\bibinfo {author} {\bibfnamefont {N.~A.}\ \bibnamefont
  {Krivova}}, \bibinfo {author} {\bibfnamefont {L.}~\bibnamefont {Balmaceda}},
  \ and\ \bibinfo {author} {\bibfnamefont {S.~K.}\ \bibnamefont {Solanki}},\
  }\bibfield  {title} {\enquote {\bibinfo {title} {Reconstruction of solar
  total irradiance since 1700 from the surface magnetic flux},}\ }\href
  {\doibase 10.1051/0004-6361:20066725} {\bibfield  {journal} {\bibinfo
  {journal} {Astronomy \& Astrophysics}\ }\textbf {\bibinfo {volume} {467}},\
  \bibinfo {pages} {335--346} (\bibinfo {year} {2007})}\BibitemShut {NoStop}%
\bibitem [{\citenamefont {Vieira}\ and\ \citenamefont
  {Solanki}(2010)}]{vieira2010}%
  \BibitemOpen
  \bibfield  {author} {\bibinfo {author} {\bibfnamefont {L.~E.~A.}\
  \bibnamefont {Vieira}}\ and\ \bibinfo {author} {\bibfnamefont {S.~K.}\
  \bibnamefont {Solanki}},\ }\bibfield  {title} {\enquote {\bibinfo {title}
  {Evolution of the solar magnetic flux on time scales of years to millenia},}\
  }\href {\doibase 10.1051/0004-6361/200913276} {\bibfield  {journal} {\bibinfo
   {journal} {Astronomy and Astrophysics}\ }\textbf {\bibinfo {volume} {509}},\
  \bibinfo {pages} {A100} (\bibinfo {year} {2010})}\BibitemShut {NoStop}%
\bibitem [{\citenamefont {Gao}, \citenamefont {Robock},\ and\ \citenamefont
  {Ammann}(2008)}]{gao2008}%
  \BibitemOpen
  \bibfield  {author} {\bibinfo {author} {\bibfnamefont {C.}~\bibnamefont
  {Gao}}, \bibinfo {author} {\bibfnamefont {A.}~\bibnamefont {Robock}}, \ and\
  \bibinfo {author} {\bibfnamefont {C.}~\bibnamefont {Ammann}},\ }\bibfield
  {title} {\enquote {\bibinfo {title} {Volcanic forcing of climate over the
  past 1500 years: {{An}} improved ice core-based index for climate models},}\
  }\href {\doibase 10.1029/2008JD010239} {\bibfield  {journal} {\bibinfo
  {journal} {Journal of Geophysical Research}\ }\textbf {\bibinfo {volume}
  {113}},\ \bibinfo {pages} {D23111} (\bibinfo {year} {2008})}\BibitemShut
  {NoStop}%
\bibitem [{\citenamefont {Ritz}, \citenamefont {Stocker},\ and\ \citenamefont
  {Joos}(2011)}]{ritz2011}%
  \BibitemOpen
  \bibfield  {author} {\bibinfo {author} {\bibfnamefont {S.~P.}\ \bibnamefont
  {Ritz}}, \bibinfo {author} {\bibfnamefont {T.~F.}\ \bibnamefont {Stocker}}, \
  and\ \bibinfo {author} {\bibfnamefont {F.}~\bibnamefont {Joos}},\ }\bibfield
  {title} {\enquote {\bibinfo {title} {A {{Coupled Dynamical
  Ocean}}\textendash{{Energy Balance Atmosphere Model}} for {{Paleoclimate
  Studies}}},}\ }\href {\doibase 10.1175/2010JCLI3351.1} {\bibfield  {journal}
  {\bibinfo  {journal} {Journal of Climate}\ }\textbf {\bibinfo {volume}
  {24}},\ \bibinfo {pages} {349--375} (\bibinfo {year} {2011})}\BibitemShut
  {NoStop}%
\bibitem [{\citenamefont {Montoya}\ \emph {et~al.}(2006)\citenamefont
  {Montoya}, \citenamefont {Griesel}, \citenamefont {Levermann}, \citenamefont
  {Mignot}, \citenamefont {Hofmann}, \citenamefont {Ganopolski},\ and\
  \citenamefont {Rahmstorf}}]{montoya2006}%
  \BibitemOpen
  \bibfield  {author} {\bibinfo {author} {\bibfnamefont {M.}~\bibnamefont
  {Montoya}}, \bibinfo {author} {\bibfnamefont {A.}~\bibnamefont {Griesel}},
  \bibinfo {author} {\bibfnamefont {A.}~\bibnamefont {Levermann}}, \bibinfo
  {author} {\bibfnamefont {J.}~\bibnamefont {Mignot}}, \bibinfo {author}
  {\bibfnamefont {M.}~\bibnamefont {Hofmann}}, \bibinfo {author} {\bibfnamefont
  {A.}~\bibnamefont {Ganopolski}}, \ and\ \bibinfo {author} {\bibfnamefont
  {S.}~\bibnamefont {Rahmstorf}},\ }\bibfield  {title} {\enquote {\bibinfo
  {title} {The earth system model of intermediate complexity
  {{CLIMBER-3$\alpha$}}. {{Part I}}: Description and performance for
  present-day conditions},}\ }\href {\doibase 10.1007/s00382-005-0061-0}
  {\bibfield  {journal} {\bibinfo  {journal} {Climate Dynamics}\ }\textbf
  {\bibinfo {volume} {26}},\ \bibinfo {pages} {327--328} (\bibinfo {year}
  {2006})}\BibitemShut {NoStop}%
\bibitem [{\citenamefont {Petoukhov}\ \emph {et~al.}(2005)\citenamefont
  {Petoukhov}, \citenamefont {Claussen}, \citenamefont {Berger}, \citenamefont
  {Crucifix}, \citenamefont {Eby}, \citenamefont {Eliseev}, \citenamefont
  {Fichefet}, \citenamefont {Ganopolski}, \citenamefont {Goosse}, \citenamefont
  {Kamenkovich}, \citenamefont {Mokhov}, \citenamefont {Montoya}, \citenamefont
  {Mysak}, \citenamefont {Sokolov}, \citenamefont {Stone}, \citenamefont
  {Wang},\ and\ \citenamefont {Weaver}}]{petoukhov2005}%
  \BibitemOpen
  \bibfield  {author} {\bibinfo {author} {\bibfnamefont {V.}~\bibnamefont
  {Petoukhov}}, \bibinfo {author} {\bibfnamefont {M.}~\bibnamefont {Claussen}},
  \bibinfo {author} {\bibfnamefont {A.}~\bibnamefont {Berger}}, \bibinfo
  {author} {\bibfnamefont {M.}~\bibnamefont {Crucifix}}, \bibinfo {author}
  {\bibfnamefont {M.}~\bibnamefont {Eby}}, \bibinfo {author} {\bibfnamefont
  {A.~V.}\ \bibnamefont {Eliseev}}, \bibinfo {author} {\bibfnamefont
  {T.}~\bibnamefont {Fichefet}}, \bibinfo {author} {\bibfnamefont
  {A.}~\bibnamefont {Ganopolski}}, \bibinfo {author} {\bibfnamefont
  {H.}~\bibnamefont {Goosse}}, \bibinfo {author} {\bibfnamefont
  {I.}~\bibnamefont {Kamenkovich}}, \bibinfo {author} {\bibfnamefont {I.~I.}\
  \bibnamefont {Mokhov}}, \bibinfo {author} {\bibfnamefont {M.}~\bibnamefont
  {Montoya}}, \bibinfo {author} {\bibfnamefont {L.~A.}\ \bibnamefont {Mysak}},
  \bibinfo {author} {\bibfnamefont {A.}~\bibnamefont {Sokolov}}, \bibinfo
  {author} {\bibfnamefont {P.}~\bibnamefont {Stone}}, \bibinfo {author}
  {\bibfnamefont {Z.}~\bibnamefont {Wang}}, \ and\ \bibinfo {author}
  {\bibfnamefont {A.~J.}\ \bibnamefont {Weaver}},\ }\bibfield  {title}
  {\enquote {\bibinfo {title} {{{EMIC Intercomparison Project}}
  ({{EMIP}}\textendash{{CO2}}): Comparative analysis of {{EMIC}} simulations of
  climate, and of equilibrium and transient responses to atmospheric {{CO2}}
  doubling},}\ }\href {\doibase 10.1007/s00382-005-0042-3} {\bibfield
  {journal} {\bibinfo  {journal} {Climate Dynamics}\ }\textbf {\bibinfo
  {volume} {25}},\ \bibinfo {pages} {363--385} (\bibinfo {year}
  {2005})}\BibitemShut {NoStop}%
\bibitem [{\citenamefont {Shaffer}, \citenamefont {Malsk{\ae}r~Olsen},\ and\
  \citenamefont {Pepke~Pedersen}(2008)}]{shaffer2008}%
  \BibitemOpen
  \bibfield  {author} {\bibinfo {author} {\bibfnamefont {G.}~\bibnamefont
  {Shaffer}}, \bibinfo {author} {\bibfnamefont {S.}~\bibnamefont
  {Malsk{\ae}r~Olsen}}, \ and\ \bibinfo {author} {\bibfnamefont {J.~O.}\
  \bibnamefont {Pepke~Pedersen}},\ }\bibfield  {title} {\enquote {\bibinfo
  {title} {Presentation, calibration and validation of the low-order, {{DCESS
  Earth System Model}} ({{Version}} 1)},}\ }\href {\doibase
  10.5194/gmd-1-17-2008} {\bibfield  {journal} {\bibinfo  {journal}
  {Geoscientific Model Development}\ }\textbf {\bibinfo {volume} {1}},\
  \bibinfo {pages} {17--51} (\bibinfo {year} {2008})}\BibitemShut {NoStop}%
\bibitem [{\citenamefont {Sokolov}\ \emph {et~al.}(2005)\citenamefont
  {Sokolov}, \citenamefont {Schlosser}, \citenamefont {Dutkiewicz},
  \citenamefont {Paltsev}, \citenamefont {Kicklighter}, \citenamefont {Jacoby},
  \citenamefont {Prinn}, \citenamefont {Forest}, \citenamefont {Reilly},
  \citenamefont {Wang} \emph {et~al.}}]{sokolov2005integrated}%
  \BibitemOpen
  \bibfield  {author} {\bibinfo {author} {\bibfnamefont {A.~P.}\ \bibnamefont
  {Sokolov}}, \bibinfo {author} {\bibfnamefont {C.~A.}\ \bibnamefont
  {Schlosser}}, \bibinfo {author} {\bibfnamefont {S.}~\bibnamefont
  {Dutkiewicz}}, \bibinfo {author} {\bibfnamefont {S.}~\bibnamefont {Paltsev}},
  \bibinfo {author} {\bibfnamefont {D.~W.}\ \bibnamefont {Kicklighter}},
  \bibinfo {author} {\bibfnamefont {H.~D.}\ \bibnamefont {Jacoby}}, \bibinfo
  {author} {\bibfnamefont {R.~G.}\ \bibnamefont {Prinn}}, \bibinfo {author}
  {\bibfnamefont {C.~E.}\ \bibnamefont {Forest}}, \bibinfo {author}
  {\bibfnamefont {J.~M.}\ \bibnamefont {Reilly}}, \bibinfo {author}
  {\bibfnamefont {C.}~\bibnamefont {Wang}},  \emph {et~al.},\ }\href
  {http://globalchange.mit.edu/publication/14579} {\enquote {\bibinfo {title}
  {{{MIT}} integrated global system model ({{IGSM}}) version 2: Model
  description and baseline evaluation},}\ }\bibinfo {type} {Tech. Rep.}\
  (\bibinfo  {institution} {{MIT Joint Program on the Science and Policy of
  Global Change}},\ \bibinfo {year} {2005})\BibitemShut {NoStop}%
\bibitem [{\citenamefont {Goosse}\ \emph {et~al.}(2010)\citenamefont {Goosse},
  \citenamefont {Brovkin}, \citenamefont {Fichefet}, \citenamefont {Haarsma},
  \citenamefont {Huybrechts}, \citenamefont {Jongma}, \citenamefont {Mouchet},
  \citenamefont {Selten}, \citenamefont {Barriat}, \citenamefont {Campin},
  \citenamefont {Deleersnijder}, \citenamefont {Driesschaert}, \citenamefont
  {Goelzer}, \citenamefont {Janssens}, \citenamefont {Loutre}, \citenamefont
  {Morales~Maqueda}, \citenamefont {Opsteegh}, \citenamefont {Mathieu},
  \citenamefont {Munhoven}, \citenamefont {Pettersson}, \citenamefont
  {Renssen}, \citenamefont {Roche}, \citenamefont {Schaeffer}, \citenamefont
  {Tartinville}, \citenamefont {Timmermann},\ and\ \citenamefont
  {Weber}}]{goosse2010}%
  \BibitemOpen
  \bibfield  {author} {\bibinfo {author} {\bibfnamefont {H.}~\bibnamefont
  {Goosse}}, \bibinfo {author} {\bibfnamefont {V.}~\bibnamefont {Brovkin}},
  \bibinfo {author} {\bibfnamefont {T.}~\bibnamefont {Fichefet}}, \bibinfo
  {author} {\bibfnamefont {R.}~\bibnamefont {Haarsma}}, \bibinfo {author}
  {\bibfnamefont {P.}~\bibnamefont {Huybrechts}}, \bibinfo {author}
  {\bibfnamefont {J.}~\bibnamefont {Jongma}}, \bibinfo {author} {\bibfnamefont
  {A.}~\bibnamefont {Mouchet}}, \bibinfo {author} {\bibfnamefont
  {F.}~\bibnamefont {Selten}}, \bibinfo {author} {\bibfnamefont {P.-Y.}\
  \bibnamefont {Barriat}}, \bibinfo {author} {\bibfnamefont {J.-M.}\
  \bibnamefont {Campin}}, \bibinfo {author} {\bibfnamefont {E.}~\bibnamefont
  {Deleersnijder}}, \bibinfo {author} {\bibfnamefont {E.}~\bibnamefont
  {Driesschaert}}, \bibinfo {author} {\bibfnamefont {H.}~\bibnamefont
  {Goelzer}}, \bibinfo {author} {\bibfnamefont {I.}~\bibnamefont {Janssens}},
  \bibinfo {author} {\bibfnamefont {M.-F.}\ \bibnamefont {Loutre}}, \bibinfo
  {author} {\bibfnamefont {M.~A.}\ \bibnamefont {Morales~Maqueda}}, \bibinfo
  {author} {\bibfnamefont {T.}~\bibnamefont {Opsteegh}}, \bibinfo {author}
  {\bibfnamefont {P.-P.}\ \bibnamefont {Mathieu}}, \bibinfo {author}
  {\bibfnamefont {G.}~\bibnamefont {Munhoven}}, \bibinfo {author}
  {\bibfnamefont {E.~J.}\ \bibnamefont {Pettersson}}, \bibinfo {author}
  {\bibfnamefont {H.}~\bibnamefont {Renssen}}, \bibinfo {author} {\bibfnamefont
  {D.~M.}\ \bibnamefont {Roche}}, \bibinfo {author} {\bibfnamefont
  {M.}~\bibnamefont {Schaeffer}}, \bibinfo {author} {\bibfnamefont
  {B.}~\bibnamefont {Tartinville}}, \bibinfo {author} {\bibfnamefont
  {A.}~\bibnamefont {Timmermann}}, \ and\ \bibinfo {author} {\bibfnamefont
  {S.~L.}\ \bibnamefont {Weber}},\ }\bibfield  {title} {\enquote {\bibinfo
  {title} {Description of the {{Earth}} system model of intermediate complexity
  {{LOVECLIM}} version 1.2},}\ }\href {\doibase 10.5194/gmd-3-603-2010}
  {\bibfield  {journal} {\bibinfo  {journal} {Geoscientific Model Development}\
  }\textbf {\bibinfo {volume} {3}},\ \bibinfo {pages} {603--633} (\bibinfo
  {year} {2010})}\BibitemShut {NoStop}%
\bibitem [{\citenamefont {Matsumoto}\ \emph {et~al.}(2008)\citenamefont
  {Matsumoto}, \citenamefont {Tokos}, \citenamefont {Price},\ and\
  \citenamefont {Cox}}]{matsumoto2008}%
  \BibitemOpen
  \bibfield  {author} {\bibinfo {author} {\bibfnamefont {K.}~\bibnamefont
  {Matsumoto}}, \bibinfo {author} {\bibfnamefont {K.~S.}\ \bibnamefont
  {Tokos}}, \bibinfo {author} {\bibfnamefont {A.~R.}\ \bibnamefont {Price}}, \
  and\ \bibinfo {author} {\bibfnamefont {S.~J.}\ \bibnamefont {Cox}},\
  }\bibfield  {title} {\enquote {\bibinfo {title} {First description of the
  {{Minnesota Earth System Model}} for {{Ocean}} biogeochemistry ({{MESMO}}
  1.0)},}\ }\href {\doibase 10.5194/gmd-1-1-2008} {\bibfield  {journal}
  {\bibinfo  {journal} {Geoscientific Model Development}\ }\textbf {\bibinfo
  {volume} {1}},\ \bibinfo {pages} {1--15} (\bibinfo {year}
  {2008})}\BibitemShut {NoStop}%
\bibitem [{\citenamefont {Zeng}(2004)}]{zeng2004}%
  \BibitemOpen
  \bibfield  {author} {\bibinfo {author} {\bibfnamefont {N.}~\bibnamefont
  {Zeng}},\ }\bibfield  {title} {\enquote {\bibinfo {title} {How strong is
  carbon cycle-climate feedback under global warming?}}\ }\href {\doibase
  10.1029/2004GL020904} {\bibfield  {journal} {\bibinfo  {journal} {Geophysical
  Research Letters}\ }\textbf {\bibinfo {volume} {31}},\ \bibinfo {pages}
  {L20203} (\bibinfo {year} {2004})}\BibitemShut {NoStop}%
\bibitem [{\citenamefont {Weaver}\ \emph {et~al.}(2001)\citenamefont {Weaver},
  \citenamefont {Eby}, \citenamefont {Wiebe}, \citenamefont {Bitz},
  \citenamefont {Duffy}, \citenamefont {Ewen}, \citenamefont {Fanning},
  \citenamefont {Holland}, \citenamefont {MacFadyen}, \citenamefont {Matthews},
  \citenamefont {Meissner}, \citenamefont {Saenko}, \citenamefont {Schmittner},
  \citenamefont {Wang},\ and\ \citenamefont {Yoshimori}}]{weaver2001}%
  \BibitemOpen
  \bibfield  {author} {\bibinfo {author} {\bibfnamefont {A.~J.}\ \bibnamefont
  {Weaver}}, \bibinfo {author} {\bibfnamefont {M.}~\bibnamefont {Eby}},
  \bibinfo {author} {\bibfnamefont {E.~C.}\ \bibnamefont {Wiebe}}, \bibinfo
  {author} {\bibfnamefont {C.~M.}\ \bibnamefont {Bitz}}, \bibinfo {author}
  {\bibfnamefont {P.~B.}\ \bibnamefont {Duffy}}, \bibinfo {author}
  {\bibfnamefont {T.~L.}\ \bibnamefont {Ewen}}, \bibinfo {author}
  {\bibfnamefont {A.~F.}\ \bibnamefont {Fanning}}, \bibinfo {author}
  {\bibfnamefont {M.~M.}\ \bibnamefont {Holland}}, \bibinfo {author}
  {\bibfnamefont {A.}~\bibnamefont {MacFadyen}}, \bibinfo {author}
  {\bibfnamefont {H.~D.}\ \bibnamefont {Matthews}}, \bibinfo {author}
  {\bibfnamefont {K.~J.}\ \bibnamefont {Meissner}}, \bibinfo {author}
  {\bibfnamefont {O.}~\bibnamefont {Saenko}}, \bibinfo {author} {\bibfnamefont
  {A.}~\bibnamefont {Schmittner}}, \bibinfo {author} {\bibfnamefont
  {H.}~\bibnamefont {Wang}}, \ and\ \bibinfo {author} {\bibfnamefont
  {M.}~\bibnamefont {Yoshimori}},\ }\bibfield  {title} {\enquote {\bibinfo
  {title} {The {{UVic}} earth system climate model: {{Model}} description,
  climatology, and applications to past, present and future climates},}\ }\href
  {\doibase 10.1080/07055900.2001.9649686} {\bibfield  {journal} {\bibinfo
  {journal} {Atmosphere-Ocean}\ }\textbf {\bibinfo {volume} {39}},\ \bibinfo
  {pages} {361--428} (\bibinfo {year} {2001})}\BibitemShut {NoStop}%
\bibitem [{\citenamefont {{Xiao-Ge}}, \citenamefont {{Tong-Wen}},\ and\
  \citenamefont {Jie}(2013)}]{xiao-ge2013}%
  \BibitemOpen
  \bibfield  {author} {\bibinfo {author} {\bibfnamefont {X.}~\bibnamefont
  {{Xiao-Ge}}}, \bibinfo {author} {\bibfnamefont {W.}~\bibnamefont
  {{Tong-Wen}}}, \ and\ \bibinfo {author} {\bibfnamefont {Z.}~\bibnamefont
  {Jie}},\ }\bibfield  {title} {\enquote {\bibinfo {title} {Introduction of
  {{CMIP5 Experiments Carried}} out with the {{Climate System Models}} of
  {{Beijing Climate Center}}},}\ }\href {\doibase 10.3724/SP.J.1248.2013.041}
  {\bibfield  {journal} {\bibinfo  {journal} {Advances in Climate Change
  Research}\ }\textbf {\bibinfo {volume} {4}},\ \bibinfo {pages} {41--49}
  (\bibinfo {year} {2013})}\BibitemShut {NoStop}%
\bibitem [{\citenamefont {Landrum}\ \emph {et~al.}(2013)\citenamefont
  {Landrum}, \citenamefont {{Otto-Bliesner}}, \citenamefont {Wahl},
  \citenamefont {Conley}, \citenamefont {Lawrence}, \citenamefont
  {Rosenbloom},\ and\ \citenamefont {Teng}}]{landrum2013}%
  \BibitemOpen
  \bibfield  {author} {\bibinfo {author} {\bibfnamefont {L.}~\bibnamefont
  {Landrum}}, \bibinfo {author} {\bibfnamefont {B.~L.}\ \bibnamefont
  {{Otto-Bliesner}}}, \bibinfo {author} {\bibfnamefont {E.~R.}\ \bibnamefont
  {Wahl}}, \bibinfo {author} {\bibfnamefont {A.}~\bibnamefont {Conley}},
  \bibinfo {author} {\bibfnamefont {P.~J.}\ \bibnamefont {Lawrence}}, \bibinfo
  {author} {\bibfnamefont {N.}~\bibnamefont {Rosenbloom}}, \ and\ \bibinfo
  {author} {\bibfnamefont {H.}~\bibnamefont {Teng}},\ }\bibfield  {title}
  {\enquote {\bibinfo {title} {Last {{Millennium Climate}} and {{Its
  Variability}} in {{CCSM4}}},}\ }\href {\doibase 10.1175/JCLI-D-11-00326.1}
  {\bibfield  {journal} {\bibinfo  {journal} {Journal of Climate}\ }\textbf
  {\bibinfo {volume} {26}},\ \bibinfo {pages} {1085--1111} (\bibinfo {year}
  {2013})}\BibitemShut {NoStop}%
\bibitem [{\citenamefont {Phipps}\ \emph {et~al.}(2012)\citenamefont {Phipps},
  \citenamefont {Rotstayn}, \citenamefont {Gordon}, \citenamefont {Roberts},
  \citenamefont {Hirst},\ and\ \citenamefont {Budd}}]{phipps2012}%
  \BibitemOpen
  \bibfield  {author} {\bibinfo {author} {\bibfnamefont {S.~J.}\ \bibnamefont
  {Phipps}}, \bibinfo {author} {\bibfnamefont {L.~D.}\ \bibnamefont
  {Rotstayn}}, \bibinfo {author} {\bibfnamefont {H.~B.}\ \bibnamefont
  {Gordon}}, \bibinfo {author} {\bibfnamefont {J.~L.}\ \bibnamefont {Roberts}},
  \bibinfo {author} {\bibfnamefont {A.~C.}\ \bibnamefont {Hirst}}, \ and\
  \bibinfo {author} {\bibfnamefont {W.~F.}\ \bibnamefont {Budd}},\ }\bibfield
  {title} {\enquote {\bibinfo {title} {The {{CSIRO Mk3L}} climate system model
  version 1.0 \textendash{} {{Part}} 2: {{Response}} to external forcings},}\
  }\href {\doibase 10.5194/gmd-5-649-2012} {\bibfield  {journal} {\bibinfo
  {journal} {Geoscientific Model Development}\ }\textbf {\bibinfo {volume}
  {5}},\ \bibinfo {pages} {649--682} (\bibinfo {year} {2012})}\BibitemShut
  {NoStop}%
\bibitem [{\citenamefont {Bao}\ \emph {et~al.}(2013)\citenamefont {Bao},
  \citenamefont {Lin}, \citenamefont {Zhou}, \citenamefont {Liu}, \citenamefont
  {Yu}, \citenamefont {Wu}, \citenamefont {He}, \citenamefont {He},
  \citenamefont {Li}, \citenamefont {Li}, \citenamefont {Li}, \citenamefont
  {Liu}, \citenamefont {Qiao}, \citenamefont {Song}, \citenamefont {Wang},
  \citenamefont {Wang}, \citenamefont {Wang}, \citenamefont {Wang},
  \citenamefont {Wang}, \citenamefont {Wu}, \citenamefont {Wu}, \citenamefont
  {Xu}, \citenamefont {Yu}, \citenamefont {Zhao}, \citenamefont {Zheng},\ and\
  \citenamefont {Zhou}}]{bao2013}%
  \BibitemOpen
  \bibfield  {author} {\bibinfo {author} {\bibfnamefont {Q.}~\bibnamefont
  {Bao}}, \bibinfo {author} {\bibfnamefont {P.}~\bibnamefont {Lin}}, \bibinfo
  {author} {\bibfnamefont {T.}~\bibnamefont {Zhou}}, \bibinfo {author}
  {\bibfnamefont {Y.}~\bibnamefont {Liu}}, \bibinfo {author} {\bibfnamefont
  {Y.}~\bibnamefont {Yu}}, \bibinfo {author} {\bibfnamefont {G.}~\bibnamefont
  {Wu}}, \bibinfo {author} {\bibfnamefont {B.}~\bibnamefont {He}}, \bibinfo
  {author} {\bibfnamefont {J.}~\bibnamefont {He}}, \bibinfo {author}
  {\bibfnamefont {L.}~\bibnamefont {Li}}, \bibinfo {author} {\bibfnamefont
  {J.}~\bibnamefont {Li}}, \bibinfo {author} {\bibfnamefont {Y.}~\bibnamefont
  {Li}}, \bibinfo {author} {\bibfnamefont {H.}~\bibnamefont {Liu}}, \bibinfo
  {author} {\bibfnamefont {F.}~\bibnamefont {Qiao}}, \bibinfo {author}
  {\bibfnamefont {Z.}~\bibnamefont {Song}}, \bibinfo {author} {\bibfnamefont
  {B.}~\bibnamefont {Wang}}, \bibinfo {author} {\bibfnamefont {J.}~\bibnamefont
  {Wang}}, \bibinfo {author} {\bibfnamefont {P.}~\bibnamefont {Wang}}, \bibinfo
  {author} {\bibfnamefont {X.}~\bibnamefont {Wang}}, \bibinfo {author}
  {\bibfnamefont {Z.}~\bibnamefont {Wang}}, \bibinfo {author} {\bibfnamefont
  {B.}~\bibnamefont {Wu}}, \bibinfo {author} {\bibfnamefont {T.}~\bibnamefont
  {Wu}}, \bibinfo {author} {\bibfnamefont {Y.}~\bibnamefont {Xu}}, \bibinfo
  {author} {\bibfnamefont {H.}~\bibnamefont {Yu}}, \bibinfo {author}
  {\bibfnamefont {W.}~\bibnamefont {Zhao}}, \bibinfo {author} {\bibfnamefont
  {W.}~\bibnamefont {Zheng}}, \ and\ \bibinfo {author} {\bibfnamefont
  {L.}~\bibnamefont {Zhou}},\ }\bibfield  {title} {\enquote {\bibinfo {title}
  {The {{Flexible Global Ocean-Atmosphere-Land}} system model, {{Spectral
  Version}} 2: {{FGOALS-s2}}},}\ }\href {\doibase 10.1007/s00376-012-2113-9}
  {\bibfield  {journal} {\bibinfo  {journal} {Advances in Atmospheric
  Sciences}\ }\textbf {\bibinfo {volume} {30}},\ \bibinfo {pages} {561--576}
  (\bibinfo {year} {2013})}\BibitemShut {NoStop}%
\bibitem [{\citenamefont {Schmidt}\ \emph {et~al.}(2006)\citenamefont
  {Schmidt}, \citenamefont {Ruedy}, \citenamefont {Hansen}, \citenamefont
  {Aleinov}, \citenamefont {Bell}, \citenamefont {Bauer}, \citenamefont
  {Bauer}, \citenamefont {Cairns}, \citenamefont {Canuto}, \citenamefont
  {Cheng}, \citenamefont {Del~Genio}, \citenamefont {Faluvegi}, \citenamefont
  {Friend}, \citenamefont {Hall}, \citenamefont {Hu}, \citenamefont {Kelley},
  \citenamefont {Kiang}, \citenamefont {Koch}, \citenamefont {Lacis},
  \citenamefont {Lerner}, \citenamefont {Lo}, \citenamefont {Miller},
  \citenamefont {Nazarenko}, \citenamefont {Oinas}, \citenamefont {Perlwitz},
  \citenamefont {Perlwitz}, \citenamefont {Rind}, \citenamefont {Romanou},
  \citenamefont {Russell}, \citenamefont {Sato}, \citenamefont {Shindell},
  \citenamefont {Stone}, \citenamefont {Sun}, \citenamefont {Tausnev},
  \citenamefont {Thresher},\ and\ \citenamefont {Yao}}]{schmidt2006}%
  \BibitemOpen
  \bibfield  {author} {\bibinfo {author} {\bibfnamefont {G.~A.}\ \bibnamefont
  {Schmidt}}, \bibinfo {author} {\bibfnamefont {R.}~\bibnamefont {Ruedy}},
  \bibinfo {author} {\bibfnamefont {J.~E.}\ \bibnamefont {Hansen}}, \bibinfo
  {author} {\bibfnamefont {I.}~\bibnamefont {Aleinov}}, \bibinfo {author}
  {\bibfnamefont {N.}~\bibnamefont {Bell}}, \bibinfo {author} {\bibfnamefont
  {M.}~\bibnamefont {Bauer}}, \bibinfo {author} {\bibfnamefont
  {S.}~\bibnamefont {Bauer}}, \bibinfo {author} {\bibfnamefont
  {B.}~\bibnamefont {Cairns}}, \bibinfo {author} {\bibfnamefont
  {V.}~\bibnamefont {Canuto}}, \bibinfo {author} {\bibfnamefont
  {Y.}~\bibnamefont {Cheng}}, \bibinfo {author} {\bibfnamefont
  {A.}~\bibnamefont {Del~Genio}}, \bibinfo {author} {\bibfnamefont
  {G.}~\bibnamefont {Faluvegi}}, \bibinfo {author} {\bibfnamefont {A.~D.}\
  \bibnamefont {Friend}}, \bibinfo {author} {\bibfnamefont {T.~M.}\
  \bibnamefont {Hall}}, \bibinfo {author} {\bibfnamefont {Y.}~\bibnamefont
  {Hu}}, \bibinfo {author} {\bibfnamefont {M.}~\bibnamefont {Kelley}}, \bibinfo
  {author} {\bibfnamefont {N.~Y.}\ \bibnamefont {Kiang}}, \bibinfo {author}
  {\bibfnamefont {D.}~\bibnamefont {Koch}}, \bibinfo {author} {\bibfnamefont
  {A.~A.}\ \bibnamefont {Lacis}}, \bibinfo {author} {\bibfnamefont
  {J.}~\bibnamefont {Lerner}}, \bibinfo {author} {\bibfnamefont {K.~K.}\
  \bibnamefont {Lo}}, \bibinfo {author} {\bibfnamefont {R.~L.}\ \bibnamefont
  {Miller}}, \bibinfo {author} {\bibfnamefont {L.}~\bibnamefont {Nazarenko}},
  \bibinfo {author} {\bibfnamefont {V.}~\bibnamefont {Oinas}}, \bibinfo
  {author} {\bibfnamefont {J.}~\bibnamefont {Perlwitz}}, \bibinfo {author}
  {\bibfnamefont {J.}~\bibnamefont {Perlwitz}}, \bibinfo {author}
  {\bibfnamefont {D.}~\bibnamefont {Rind}}, \bibinfo {author} {\bibfnamefont
  {A.}~\bibnamefont {Romanou}}, \bibinfo {author} {\bibfnamefont {G.~L.}\
  \bibnamefont {Russell}}, \bibinfo {author} {\bibfnamefont {M.}~\bibnamefont
  {Sato}}, \bibinfo {author} {\bibfnamefont {D.~T.}\ \bibnamefont {Shindell}},
  \bibinfo {author} {\bibfnamefont {P.~H.}\ \bibnamefont {Stone}}, \bibinfo
  {author} {\bibfnamefont {S.}~\bibnamefont {Sun}}, \bibinfo {author}
  {\bibfnamefont {N.}~\bibnamefont {Tausnev}}, \bibinfo {author} {\bibfnamefont
  {D.}~\bibnamefont {Thresher}}, \ and\ \bibinfo {author} {\bibfnamefont
  {M.-S.}\ \bibnamefont {Yao}},\ }\bibfield  {title} {\enquote {\bibinfo
  {title} {Present-{{Day Atmospheric Simulations Using GISS ModelE}}:
  {{Comparison}} to {{In Situ}}, {{Satellite}}, and {{Reanalysis Data}}},}\
  }\href {\doibase 10.1175/JCLI3612.1} {\bibfield  {journal} {\bibinfo
  {journal} {Journal of Climate}\ }\textbf {\bibinfo {volume} {19}},\ \bibinfo
  {pages} {153--192} (\bibinfo {year} {2006})}\BibitemShut {NoStop}%
\bibitem [{\citenamefont {Schurer}, \citenamefont {Tett},\ and\ \citenamefont
  {Hegerl}(2014)}]{schurer2014}%
  \BibitemOpen
  \bibfield  {author} {\bibinfo {author} {\bibfnamefont {A.~P.}\ \bibnamefont
  {Schurer}}, \bibinfo {author} {\bibfnamefont {S.~F.~B.}\ \bibnamefont
  {Tett}}, \ and\ \bibinfo {author} {\bibfnamefont {G.~C.}\ \bibnamefont
  {Hegerl}},\ }\bibfield  {title} {\enquote {\bibinfo {title} {Small influence
  of solar variability on climate over the past millennium},}\ }\href {\doibase
  10.1038/ngeo2040} {\bibfield  {journal} {\bibinfo  {journal} {Nature
  Geoscience}\ }\textbf {\bibinfo {volume} {7}},\ \bibinfo {pages} {104--108}
  (\bibinfo {year} {2014})}\BibitemShut {NoStop}%
\bibitem [{\citenamefont {Jones}\ \emph {et~al.}(2011)\citenamefont {Jones},
  \citenamefont {Hughes}, \citenamefont {Bellouin}, \citenamefont {Hardiman},
  \citenamefont {Jones}, \citenamefont {Knight}, \citenamefont {Liddicoat},
  \citenamefont {O'Connor}, \citenamefont {Andres}, \citenamefont {Bell},
  \citenamefont {Boo}, \citenamefont {Bozzo}, \citenamefont {Butchart},
  \citenamefont {Cadule}, \citenamefont {Corbin}, \citenamefont
  {{Doutriaux-Boucher}}, \citenamefont {Friedlingstein}, \citenamefont
  {Gornall}, \citenamefont {Gray}, \citenamefont {Halloran}, \citenamefont
  {Hurtt}, \citenamefont {Ingram}, \citenamefont {Lamarque}, \citenamefont
  {Law}, \citenamefont {Meinshausen}, \citenamefont {Osprey}, \citenamefont
  {Palin}, \citenamefont {Parsons~Chini}, \citenamefont {Raddatz},
  \citenamefont {Sanderson}, \citenamefont {Sellar}, \citenamefont {Schurer},
  \citenamefont {Valdes}, \citenamefont {Wood}, \citenamefont {Woodward},
  \citenamefont {Yoshioka},\ and\ \citenamefont {Zerroukat}}]{jones2011}%
  \BibitemOpen
  \bibfield  {author} {\bibinfo {author} {\bibfnamefont {C.~D.}\ \bibnamefont
  {Jones}}, \bibinfo {author} {\bibfnamefont {J.~K.}\ \bibnamefont {Hughes}},
  \bibinfo {author} {\bibfnamefont {N.}~\bibnamefont {Bellouin}}, \bibinfo
  {author} {\bibfnamefont {S.~C.}\ \bibnamefont {Hardiman}}, \bibinfo {author}
  {\bibfnamefont {G.~S.}\ \bibnamefont {Jones}}, \bibinfo {author}
  {\bibfnamefont {J.}~\bibnamefont {Knight}}, \bibinfo {author} {\bibfnamefont
  {S.}~\bibnamefont {Liddicoat}}, \bibinfo {author} {\bibfnamefont {F.~M.}\
  \bibnamefont {O'Connor}}, \bibinfo {author} {\bibfnamefont {R.~J.}\
  \bibnamefont {Andres}}, \bibinfo {author} {\bibfnamefont {C.}~\bibnamefont
  {Bell}}, \bibinfo {author} {\bibfnamefont {K.-O.}\ \bibnamefont {Boo}},
  \bibinfo {author} {\bibfnamefont {A.}~\bibnamefont {Bozzo}}, \bibinfo
  {author} {\bibfnamefont {N.}~\bibnamefont {Butchart}}, \bibinfo {author}
  {\bibfnamefont {P.}~\bibnamefont {Cadule}}, \bibinfo {author} {\bibfnamefont
  {K.~D.}\ \bibnamefont {Corbin}}, \bibinfo {author} {\bibfnamefont
  {M.}~\bibnamefont {{Doutriaux-Boucher}}}, \bibinfo {author} {\bibfnamefont
  {P.}~\bibnamefont {Friedlingstein}}, \bibinfo {author} {\bibfnamefont
  {J.}~\bibnamefont {Gornall}}, \bibinfo {author} {\bibfnamefont
  {L.}~\bibnamefont {Gray}}, \bibinfo {author} {\bibfnamefont {P.~R.}\
  \bibnamefont {Halloran}}, \bibinfo {author} {\bibfnamefont {G.}~\bibnamefont
  {Hurtt}}, \bibinfo {author} {\bibfnamefont {W.~J.}\ \bibnamefont {Ingram}},
  \bibinfo {author} {\bibfnamefont {J.-F.}\ \bibnamefont {Lamarque}}, \bibinfo
  {author} {\bibfnamefont {R.~M.}\ \bibnamefont {Law}}, \bibinfo {author}
  {\bibfnamefont {M.}~\bibnamefont {Meinshausen}}, \bibinfo {author}
  {\bibfnamefont {S.}~\bibnamefont {Osprey}}, \bibinfo {author} {\bibfnamefont
  {E.~J.}\ \bibnamefont {Palin}}, \bibinfo {author} {\bibfnamefont
  {L.}~\bibnamefont {Parsons~Chini}}, \bibinfo {author} {\bibfnamefont
  {T.}~\bibnamefont {Raddatz}}, \bibinfo {author} {\bibfnamefont {M.~G.}\
  \bibnamefont {Sanderson}}, \bibinfo {author} {\bibfnamefont {A.~A.}\
  \bibnamefont {Sellar}}, \bibinfo {author} {\bibfnamefont {A.}~\bibnamefont
  {Schurer}}, \bibinfo {author} {\bibfnamefont {P.}~\bibnamefont {Valdes}},
  \bibinfo {author} {\bibfnamefont {N.}~\bibnamefont {Wood}}, \bibinfo {author}
  {\bibfnamefont {S.}~\bibnamefont {Woodward}}, \bibinfo {author}
  {\bibfnamefont {M.}~\bibnamefont {Yoshioka}}, \ and\ \bibinfo {author}
  {\bibfnamefont {M.}~\bibnamefont {Zerroukat}},\ }\bibfield  {title} {\enquote
  {\bibinfo {title} {The {{HadGEM2-ES}} implementation of {{CMIP5}} centennial
  simulations},}\ }\href {\doibase 10.5194/gmd-4-543-2011} {\bibfield
  {journal} {\bibinfo  {journal} {Geoscientific Model Development}\ }\textbf
  {\bibinfo {volume} {4}},\ \bibinfo {pages} {543--570} (\bibinfo {year}
  {2011})}\BibitemShut {NoStop}%
\bibitem [{\citenamefont {Dufresne}\ \emph {et~al.}(2013)\citenamefont
  {Dufresne}, \citenamefont {Foujols}, \citenamefont {Denvil}, \citenamefont
  {Caubel}, \citenamefont {Marti}, \citenamefont {Aumont}, \citenamefont
  {Balkanski}, \citenamefont {Bekki}, \citenamefont {Bellenger}, \citenamefont
  {Benshila}, \citenamefont {Bony}, \citenamefont {Bopp}, \citenamefont
  {Braconnot}, \citenamefont {Brockmann}, \citenamefont {Cadule}, \citenamefont
  {Cheruy}, \citenamefont {Codron}, \citenamefont {Cozic}, \citenamefont
  {Cugnet}, \citenamefont {{de Noblet}}, \citenamefont {Duvel}, \citenamefont
  {Eth{\'e}}, \citenamefont {Fairhead}, \citenamefont {Fichefet}, \citenamefont
  {Flavoni}, \citenamefont {Friedlingstein}, \citenamefont {Grandpeix},
  \citenamefont {Guez}, \citenamefont {Guilyardi}, \citenamefont
  {Hauglustaine}, \citenamefont {Hourdin}, \citenamefont {Idelkadi},
  \citenamefont {Ghattas}, \citenamefont {Joussaume}, \citenamefont {Kageyama},
  \citenamefont {Krinner}, \citenamefont {Labetoulle}, \citenamefont
  {Lahellec}, \citenamefont {Lefebvre}, \citenamefont {Lefevre}, \citenamefont
  {Levy}, \citenamefont {Li}, \citenamefont {Lloyd}, \citenamefont {Lott},
  \citenamefont {Madec}, \citenamefont {Mancip}, \citenamefont {Marchand},
  \citenamefont {Masson}, \citenamefont {Meurdesoif}, \citenamefont {Mignot},
  \citenamefont {Musat}, \citenamefont {Parouty}, \citenamefont {Polcher},
  \citenamefont {Rio}, \citenamefont {Schulz}, \citenamefont {Swingedouw},
  \citenamefont {Szopa}, \citenamefont {Talandier}, \citenamefont {Terray},
  \citenamefont {Viovy},\ and\ \citenamefont {Vuichard}}]{dufresne2013}%
  \BibitemOpen
  \bibfield  {author} {\bibinfo {author} {\bibfnamefont {J.-L.}\ \bibnamefont
  {Dufresne}}, \bibinfo {author} {\bibfnamefont {M.-A.}\ \bibnamefont
  {Foujols}}, \bibinfo {author} {\bibfnamefont {S.}~\bibnamefont {Denvil}},
  \bibinfo {author} {\bibfnamefont {A.}~\bibnamefont {Caubel}}, \bibinfo
  {author} {\bibfnamefont {O.}~\bibnamefont {Marti}}, \bibinfo {author}
  {\bibfnamefont {O.}~\bibnamefont {Aumont}}, \bibinfo {author} {\bibfnamefont
  {Y.}~\bibnamefont {Balkanski}}, \bibinfo {author} {\bibfnamefont
  {S.}~\bibnamefont {Bekki}}, \bibinfo {author} {\bibfnamefont
  {H.}~\bibnamefont {Bellenger}}, \bibinfo {author} {\bibfnamefont
  {R.}~\bibnamefont {Benshila}}, \bibinfo {author} {\bibfnamefont
  {S.}~\bibnamefont {Bony}}, \bibinfo {author} {\bibfnamefont {L.}~\bibnamefont
  {Bopp}}, \bibinfo {author} {\bibfnamefont {P.}~\bibnamefont {Braconnot}},
  \bibinfo {author} {\bibfnamefont {P.}~\bibnamefont {Brockmann}}, \bibinfo
  {author} {\bibfnamefont {P.}~\bibnamefont {Cadule}}, \bibinfo {author}
  {\bibfnamefont {F.}~\bibnamefont {Cheruy}}, \bibinfo {author} {\bibfnamefont
  {F.}~\bibnamefont {Codron}}, \bibinfo {author} {\bibfnamefont
  {A.}~\bibnamefont {Cozic}}, \bibinfo {author} {\bibfnamefont
  {D.}~\bibnamefont {Cugnet}}, \bibinfo {author} {\bibfnamefont
  {N.}~\bibnamefont {{de Noblet}}}, \bibinfo {author} {\bibfnamefont {J.-P.}\
  \bibnamefont {Duvel}}, \bibinfo {author} {\bibfnamefont {C.}~\bibnamefont
  {Eth{\'e}}}, \bibinfo {author} {\bibfnamefont {L.}~\bibnamefont {Fairhead}},
  \bibinfo {author} {\bibfnamefont {T.}~\bibnamefont {Fichefet}}, \bibinfo
  {author} {\bibfnamefont {S.}~\bibnamefont {Flavoni}}, \bibinfo {author}
  {\bibfnamefont {P.}~\bibnamefont {Friedlingstein}}, \bibinfo {author}
  {\bibfnamefont {J.-Y.}\ \bibnamefont {Grandpeix}}, \bibinfo {author}
  {\bibfnamefont {L.}~\bibnamefont {Guez}}, \bibinfo {author} {\bibfnamefont
  {E.}~\bibnamefont {Guilyardi}}, \bibinfo {author} {\bibfnamefont
  {D.}~\bibnamefont {Hauglustaine}}, \bibinfo {author} {\bibfnamefont
  {F.}~\bibnamefont {Hourdin}}, \bibinfo {author} {\bibfnamefont
  {A.}~\bibnamefont {Idelkadi}}, \bibinfo {author} {\bibfnamefont
  {J.}~\bibnamefont {Ghattas}}, \bibinfo {author} {\bibfnamefont
  {S.}~\bibnamefont {Joussaume}}, \bibinfo {author} {\bibfnamefont
  {M.}~\bibnamefont {Kageyama}}, \bibinfo {author} {\bibfnamefont
  {G.}~\bibnamefont {Krinner}}, \bibinfo {author} {\bibfnamefont
  {S.}~\bibnamefont {Labetoulle}}, \bibinfo {author} {\bibfnamefont
  {A.}~\bibnamefont {Lahellec}}, \bibinfo {author} {\bibfnamefont {M.-P.}\
  \bibnamefont {Lefebvre}}, \bibinfo {author} {\bibfnamefont {F.}~\bibnamefont
  {Lefevre}}, \bibinfo {author} {\bibfnamefont {C.}~\bibnamefont {Levy}},
  \bibinfo {author} {\bibfnamefont {Z.~X.}\ \bibnamefont {Li}}, \bibinfo
  {author} {\bibfnamefont {J.}~\bibnamefont {Lloyd}}, \bibinfo {author}
  {\bibfnamefont {F.}~\bibnamefont {Lott}}, \bibinfo {author} {\bibfnamefont
  {G.}~\bibnamefont {Madec}}, \bibinfo {author} {\bibfnamefont
  {M.}~\bibnamefont {Mancip}}, \bibinfo {author} {\bibfnamefont
  {M.}~\bibnamefont {Marchand}}, \bibinfo {author} {\bibfnamefont
  {S.}~\bibnamefont {Masson}}, \bibinfo {author} {\bibfnamefont
  {Y.}~\bibnamefont {Meurdesoif}}, \bibinfo {author} {\bibfnamefont
  {J.}~\bibnamefont {Mignot}}, \bibinfo {author} {\bibfnamefont
  {I.}~\bibnamefont {Musat}}, \bibinfo {author} {\bibfnamefont
  {S.}~\bibnamefont {Parouty}}, \bibinfo {author} {\bibfnamefont
  {J.}~\bibnamefont {Polcher}}, \bibinfo {author} {\bibfnamefont
  {C.}~\bibnamefont {Rio}}, \bibinfo {author} {\bibfnamefont {M.}~\bibnamefont
  {Schulz}}, \bibinfo {author} {\bibfnamefont {D.}~\bibnamefont {Swingedouw}},
  \bibinfo {author} {\bibfnamefont {S.}~\bibnamefont {Szopa}}, \bibinfo
  {author} {\bibfnamefont {C.}~\bibnamefont {Talandier}}, \bibinfo {author}
  {\bibfnamefont {P.}~\bibnamefont {Terray}}, \bibinfo {author} {\bibfnamefont
  {N.}~\bibnamefont {Viovy}}, \ and\ \bibinfo {author} {\bibfnamefont
  {N.}~\bibnamefont {Vuichard}},\ }\bibfield  {title} {\enquote {\bibinfo
  {title} {Climate change projections using the {{IPSL-CM5 Earth System
  Model}}: From {{CMIP3}} to {{CMIP5}}},}\ }\href {\doibase
  10.1007/s00382-012-1636-1} {\bibfield  {journal} {\bibinfo  {journal}
  {Climate Dynamics}\ }\textbf {\bibinfo {volume} {40}},\ \bibinfo {pages}
  {2123--2165} (\bibinfo {year} {2013})}\BibitemShut {NoStop}%
\bibitem [{\citenamefont {Hourdin}\ \emph {et~al.}(2013)\citenamefont
  {Hourdin}, \citenamefont {Foujols}, \citenamefont {Codron}, \citenamefont
  {Guemas}, \citenamefont {Dufresne}, \citenamefont {Bony}, \citenamefont
  {Denvil}, \citenamefont {Guez}, \citenamefont {Lott}, \citenamefont
  {Ghattas}, \citenamefont {Braconnot}, \citenamefont {Marti}, \citenamefont
  {Meurdesoif},\ and\ \citenamefont {Bopp}}]{hourdin2013}%
  \BibitemOpen
  \bibfield  {author} {\bibinfo {author} {\bibfnamefont {F.}~\bibnamefont
  {Hourdin}}, \bibinfo {author} {\bibfnamefont {M.-A.}\ \bibnamefont
  {Foujols}}, \bibinfo {author} {\bibfnamefont {F.}~\bibnamefont {Codron}},
  \bibinfo {author} {\bibfnamefont {V.}~\bibnamefont {Guemas}}, \bibinfo
  {author} {\bibfnamefont {J.-L.}\ \bibnamefont {Dufresne}}, \bibinfo {author}
  {\bibfnamefont {S.}~\bibnamefont {Bony}}, \bibinfo {author} {\bibfnamefont
  {S.}~\bibnamefont {Denvil}}, \bibinfo {author} {\bibfnamefont
  {L.}~\bibnamefont {Guez}}, \bibinfo {author} {\bibfnamefont {F.}~\bibnamefont
  {Lott}}, \bibinfo {author} {\bibfnamefont {J.}~\bibnamefont {Ghattas}},
  \bibinfo {author} {\bibfnamefont {P.}~\bibnamefont {Braconnot}}, \bibinfo
  {author} {\bibfnamefont {O.}~\bibnamefont {Marti}}, \bibinfo {author}
  {\bibfnamefont {Y.}~\bibnamefont {Meurdesoif}}, \ and\ \bibinfo {author}
  {\bibfnamefont {L.}~\bibnamefont {Bopp}},\ }\bibfield  {title} {\enquote
  {\bibinfo {title} {Impact of the {{LMDZ}} atmospheric grid configuration on
  the climate and sensitivity of the {{IPSL-CM5A}} coupled model},}\ }\href
  {\doibase 10.1007/s00382-012-1411-3} {\bibfield  {journal} {\bibinfo
  {journal} {Climate Dynamics}\ }\textbf {\bibinfo {volume} {40}},\ \bibinfo
  {pages} {2167--2192} (\bibinfo {year} {2013})}\BibitemShut {NoStop}%
\bibitem [{\citenamefont {Sueyoshi}\ \emph {et~al.}(2013)\citenamefont
  {Sueyoshi}, \citenamefont {Ohgaito}, \citenamefont {Yamamoto}, \citenamefont
  {Chikamoto}, \citenamefont {Hajima}, \citenamefont {Okajima}, \citenamefont
  {Yoshimori}, \citenamefont {Abe}, \citenamefont {O'ishi}, \citenamefont
  {Saito}, \citenamefont {Watanabe}, \citenamefont {Kawamiya},\ and\
  \citenamefont {{Abe-Ouchi}}}]{sueyoshi2013}%
  \BibitemOpen
  \bibfield  {author} {\bibinfo {author} {\bibfnamefont {T.}~\bibnamefont
  {Sueyoshi}}, \bibinfo {author} {\bibfnamefont {R.}~\bibnamefont {Ohgaito}},
  \bibinfo {author} {\bibfnamefont {A.}~\bibnamefont {Yamamoto}}, \bibinfo
  {author} {\bibfnamefont {M.~O.}\ \bibnamefont {Chikamoto}}, \bibinfo {author}
  {\bibfnamefont {T.}~\bibnamefont {Hajima}}, \bibinfo {author} {\bibfnamefont
  {H.}~\bibnamefont {Okajima}}, \bibinfo {author} {\bibfnamefont
  {M.}~\bibnamefont {Yoshimori}}, \bibinfo {author} {\bibfnamefont
  {M.}~\bibnamefont {Abe}}, \bibinfo {author} {\bibfnamefont {R.}~\bibnamefont
  {O'ishi}}, \bibinfo {author} {\bibfnamefont {F.}~\bibnamefont {Saito}},
  \bibinfo {author} {\bibfnamefont {S.}~\bibnamefont {Watanabe}}, \bibinfo
  {author} {\bibfnamefont {M.}~\bibnamefont {Kawamiya}}, \ and\ \bibinfo
  {author} {\bibfnamefont {A.}~\bibnamefont {{Abe-Ouchi}}},\ }\bibfield
  {title} {\enquote {\bibinfo {title} {Set-up of the {{PMIP3}} paleoclimate
  experiments conducted using an {{Earth}} system model, {{MIROC-ESM}}},}\
  }\href {\doibase 10.5194/gmd-6-819-2013} {\bibfield  {journal} {\bibinfo
  {journal} {Geoscientific Model Development}\ }\textbf {\bibinfo {volume}
  {6}},\ \bibinfo {pages} {819--836} (\bibinfo {year} {2013})}\BibitemShut
  {NoStop}%
\bibitem [{\citenamefont {Giorgetta}\ \emph {et~al.}(2013)\citenamefont
  {Giorgetta}, \citenamefont {Jungclaus}, \citenamefont {Reick}, \citenamefont
  {Legutke}, \citenamefont {Bader}, \citenamefont {B{\"o}ttinger},
  \citenamefont {Brovkin}, \citenamefont {Crueger}, \citenamefont {Esch},
  \citenamefont {Fieg}, \citenamefont {Glushak}, \citenamefont {Gayler},
  \citenamefont {Haak}, \citenamefont {Hollweg}, \citenamefont {Ilyina},
  \citenamefont {Kinne}, \citenamefont {Kornblueh}, \citenamefont {Matei},
  \citenamefont {Mauritsen}, \citenamefont {Mikolajewicz}, \citenamefont
  {Mueller}, \citenamefont {Notz}, \citenamefont {Pithan}, \citenamefont
  {Raddatz}, \citenamefont {Rast}, \citenamefont {Redler}, \citenamefont
  {Roeckner}, \citenamefont {Schmidt}, \citenamefont {Schnur}, \citenamefont
  {Segschneider}, \citenamefont {Six}, \citenamefont {Stockhause},
  \citenamefont {Timmreck}, \citenamefont {Wegner}, \citenamefont {Widmann},
  \citenamefont {Wieners}, \citenamefont {Claussen}, \citenamefont {Marotzke},\
  and\ \citenamefont {Stevens}}]{giorgetta2013}%
  \BibitemOpen
  \bibfield  {author} {\bibinfo {author} {\bibfnamefont {M.~A.}\ \bibnamefont
  {Giorgetta}}, \bibinfo {author} {\bibfnamefont {J.}~\bibnamefont
  {Jungclaus}}, \bibinfo {author} {\bibfnamefont {C.~H.}\ \bibnamefont
  {Reick}}, \bibinfo {author} {\bibfnamefont {S.}~\bibnamefont {Legutke}},
  \bibinfo {author} {\bibfnamefont {J.}~\bibnamefont {Bader}}, \bibinfo
  {author} {\bibfnamefont {M.}~\bibnamefont {B{\"o}ttinger}}, \bibinfo {author}
  {\bibfnamefont {V.}~\bibnamefont {Brovkin}}, \bibinfo {author} {\bibfnamefont
  {T.}~\bibnamefont {Crueger}}, \bibinfo {author} {\bibfnamefont
  {M.}~\bibnamefont {Esch}}, \bibinfo {author} {\bibfnamefont {K.}~\bibnamefont
  {Fieg}}, \bibinfo {author} {\bibfnamefont {K.}~\bibnamefont {Glushak}},
  \bibinfo {author} {\bibfnamefont {V.}~\bibnamefont {Gayler}}, \bibinfo
  {author} {\bibfnamefont {H.}~\bibnamefont {Haak}}, \bibinfo {author}
  {\bibfnamefont {H.-D.}\ \bibnamefont {Hollweg}}, \bibinfo {author}
  {\bibfnamefont {T.}~\bibnamefont {Ilyina}}, \bibinfo {author} {\bibfnamefont
  {S.}~\bibnamefont {Kinne}}, \bibinfo {author} {\bibfnamefont
  {L.}~\bibnamefont {Kornblueh}}, \bibinfo {author} {\bibfnamefont
  {D.}~\bibnamefont {Matei}}, \bibinfo {author} {\bibfnamefont
  {T.}~\bibnamefont {Mauritsen}}, \bibinfo {author} {\bibfnamefont
  {U.}~\bibnamefont {Mikolajewicz}}, \bibinfo {author} {\bibfnamefont
  {W.}~\bibnamefont {Mueller}}, \bibinfo {author} {\bibfnamefont
  {D.}~\bibnamefont {Notz}}, \bibinfo {author} {\bibfnamefont {F.}~\bibnamefont
  {Pithan}}, \bibinfo {author} {\bibfnamefont {T.}~\bibnamefont {Raddatz}},
  \bibinfo {author} {\bibfnamefont {S.}~\bibnamefont {Rast}}, \bibinfo {author}
  {\bibfnamefont {R.}~\bibnamefont {Redler}}, \bibinfo {author} {\bibfnamefont
  {E.}~\bibnamefont {Roeckner}}, \bibinfo {author} {\bibfnamefont
  {H.}~\bibnamefont {Schmidt}}, \bibinfo {author} {\bibfnamefont
  {R.}~\bibnamefont {Schnur}}, \bibinfo {author} {\bibfnamefont
  {J.}~\bibnamefont {Segschneider}}, \bibinfo {author} {\bibfnamefont {K.~D.}\
  \bibnamefont {Six}}, \bibinfo {author} {\bibfnamefont {M.}~\bibnamefont
  {Stockhause}}, \bibinfo {author} {\bibfnamefont {C.}~\bibnamefont
  {Timmreck}}, \bibinfo {author} {\bibfnamefont {J.}~\bibnamefont {Wegner}},
  \bibinfo {author} {\bibfnamefont {H.}~\bibnamefont {Widmann}}, \bibinfo
  {author} {\bibfnamefont {K.-H.}\ \bibnamefont {Wieners}}, \bibinfo {author}
  {\bibfnamefont {M.}~\bibnamefont {Claussen}}, \bibinfo {author}
  {\bibfnamefont {J.}~\bibnamefont {Marotzke}}, \ and\ \bibinfo {author}
  {\bibfnamefont {B.}~\bibnamefont {Stevens}},\ }\bibfield  {title} {\enquote
  {\bibinfo {title} {Climate and carbon cycle changes from 1850 to 2100 in
  {{MPI-ESM}} simulations for the {{Coupled Model Intercomparison Project}}
  phase 5: {{Climate Changes}} in {{MPI-ESM}}},}\ }\href {\doibase
  10.1002/jame.20038} {\bibfield  {journal} {\bibinfo  {journal} {Journal of
  Advances in Modeling Earth Systems}\ }\textbf {\bibinfo {volume} {5}},\
  \bibinfo {pages} {572--597} (\bibinfo {year} {2013})}\BibitemShut {NoStop}%
\bibitem [{\citenamefont {Jungclaus}\ \emph {et~al.}(2013)\citenamefont
  {Jungclaus}, \citenamefont {Fischer}, \citenamefont {Haak}, \citenamefont
  {Lohmann}, \citenamefont {Marotzke}, \citenamefont {Matei}, \citenamefont
  {Mikolajewicz}, \citenamefont {Notz},\ and\ \citenamefont
  {Storch}}]{jungclaus2013}%
  \BibitemOpen
  \bibfield  {author} {\bibinfo {author} {\bibfnamefont {J.~H.}\ \bibnamefont
  {Jungclaus}}, \bibinfo {author} {\bibfnamefont {N.}~\bibnamefont {Fischer}},
  \bibinfo {author} {\bibfnamefont {H.}~\bibnamefont {Haak}}, \bibinfo {author}
  {\bibfnamefont {K.}~\bibnamefont {Lohmann}}, \bibinfo {author} {\bibfnamefont
  {J.}~\bibnamefont {Marotzke}}, \bibinfo {author} {\bibfnamefont
  {D.}~\bibnamefont {Matei}}, \bibinfo {author} {\bibfnamefont
  {U.}~\bibnamefont {Mikolajewicz}}, \bibinfo {author} {\bibfnamefont
  {D.}~\bibnamefont {Notz}}, \ and\ \bibinfo {author} {\bibfnamefont {J.~S.}\
  \bibnamefont {Storch}},\ }\bibfield  {title} {\enquote {\bibinfo {title}
  {Characteristics of the ocean simulations in the {{Max Planck Institute Ocean
  Model}} ({{MPIOM}}) the ocean component of the {{MPI}}-{{Earth}} system
  model},}\ }\href {\doibase 10.1002/jame.20023} {\bibfield  {journal}
  {\bibinfo  {journal} {Journal of Advances in Modeling Earth Systems}\
  }\textbf {\bibinfo {volume} {5}},\ \bibinfo {pages} {422--446} (\bibinfo
  {year} {2013})}\BibitemShut {NoStop}%
\bibitem [{\citenamefont {Cummins}, \citenamefont {Stephenson},\ and\
  \citenamefont {Stott}(2020)}]{cummins2020a}%
  \BibitemOpen
  \bibfield  {author} {\bibinfo {author} {\bibfnamefont {D.~P.}\ \bibnamefont
  {Cummins}}, \bibinfo {author} {\bibfnamefont {D.~B.}\ \bibnamefont
  {Stephenson}}, \ and\ \bibinfo {author} {\bibfnamefont {P.~A.}\ \bibnamefont
  {Stott}},\ }\bibfield  {title} {\enquote {\bibinfo {title} {Optimal
  {{Estimation}} of {{Stochastic Energy Balance Model Parameters}}},}\ }\href
  {\doibase 10.1175/JCLI-D-19-0589.1} {\bibfield  {journal} {\bibinfo
  {journal} {Journal of Climate}\ }\textbf {\bibinfo {volume} {33}},\ \bibinfo
  {pages} {7909--7926} (\bibinfo {year} {2020})}\BibitemShut {NoStop}%
\bibitem [{\citenamefont {Geoffroy}\ \emph
  {et~al.}(2013{\natexlab{b}})\citenamefont {Geoffroy}, \citenamefont
  {{Saint-Martin}}, \citenamefont {Bellon}, \citenamefont {Voldoire},
  \citenamefont {Olivi{\'e}},\ and\ \citenamefont {Tyt{\'e}ca}}]{geoffroy2013}%
  \BibitemOpen
  \bibfield  {author} {\bibinfo {author} {\bibfnamefont {O.}~\bibnamefont
  {Geoffroy}}, \bibinfo {author} {\bibfnamefont {D.}~\bibnamefont
  {{Saint-Martin}}}, \bibinfo {author} {\bibfnamefont {G.}~\bibnamefont
  {Bellon}}, \bibinfo {author} {\bibfnamefont {A.}~\bibnamefont {Voldoire}},
  \bibinfo {author} {\bibfnamefont {D.~J.~L.}\ \bibnamefont {Olivi{\'e}}}, \
  and\ \bibinfo {author} {\bibfnamefont {S.}~\bibnamefont {Tyt{\'e}ca}},\
  }\bibfield  {title} {\enquote {\bibinfo {title} {Transient {{Climate
  Response}} in a {{Two-Layer Energy-Balance Model}}. {{Part II}}:
  {{Representation}} of the {{Efficacy}} of {{Deep-Ocean Heat Uptake}} and
  {{Validation}} for {{CMIP5 AOGCMs}}},}\ }\href {\doibase
  10.1175/JCLI-D-12-00196.1} {\bibfield  {journal} {\bibinfo  {journal}
  {Journal of Climate}\ }\textbf {\bibinfo {volume} {26}},\ \bibinfo {pages}
  {1859--1876} (\bibinfo {year} {2013}{\natexlab{b}})}\BibitemShut {NoStop}%
\bibitem [{\citenamefont {Gelman}\ and\ \citenamefont
  {Rubin}(1992)}]{gelman1992}%
  \BibitemOpen
  \bibfield  {author} {\bibinfo {author} {\bibfnamefont {A.}~\bibnamefont
  {Gelman}}\ and\ \bibinfo {author} {\bibfnamefont {D.~B.}\ \bibnamefont
  {Rubin}},\ }\bibfield  {title} {\enquote {\bibinfo {title} {Inference from
  {{Iterative Simulation Using Multiple Sequences}}},}\ }\href {\doibase
  10.1214/ss/1177011136} {\bibfield  {journal} {\bibinfo  {journal}
  {Statistical Science}\ }\textbf {\bibinfo {volume} {7}},\ \bibinfo {pages}
  {457--472} (\bibinfo {year} {1992})}\BibitemShut {NoStop}%
\bibitem [{\citenamefont {Brooks}\ and\ \citenamefont
  {Gelman}(1998)}]{brooks1998}%
  \BibitemOpen
  \bibfield  {author} {\bibinfo {author} {\bibfnamefont {S.~P.}\ \bibnamefont
  {Brooks}}\ and\ \bibinfo {author} {\bibfnamefont {A.}~\bibnamefont
  {Gelman}},\ }\bibfield  {title} {\enquote {\bibinfo {title} {General
  {{Methods}} for {{Monitoring Convergence}} of {{Iterative Simulations}}},}\
  }\href {\doibase 10.1080/10618600.1998.10474787} {\bibfield  {journal}
  {\bibinfo  {journal} {Journal of Computational and Graphical Statistics}\
  }\textbf {\bibinfo {volume} {7}},\ \bibinfo {pages} {434--455} (\bibinfo
  {year} {1998})}\BibitemShut {NoStop}%
\bibitem [{\citenamefont {Gelman}(2014)}]{gelman2014}%
  \BibitemOpen
  \bibfield  {author} {\bibinfo {author} {\bibfnamefont {A.}~\bibnamefont
  {Gelman}},\ }\href {\doibase https://doi.org/10.1201/b16018} {\emph {\bibinfo
  {title} {Bayesian Data Analysis}}},\ \bibinfo {edition} {third edition}\
  ed.,\ Chapman \& {{Hall}}/{{CRC}} Texts in Statistical Science\ (\bibinfo
  {publisher} {{CRC Press}},\ \bibinfo {address} {{Boca Raton}},\ \bibinfo
  {year} {2014})\BibitemShut {NoStop}%
\bibitem [{\citenamefont {Plummer}\ \emph {et~al.}(2020)\citenamefont
  {Plummer}, \citenamefont {Best}, \citenamefont {Cowles}, \citenamefont
  {Vines},\ and\ \citenamefont {Plummer}}]{plummer2020}%
  \BibitemOpen
  \bibfield  {author} {\bibinfo {author} {\bibfnamefont {M.}~\bibnamefont
  {Plummer}}, \bibinfo {author} {\bibfnamefont {N.}~\bibnamefont {Best}},
  \bibinfo {author} {\bibfnamefont {K.}~\bibnamefont {Cowles}}, \bibinfo
  {author} {\bibfnamefont {K.}~\bibnamefont {Vines}}, \ and\ \bibinfo {author}
  {\bibfnamefont {M.~M.}\ \bibnamefont {Plummer}},\ }\href
  {https://cran.r-project.org/web/packages/coda/index.html} {\enquote {\bibinfo
  {title} {R package "{{CODA}}"},}\ } (\bibinfo {year} {2020})\BibitemShut
  {NoStop}%
\bibitem [{\citenamefont {Flegal}, \citenamefont {Haran},\ and\ \citenamefont
  {Jones}(2008)}]{flegal2008}%
  \BibitemOpen
  \bibfield  {author} {\bibinfo {author} {\bibfnamefont {J.~M.}\ \bibnamefont
  {Flegal}}, \bibinfo {author} {\bibfnamefont {M.}~\bibnamefont {Haran}}, \
  and\ \bibinfo {author} {\bibfnamefont {G.~L.}\ \bibnamefont {Jones}},\
  }\bibfield  {title} {\enquote {\bibinfo {title} {Markov {{Chain Monte
  Carlo}}: {{Can We Trust}} the {{Third Significant Figure}}?}}\ }\href
  {\doibase 10.1214/08-STS257} {\bibfield  {journal} {\bibinfo  {journal}
  {Statistical Science}\ }\textbf {\bibinfo {volume} {23}} (\bibinfo {year}
  {2008}),\ 10.1214/08-STS257}\BibitemShut {NoStop}%
\bibitem [{\citenamefont {Haran}\ and\ \citenamefont
  {Hughes}(2020)}]{haran2020}%
  \BibitemOpen
  \bibfield  {author} {\bibinfo {author} {\bibfnamefont {M.}~\bibnamefont
  {Haran}}\ and\ \bibinfo {author} {\bibfnamefont {J.}~\bibnamefont {Hughes}},\
  }\href {https://cran.r-project.org/web/packages/batchmeans/index.html}
  {\enquote {\bibinfo {title} {R package "batchmeans: {{Consistent Batch Means
  Estimation}} of {{Monte Carlo}}"},}\ } (\bibinfo {year} {2020})\BibitemShut
  {NoStop}%
\bibitem [{\citenamefont {Laepple}\ and\ \citenamefont
  {Huybers}(2014{\natexlab{b}})}]{laepple2014a}%
  \BibitemOpen
  \bibfield  {author} {\bibinfo {author} {\bibfnamefont {T.}~\bibnamefont
  {Laepple}}\ and\ \bibinfo {author} {\bibfnamefont {P.}~\bibnamefont
  {Huybers}},\ }\bibfield  {title} {\enquote {\bibinfo {title} {Ocean surface
  temperature variability: {{Large}} model data differences at decadal and
  longer periods},}\ }\href {\doibase 10.1073/pnas.1412077111} {\bibfield
  {journal} {\bibinfo  {journal} {Proceedings of the National Academy of
  Sciences}\ }\textbf {\bibinfo {volume} {111}},\ \bibinfo {pages}
  {16682--16687} (\bibinfo {year} {2014}{\natexlab{b}})}\BibitemShut {NoStop}%
\bibitem [{\citenamefont {Bothe}, \citenamefont {Jungclaus},\ and\
  \citenamefont {Zanchettin}(2013)}]{bothe2013}%
  \BibitemOpen
  \bibfield  {author} {\bibinfo {author} {\bibfnamefont {O.}~\bibnamefont
  {Bothe}}, \bibinfo {author} {\bibfnamefont {J.~H.}\ \bibnamefont
  {Jungclaus}}, \ and\ \bibinfo {author} {\bibfnamefont {D.}~\bibnamefont
  {Zanchettin}},\ }\bibfield  {title} {\enquote {\bibinfo {title} {Consistency
  of the multi-model {{CMIP5}}/{{PMIP3-past1000}} ensemble},}\ }\href {\doibase
  10.5194/cp-9-2471-2013} {\bibfield  {journal} {\bibinfo  {journal} {Climate
  of the Past}\ }\textbf {\bibinfo {volume} {9}},\ \bibinfo {pages}
  {2471--2487} (\bibinfo {year} {2013})}\BibitemShut {NoStop}%
\bibitem [{\citenamefont {Zhang}\ \emph {et~al.}(2013)\citenamefont {Zhang},
  \citenamefont {Delworth}, \citenamefont {Sutton}, \citenamefont {Hodson},
  \citenamefont {Dixon}, \citenamefont {Held}, \citenamefont {Kushnir},
  \citenamefont {Marshall}, \citenamefont {Ming}, \citenamefont {Msadek},
  \citenamefont {Robson}, \citenamefont {Rosati}, \citenamefont {Ting},\ and\
  \citenamefont {Vecchi}}]{zhang2013}%
  \BibitemOpen
  \bibfield  {author} {\bibinfo {author} {\bibfnamefont {R.}~\bibnamefont
  {Zhang}}, \bibinfo {author} {\bibfnamefont {T.~L.}\ \bibnamefont {Delworth}},
  \bibinfo {author} {\bibfnamefont {R.}~\bibnamefont {Sutton}}, \bibinfo
  {author} {\bibfnamefont {D.~L.~R.}\ \bibnamefont {Hodson}}, \bibinfo {author}
  {\bibfnamefont {K.~W.}\ \bibnamefont {Dixon}}, \bibinfo {author}
  {\bibfnamefont {I.~M.}\ \bibnamefont {Held}}, \bibinfo {author}
  {\bibfnamefont {Y.}~\bibnamefont {Kushnir}}, \bibinfo {author} {\bibfnamefont
  {J.}~\bibnamefont {Marshall}}, \bibinfo {author} {\bibfnamefont
  {Y.}~\bibnamefont {Ming}}, \bibinfo {author} {\bibfnamefont {R.}~\bibnamefont
  {Msadek}}, \bibinfo {author} {\bibfnamefont {J.}~\bibnamefont {Robson}},
  \bibinfo {author} {\bibfnamefont {A.~J.}\ \bibnamefont {Rosati}}, \bibinfo
  {author} {\bibfnamefont {M.}~\bibnamefont {Ting}}, \ and\ \bibinfo {author}
  {\bibfnamefont {G.~A.}\ \bibnamefont {Vecchi}},\ }\bibfield  {title}
  {\enquote {\bibinfo {title} {Have {{Aerosols Caused}} the {{Observed Atlantic
  Multidecadal Variability}}?}}\ }\href {\doibase 10.1175/JAS-D-12-0331.1}
  {\bibfield  {journal} {\bibinfo  {journal} {Journal of the Atmospheric
  Sciences}\ }\textbf {\bibinfo {volume} {70}},\ \bibinfo {pages} {1135--1144}
  (\bibinfo {year} {2013})}\BibitemShut {NoStop}%
\bibitem [{\citenamefont {Olonscheck}\ and\ \citenamefont
  {Notz}(2017)}]{olonscheck2017}%
  \BibitemOpen
  \bibfield  {author} {\bibinfo {author} {\bibfnamefont {D.}~\bibnamefont
  {Olonscheck}}\ and\ \bibinfo {author} {\bibfnamefont {D.}~\bibnamefont
  {Notz}},\ }\bibfield  {title} {\enquote {\bibinfo {title} {Consistently
  {{Estimating Internal Climate Variability}} from {{Climate Model
  Simulations}}},}\ }\href {\doibase 10.1175/JCLI-D-16-0428.1} {\bibfield
  {journal} {\bibinfo  {journal} {Journal of Climate}\ }\textbf {\bibinfo
  {volume} {30}},\ \bibinfo {pages} {9555--9573} (\bibinfo {year}
  {2017})}\BibitemShut {NoStop}%
\bibitem [{\citenamefont {Lehner}\ \emph {et~al.}(2020)\citenamefont {Lehner},
  \citenamefont {Deser}, \citenamefont {Maher}, \citenamefont {Marotzke},
  \citenamefont {Fischer}, \citenamefont {Brunner}, \citenamefont {Knutti},\
  and\ \citenamefont {Hawkins}}]{lehner2020}%
  \BibitemOpen
  \bibfield  {author} {\bibinfo {author} {\bibfnamefont {F.}~\bibnamefont
  {Lehner}}, \bibinfo {author} {\bibfnamefont {C.}~\bibnamefont {Deser}},
  \bibinfo {author} {\bibfnamefont {N.}~\bibnamefont {Maher}}, \bibinfo
  {author} {\bibfnamefont {J.}~\bibnamefont {Marotzke}}, \bibinfo {author}
  {\bibfnamefont {E.~M.}\ \bibnamefont {Fischer}}, \bibinfo {author}
  {\bibfnamefont {L.}~\bibnamefont {Brunner}}, \bibinfo {author} {\bibfnamefont
  {R.}~\bibnamefont {Knutti}}, \ and\ \bibinfo {author} {\bibfnamefont
  {E.}~\bibnamefont {Hawkins}},\ }\bibfield  {title} {\enquote {\bibinfo
  {title} {Partitioning climate projection uncertainty with multiple large
  ensembles and {{CMIP5}}/6},}\ }\href {\doibase 10.5194/esd-11-491-2020}
  {\bibfield  {journal} {\bibinfo  {journal} {Earth System Dynamics}\ }\textbf
  {\bibinfo {volume} {11}},\ \bibinfo {pages} {491--508} (\bibinfo {year}
  {2020})}\BibitemShut {NoStop}%
\bibitem [{\citenamefont {Mann}\ \emph {et~al.}(2022)\citenamefont {Mann},
  \citenamefont {Steinman}, \citenamefont {Brouillette}, \citenamefont
  {Fernandez},\ and\ \citenamefont {Miller}}]{mann2022}%
  \BibitemOpen
  \bibfield  {author} {\bibinfo {author} {\bibfnamefont {M.~E.}\ \bibnamefont
  {Mann}}, \bibinfo {author} {\bibfnamefont {B.~A.}\ \bibnamefont {Steinman}},
  \bibinfo {author} {\bibfnamefont {D.~J.}\ \bibnamefont {Brouillette}},
  \bibinfo {author} {\bibfnamefont {A.}~\bibnamefont {Fernandez}}, \ and\
  \bibinfo {author} {\bibfnamefont {S.~K.}\ \bibnamefont {Miller}},\ }\bibfield
   {title} {\enquote {\bibinfo {title} {On the {{Estimation}} of {{Internal
  Climate Variability During}} the {{Preindustrial Past Millennium}}},}\ }\href
  {\doibase 10.1029/2021GL096596} {\bibfield  {journal} {\bibinfo  {journal}
  {Geophysical Research Letters}\ }\textbf {\bibinfo {volume} {49}} (\bibinfo
  {year} {2022}),\ 10.1029/2021GL096596}\BibitemShut {NoStop}%
\bibitem [{\citenamefont {Mann}, \citenamefont {Steinman},\ and\ \citenamefont
  {Miller}(2014)}]{mann2014}%
  \BibitemOpen
  \bibfield  {author} {\bibinfo {author} {\bibfnamefont {M.~E.}\ \bibnamefont
  {Mann}}, \bibinfo {author} {\bibfnamefont {B.~A.}\ \bibnamefont {Steinman}},
  \ and\ \bibinfo {author} {\bibfnamefont {S.~K.}\ \bibnamefont {Miller}},\
  }\bibfield  {title} {\enquote {\bibinfo {title} {On forced temperature
  changes, internal variability, and the {{AMO}}},}\ }\href {\doibase
  10.1002/2014GL059233} {\bibfield  {journal} {\bibinfo  {journal} {Geophysical
  Research Letters}\ }\textbf {\bibinfo {volume} {41}},\ \bibinfo {pages}
  {3211--3219} (\bibinfo {year} {2014})}\BibitemShut {NoStop}%
\bibitem [{\citenamefont {Crowley}(2000)}]{crowley2000}%
  \BibitemOpen
  \bibfield  {author} {\bibinfo {author} {\bibfnamefont {T.~J.}\ \bibnamefont
  {Crowley}},\ }\bibfield  {title} {\enquote {\bibinfo {title} {Causes of
  {{Climate Change Over}} the {{Past}} 1000 {{Years}}},}\ }\href {\doibase
  10.1126/science.289.5477.270} {\bibfield  {journal} {\bibinfo  {journal}
  {Science}\ }\textbf {\bibinfo {volume} {289}},\ \bibinfo {pages} {270--277}
  (\bibinfo {year} {2000})}\BibitemShut {NoStop}%
\bibitem [{\citenamefont {Kay}\ \emph {et~al.}(2015)\citenamefont {Kay},
  \citenamefont {Deser}, \citenamefont {Phillips}, \citenamefont {Mai},
  \citenamefont {Hannay}, \citenamefont {Strand}, \citenamefont {Arblaster},
  \citenamefont {Bates}, \citenamefont {Danabasoglu}, \citenamefont {Edwards},
  \citenamefont {Holland}, \citenamefont {Kushner}, \citenamefont {Lamarque},
  \citenamefont {Lawrence}, \citenamefont {Lindsay}, \citenamefont {Middleton},
  \citenamefont {Munoz}, \citenamefont {Neale}, \citenamefont {Oleson},
  \citenamefont {Polvani},\ and\ \citenamefont {Vertenstein}}]{kay2015}%
  \BibitemOpen
  \bibfield  {author} {\bibinfo {author} {\bibfnamefont {J.~E.}\ \bibnamefont
  {Kay}}, \bibinfo {author} {\bibfnamefont {C.}~\bibnamefont {Deser}}, \bibinfo
  {author} {\bibfnamefont {A.}~\bibnamefont {Phillips}}, \bibinfo {author}
  {\bibfnamefont {A.}~\bibnamefont {Mai}}, \bibinfo {author} {\bibfnamefont
  {C.}~\bibnamefont {Hannay}}, \bibinfo {author} {\bibfnamefont
  {G.}~\bibnamefont {Strand}}, \bibinfo {author} {\bibfnamefont {J.~M.}\
  \bibnamefont {Arblaster}}, \bibinfo {author} {\bibfnamefont {S.~C.}\
  \bibnamefont {Bates}}, \bibinfo {author} {\bibfnamefont {G.}~\bibnamefont
  {Danabasoglu}}, \bibinfo {author} {\bibfnamefont {J.}~\bibnamefont
  {Edwards}}, \bibinfo {author} {\bibfnamefont {M.}~\bibnamefont {Holland}},
  \bibinfo {author} {\bibfnamefont {P.}~\bibnamefont {Kushner}}, \bibinfo
  {author} {\bibfnamefont {J.-F.}\ \bibnamefont {Lamarque}}, \bibinfo {author}
  {\bibfnamefont {D.}~\bibnamefont {Lawrence}}, \bibinfo {author}
  {\bibfnamefont {K.}~\bibnamefont {Lindsay}}, \bibinfo {author} {\bibfnamefont
  {A.}~\bibnamefont {Middleton}}, \bibinfo {author} {\bibfnamefont
  {E.}~\bibnamefont {Munoz}}, \bibinfo {author} {\bibfnamefont
  {R.}~\bibnamefont {Neale}}, \bibinfo {author} {\bibfnamefont
  {K.}~\bibnamefont {Oleson}}, \bibinfo {author} {\bibfnamefont
  {L.}~\bibnamefont {Polvani}}, \ and\ \bibinfo {author} {\bibfnamefont
  {M.}~\bibnamefont {Vertenstein}},\ }\bibfield  {title} {\enquote {\bibinfo
  {title} {The {{Community Earth System Model}} ({{CESM}}) {{Large Ensemble
  Project}}: {{A Community Resource}} for {{Studying Climate Change}} in the
  {{Presence}} of {{Internal Climate Variability}}},}\ }\href {\doibase
  10.1175/BAMS-D-13-00255.1} {\bibfield  {journal} {\bibinfo  {journal}
  {Bulletin of the American Meteorological Society}\ }\textbf {\bibinfo
  {volume} {96}},\ \bibinfo {pages} {1333--1349} (\bibinfo {year}
  {2015})}\BibitemShut {NoStop}%
\bibitem [{\citenamefont {MacMynowski}, \citenamefont {Shin},\ and\
  \citenamefont {Caldeira}(2011)}]{macmynowski2011}%
  \BibitemOpen
  \bibfield  {author} {\bibinfo {author} {\bibfnamefont {D.~G.}\ \bibnamefont
  {MacMynowski}}, \bibinfo {author} {\bibfnamefont {H.-J.}\ \bibnamefont
  {Shin}}, \ and\ \bibinfo {author} {\bibfnamefont {K.}~\bibnamefont
  {Caldeira}},\ }\bibfield  {title} {\enquote {\bibinfo {title} {The frequency
  response of temperature and precipitation in a climate model},}\ }\href
  {\doibase 10.1029/2011GL048623} {\bibfield  {journal} {\bibinfo  {journal}
  {Geophysical Research Letters}\ }\textbf {\bibinfo {volume} {38}},\ \bibinfo
  {pages} {n/a--n/a} (\bibinfo {year} {2011})}\BibitemShut {NoStop}%
\bibitem [{\citenamefont {Maher}\ \emph {et~al.}(2018)\citenamefont {Maher},
  \citenamefont {Matei}, \citenamefont {Milinski},\ and\ \citenamefont
  {Marotzke}}]{maher2018}%
  \BibitemOpen
  \bibfield  {author} {\bibinfo {author} {\bibfnamefont {N.}~\bibnamefont
  {Maher}}, \bibinfo {author} {\bibfnamefont {D.}~\bibnamefont {Matei}},
  \bibinfo {author} {\bibfnamefont {S.}~\bibnamefont {Milinski}}, \ and\
  \bibinfo {author} {\bibfnamefont {J.}~\bibnamefont {Marotzke}},\ }\bibfield
  {title} {\enquote {\bibinfo {title} {{{ENSO Change}} in {{Climate
  Projections}}: {{Forced Response}} or {{Internal Variability}}?}}\ }\href
  {\doibase 10.1029/2018GL079764} {\bibfield  {journal} {\bibinfo  {journal}
  {Geophysical Research Letters}\ }\textbf {\bibinfo {volume} {45}} (\bibinfo
  {year} {2018}),\ 10.1029/2018GL079764}\BibitemShut {NoStop}%
\bibitem [{\citenamefont {Bonnet}\ \emph {et~al.}(2021)\citenamefont {Bonnet},
  \citenamefont {Swingedouw}, \citenamefont {Gastineau}, \citenamefont
  {Boucher}, \citenamefont {Deshayes}, \citenamefont {Hourdin}, \citenamefont
  {Mignot}, \citenamefont {Servonnat},\ and\ \citenamefont
  {Sima}}]{bonnet2021}%
  \BibitemOpen
  \bibfield  {author} {\bibinfo {author} {\bibfnamefont {R.}~\bibnamefont
  {Bonnet}}, \bibinfo {author} {\bibfnamefont {D.}~\bibnamefont {Swingedouw}},
  \bibinfo {author} {\bibfnamefont {G.}~\bibnamefont {Gastineau}}, \bibinfo
  {author} {\bibfnamefont {O.}~\bibnamefont {Boucher}}, \bibinfo {author}
  {\bibfnamefont {J.}~\bibnamefont {Deshayes}}, \bibinfo {author}
  {\bibfnamefont {F.}~\bibnamefont {Hourdin}}, \bibinfo {author} {\bibfnamefont
  {J.}~\bibnamefont {Mignot}}, \bibinfo {author} {\bibfnamefont
  {J.}~\bibnamefont {Servonnat}}, \ and\ \bibinfo {author} {\bibfnamefont
  {A.}~\bibnamefont {Sima}},\ }\bibfield  {title} {\enquote {\bibinfo {title}
  {Increased risk of near term global warming due to a recent {{AMOC}}
  weakening},}\ }\href {\doibase 10.1038/s41467-021-26370-0} {\bibfield
  {journal} {\bibinfo  {journal} {Nature Communications}\ }\textbf {\bibinfo
  {volume} {12}},\ \bibinfo {pages} {6108} (\bibinfo {year}
  {2021})}\BibitemShut {NoStop}%
\bibitem [{\citenamefont {Ellerhoff}\ \emph {et~al.}(2022)\citenamefont
  {Ellerhoff}, \citenamefont {Kirschner}, \citenamefont {Ziegler},
  \citenamefont {Holloway}, \citenamefont {Sime},\ and\ \citenamefont
  {Rehfeld}}]{ellerhoff2022a}%
  \BibitemOpen
  \bibfield  {author} {\bibinfo {author} {\bibfnamefont {B.}~\bibnamefont
  {Ellerhoff}}, \bibinfo {author} {\bibfnamefont {M.~J.}\ \bibnamefont
  {Kirschner}}, \bibinfo {author} {\bibfnamefont {E.}~\bibnamefont {Ziegler}},
  \bibinfo {author} {\bibfnamefont {M.~D.}\ \bibnamefont {Holloway}}, \bibinfo
  {author} {\bibfnamefont {L.}~\bibnamefont {Sime}}, \ and\ \bibinfo {author}
  {\bibfnamefont {K.}~\bibnamefont {Rehfeld}},\ }\bibfield  {title} {\enquote
  {\bibinfo {title} {Contrasting {{State}}-{{Dependent Effects}} of {{Natural
  Forcing}} on {{Global}} and {{Local Climate Variability}}},}\ }\href
  {\doibase 10.1029/2022GL098335} {\bibfield  {journal} {\bibinfo  {journal}
  {Geophysical Research Letters}\ }\textbf {\bibinfo {volume} {49}} (\bibinfo
  {year} {2022}),\ 10.1029/2022GL098335}\BibitemShut {NoStop}%
\bibitem [{\citenamefont {Lovejoy}(2021)}]{lovejoy2021}%
  \BibitemOpen
  \bibfield  {author} {\bibinfo {author} {\bibfnamefont {S.}~\bibnamefont
  {Lovejoy}},\ }\bibfield  {title} {\enquote {\bibinfo {title} {The half-order
  energy balance equation \textendash{} {{Part}} 1: {{The}} homogeneous
  {{HEBE}} and long memories},}\ }\href {\doibase 10.5194/esd-12-469-2021}
  {\bibfield  {journal} {\bibinfo  {journal} {Earth System Dynamics}\ }\textbf
  {\bibinfo {volume} {12}},\ \bibinfo {pages} {469--487} (\bibinfo {year}
  {2021})}\BibitemShut {NoStop}%
\bibitem [{\citenamefont {Yan}, \citenamefont {Zhang},\ and\ \citenamefont
  {Knutson}(2018)}]{yan2018}%
  \BibitemOpen
  \bibfield  {author} {\bibinfo {author} {\bibfnamefont {X.}~\bibnamefont
  {Yan}}, \bibinfo {author} {\bibfnamefont {R.}~\bibnamefont {Zhang}}, \ and\
  \bibinfo {author} {\bibfnamefont {T.~R.}\ \bibnamefont {Knutson}},\
  }\bibfield  {title} {\enquote {\bibinfo {title} {Underestimated {{AMOC
  Variability}} and {{Implications}} for {{AMV}} and {{Predictability}} in
  {{CMIP Models}}},}\ }\href {\doibase 10.1029/2018GL077378} {\bibfield
  {journal} {\bibinfo  {journal} {Geophysical Research Letters}\ }\textbf
  {\bibinfo {volume} {45}},\ \bibinfo {pages} {4319--4328} (\bibinfo {year}
  {2018})}\BibitemShut {NoStop}%
\bibitem [{\citenamefont {Schurer}\ \emph {et~al.}(2013)\citenamefont
  {Schurer}, \citenamefont {Hegerl}, \citenamefont {Mann}, \citenamefont
  {Tett},\ and\ \citenamefont {Phipps}}]{schurer2013}%
  \BibitemOpen
  \bibfield  {author} {\bibinfo {author} {\bibfnamefont {A.~P.}\ \bibnamefont
  {Schurer}}, \bibinfo {author} {\bibfnamefont {G.~C.}\ \bibnamefont {Hegerl}},
  \bibinfo {author} {\bibfnamefont {M.~E.}\ \bibnamefont {Mann}}, \bibinfo
  {author} {\bibfnamefont {S.~F.~B.}\ \bibnamefont {Tett}}, \ and\ \bibinfo
  {author} {\bibfnamefont {S.~J.}\ \bibnamefont {Phipps}},\ }\bibfield  {title}
  {\enquote {\bibinfo {title} {Separating {{Forced}} from {{Chaotic Climate
  Variability}} over the {{Past Millennium}}},}\ }\href {\doibase
  10.1175/JCLI-D-12-00826.1} {\bibfield  {journal} {\bibinfo  {journal}
  {Journal of Climate}\ }\textbf {\bibinfo {volume} {26}},\ \bibinfo {pages}
  {6954--6973} (\bibinfo {year} {2013})}\BibitemShut {NoStop}%
\bibitem [{\citenamefont {Chylek}\ \emph {et~al.}(2020)\citenamefont {Chylek},
  \citenamefont {Folland}, \citenamefont {Klett},\ and\ \citenamefont
  {Dubey}}]{chylek2020}%
  \BibitemOpen
  \bibfield  {author} {\bibinfo {author} {\bibfnamefont {P.}~\bibnamefont
  {Chylek}}, \bibinfo {author} {\bibfnamefont {C.}~\bibnamefont {Folland}},
  \bibinfo {author} {\bibfnamefont {J.~D.}\ \bibnamefont {Klett}}, \ and\
  \bibinfo {author} {\bibfnamefont {M.~K.}\ \bibnamefont {Dubey}},\ }\bibfield
  {title} {\enquote {\bibinfo {title} {{{CMIP5 Climate Models Overestimate
  Cooling}} by {{Volcanic Aerosols}}},}\ }\href {\doibase 10.1029/2020GL087047}
  {\bibfield  {journal} {\bibinfo  {journal} {Geophysical Research Letters}\
  }\textbf {\bibinfo {volume} {47}} (\bibinfo {year} {2020}),\
  10.1029/2020GL087047}\BibitemShut {NoStop}%
\bibitem [{\citenamefont {Hopcroft}\ \emph {et~al.}(2018)\citenamefont
  {Hopcroft}, \citenamefont {Kandlbauer}, \citenamefont {Valdes},\ and\
  \citenamefont {Sparks}}]{hopcroft2018}%
  \BibitemOpen
  \bibfield  {author} {\bibinfo {author} {\bibfnamefont {P.~O.}\ \bibnamefont
  {Hopcroft}}, \bibinfo {author} {\bibfnamefont {J.}~\bibnamefont
  {Kandlbauer}}, \bibinfo {author} {\bibfnamefont {P.~J.}\ \bibnamefont
  {Valdes}}, \ and\ \bibinfo {author} {\bibfnamefont {R.~S.~J.}\ \bibnamefont
  {Sparks}},\ }\bibfield  {title} {\enquote {\bibinfo {title} {Reduced cooling
  following future volcanic eruptions},}\ }\href {\doibase
  10.1007/s00382-017-3964-7} {\bibfield  {journal} {\bibinfo  {journal}
  {Climate Dynamics}\ }\textbf {\bibinfo {volume} {51}},\ \bibinfo {pages}
  {1449--1463} (\bibinfo {year} {2018})}\BibitemShut {NoStop}%
\bibitem [{\citenamefont {Kunz}\ and\ \citenamefont
  {Laepple}(2021)}]{kunz2021}%
  \BibitemOpen
  \bibfield  {author} {\bibinfo {author} {\bibfnamefont {T.}~\bibnamefont
  {Kunz}}\ and\ \bibinfo {author} {\bibfnamefont {T.}~\bibnamefont {Laepple}},\
  }\bibfield  {title} {\enquote {\bibinfo {title} {Frequency-{{Dependent
  Estimation}} of {{Effective Spatial Degrees}} of {{Freedom}}},}\ }\href
  {\doibase 10.1175/JCLI-D-20-0228.1} {\bibfield  {journal} {\bibinfo
  {journal} {Journal of Climate}\ }\textbf {\bibinfo {volume} {34}},\ \bibinfo
  {pages} {7373--7388} (\bibinfo {year} {2021})}\BibitemShut {NoStop}%
\bibitem [{\citenamefont {Smith}, \citenamefont {Gregory},\ and\ \citenamefont
  {Osprey}(2008)}]{smith2008}%
  \BibitemOpen
  \bibfield  {author} {\bibinfo {author} {\bibfnamefont {R.~S.}\ \bibnamefont
  {Smith}}, \bibinfo {author} {\bibfnamefont {J.~M.}\ \bibnamefont {Gregory}},
  \ and\ \bibinfo {author} {\bibfnamefont {A.}~\bibnamefont {Osprey}},\
  }\bibfield  {title} {\enquote {\bibinfo {title} {A description of the
  {{FAMOUS}} (version {{XDBUA}}) climate model and control run},}\ }\href
  {\doibase 10.5194/gmd-1-53-2008} {\bibfield  {journal} {\bibinfo  {journal}
  {Geoscientific Model Development}\ }\textbf {\bibinfo {volume} {1}},\
  \bibinfo {pages} {53--68} (\bibinfo {year} {2008})}\BibitemShut {NoStop}%
\bibitem [{\citenamefont {Guo}\ \emph {et~al.}(2016)\citenamefont {Guo},
  \citenamefont {Lee}, \citenamefont {Sakrejda}, \citenamefont {Gabry},
  \citenamefont {Goodrich}, \citenamefont {De~Guzman}, \citenamefont {Niebler},
  \citenamefont {Heller},\ and\ \citenamefont {Fletcher}}]{guo2016rstan}%
  \BibitemOpen
  \bibfield  {author} {\bibinfo {author} {\bibfnamefont {J.}~\bibnamefont
  {Guo}}, \bibinfo {author} {\bibfnamefont {D.}~\bibnamefont {Lee}}, \bibinfo
  {author} {\bibfnamefont {K.}~\bibnamefont {Sakrejda}}, \bibinfo {author}
  {\bibfnamefont {J.}~\bibnamefont {Gabry}}, \bibinfo {author} {\bibfnamefont
  {B.}~\bibnamefont {Goodrich}}, \bibinfo {author} {\bibfnamefont
  {J.}~\bibnamefont {De~Guzman}}, \bibinfo {author} {\bibfnamefont
  {E.}~\bibnamefont {Niebler}}, \bibinfo {author} {\bibfnamefont
  {T.}~\bibnamefont {Heller}}, \ and\ \bibinfo {author} {\bibfnamefont
  {J.}~\bibnamefont {Fletcher}},\ }\bibfield  {title} {\enquote {\bibinfo
  {title} {Rstan: {{R}} interface to stan},}\ }\href
  {https://mc-stan.org/rstan/} {\bibfield  {journal} {\bibinfo  {journal} {R}\
  }\textbf {\bibinfo {volume} {534}},\ \bibinfo {pages} {0--3} (\bibinfo {year}
  {2016})}\BibitemShut {NoStop}%
\bibitem [{\citenamefont {Lykkegaard}\ \emph {et~al.}(2022)\citenamefont
  {Lykkegaard}, \citenamefont {Dodwell}, \citenamefont {Fox}, \citenamefont
  {Mingas},\ and\ \citenamefont {Scheichl}}]{lykkegaard2022}%
  \BibitemOpen
  \bibfield  {author} {\bibinfo {author} {\bibfnamefont {M.~B.}\ \bibnamefont
  {Lykkegaard}}, \bibinfo {author} {\bibfnamefont {T.~J.}\ \bibnamefont
  {Dodwell}}, \bibinfo {author} {\bibfnamefont {C.}~\bibnamefont {Fox}},
  \bibinfo {author} {\bibfnamefont {G.}~\bibnamefont {Mingas}}, \ and\ \bibinfo
  {author} {\bibfnamefont {R.}~\bibnamefont {Scheichl}},\ }\href
  {https://arxiv.org/abs/2202.03876} {\enquote {\bibinfo {title} {Multilevel
  {{Delayed Acceptance MCMC}}},}\ } (\bibinfo {year} {2022})\BibitemShut
  {NoStop}%
\bibitem [{\citenamefont {Lucarini}(2018)}]{lucarini2018}%
  \BibitemOpen
  \bibfield  {author} {\bibinfo {author} {\bibfnamefont {V.}~\bibnamefont
  {Lucarini}},\ }\bibfield  {title} {\enquote {\bibinfo {title} {Revising and
  {{Extending}} the {{Linear Response Theory}} for {{Statistical Mechanical
  Systems}}: {{Evaluating Observables}} as {{Predictors}} and
  {{Predictands}}},}\ }\href {\doibase 10.1007/s10955-018-2151-5} {\bibfield
  {journal} {\bibinfo  {journal} {Journal of Statistical Physics}\ }\textbf
  {\bibinfo {volume} {173}},\ \bibinfo {pages} {1698--1721} (\bibinfo {year}
  {2018})}\BibitemShut {NoStop}%
\bibitem [{\citenamefont {Torres~Mendon{\c c}a}, \citenamefont {Pongratz},\
  and\ \citenamefont {Reick}(2021)}]{torresmendonca2021}%
  \BibitemOpen
  \bibfield  {author} {\bibinfo {author} {\bibfnamefont {G.~L.}\ \bibnamefont
  {Torres~Mendon{\c c}a}}, \bibinfo {author} {\bibfnamefont {J.}~\bibnamefont
  {Pongratz}}, \ and\ \bibinfo {author} {\bibfnamefont {C.~H.}\ \bibnamefont
  {Reick}},\ }\bibfield  {title} {\enquote {\bibinfo {title} {Identification of
  linear response functions from arbitrary perturbation experiments in the
  presence of noise \textendash{} {{Part}} 1: {{Method}} development and toy
  model demonstration},}\ }\href {\doibase 10.5194/npg-28-501-2021} {\bibfield
  {journal} {\bibinfo  {journal} {Nonlinear Processes in Geophysics}\ }\textbf
  {\bibinfo {volume} {28}},\ \bibinfo {pages} {501--532} (\bibinfo {year}
  {2021})}\BibitemShut {NoStop}%
\bibitem [{\citenamefont {{Otto-Bliesner}}\ \emph {et~al.}(2016)\citenamefont
  {{Otto-Bliesner}}, \citenamefont {Brady}, \citenamefont {Fasullo},
  \citenamefont {Jahn}, \citenamefont {Landrum}, \citenamefont {Stevenson},
  \citenamefont {Rosenbloom}, \citenamefont {Mai},\ and\ \citenamefont
  {Strand}}]{otto-bliesner2016}%
  \BibitemOpen
  \bibfield  {author} {\bibinfo {author} {\bibfnamefont {B.~L.}\ \bibnamefont
  {{Otto-Bliesner}}}, \bibinfo {author} {\bibfnamefont {E.~C.}\ \bibnamefont
  {Brady}}, \bibinfo {author} {\bibfnamefont {J.}~\bibnamefont {Fasullo}},
  \bibinfo {author} {\bibfnamefont {A.}~\bibnamefont {Jahn}}, \bibinfo {author}
  {\bibfnamefont {L.}~\bibnamefont {Landrum}}, \bibinfo {author} {\bibfnamefont
  {S.}~\bibnamefont {Stevenson}}, \bibinfo {author} {\bibfnamefont
  {N.}~\bibnamefont {Rosenbloom}}, \bibinfo {author} {\bibfnamefont
  {A.}~\bibnamefont {Mai}}, \ and\ \bibinfo {author} {\bibfnamefont
  {G.}~\bibnamefont {Strand}},\ }\bibfield  {title} {\enquote {\bibinfo {title}
  {Climate {{Variability}} and {{Change}} since 850 {{CE}}: {{An Ensemble
  Approach}} with the {{Community Earth System Model}}},}\ }\href {\doibase
  10.1175/BAMS-D-14-00233.1} {\bibfield  {journal} {\bibinfo  {journal}
  {Bulletin of the American Meteorological Society}\ }\textbf {\bibinfo
  {volume} {97}},\ \bibinfo {pages} {735--754} (\bibinfo {year}
  {2016})}\BibitemShut {NoStop}%
\end{thebibliography}
\end{document}